\documentclass[onecolumn]{IEEEtran}
\usepackage{cite}
\ifCLASSINFOpdf
\usepackage[pdftex]{graphicx}
\graphicspath{./figures}
\DeclareGraphicsExtensions{.pdf,.jpeg,.png}
\else
\fi
\usepackage[cmex10]{amsmath}
\usepackage{array}
\usepackage{url}
\usepackage[utf8]{inputenc}  
\usepackage[T1]{fontenc} 
\usepackage{mathtools}
\usepackage{amssymb,amsfonts,amsmath}
\usepackage{dsfont}
\usepackage{stmaryrd}
\usepackage{bm}
\usepackage{hyperref}
\DeclareUnicodeCharacter{00A0}{~}

\def \({\left(}
\def \){\right)}
\def \[{\left[}
\def \]{\right]}
\newcommand{\tbf}[1]{{\textbf{#1}}}

\newcommand{\defeq}{\vcentcolon=}

\newcommand{\bY}{{\textbf {Y}}}

\newcommand{\bZ}{{\textbf {Z}}}

\newcommand{\bX}{{\textbf {X}}}
\newcommand{\bx}{{\textbf {x}}}

\newcommand{\by}{{\textbf {y}}}
\newcommand{\bz}{{\textbf {z}}}

\newcommand{\bs}{{\textbf {s}}}
\newcommand{\bS}{{\textbf {S}}}

\newcommand{\bR}{{\textbf {R}}}

\newcommand{\ba}{{\textbf {a}}}

\newcommand{\be}{\begin{equation}}
\newcommand{\ee}{\end{equation}}

\newcommand\smallO{
  \mathchoice
    {{\scriptstyle\mathcal{O}}}
    {{\scriptstyle\mathcal{O}}}
    {{\scriptscriptstyle\mathcal{O}}}
    {\scalebox{.7}{$\scriptscriptstyle\mathcal{O}$}}
  }
\newcommand{\bea}{\begin{align}}
\newcommand{\eea}{\end{align}}

\newtheorem{theorem}{Theorem}[section]
\newtheorem{lemma}[theorem]{\textbf{Lemma}}
\newtheorem{thm}[theorem]{\textbf{Theorem}}
\newtheorem{remark}[theorem]{\textbf{Remark}}
\newtheorem{proposition}[theorem]{\textbf{Proposition}}
\newtheorem{corollary}[theorem]{\textbf{Corollary}}
\newtheorem{definition}[theorem]{\textbf{Definition}}

\DeclareMathAlphabet{\varmathbb}{U}{bbold}{m}{n}

\newcommand{\EE}{\mathbb{E}}

\begin{document}
\title{Mutual Information and \\Optimality of Approximate Message-Passing \\in Random Linear Estimation}

\author{Jean~Barbier,
		Nicolas~Macris,
		Mohamad~Dia
        and~Florent~Krzakala
\thanks{J. Barbier is with The Abdus Salam International Center for Theoretical Physics, Trieste, Italy. jbarbier@ictp.it}
\thanks{N. Macris is with the Laboratoire de Théorie des Communications, Facult\'e Informatique et Communications, Ecole Polytechnique F\'ed\'erale de Lausanne, Switzerland. nicolas.macris@epfl.ch}
\thanks{M. Dia is with the Institute for Data Science (i4DS), University of Applied Sciences Northwestern, Switzerland. dia.mohamad@gmail.com}
\thanks{F. Krzakala is with the Laboratoire de Physique de l'\'Ecole normale supérieure, PSL Reseach University, Sorbonne Universités, UMR 8550 CNRS \& UPMC, Universit\'e Pierre et Marie Curie, CNRS, France. florent.krzakala@ens.fr}
\thanks{Copyright (c) 2017 IEEE. Personal use of this material is permitted. However, permission to use this material for any other purposes must be obtained from the IEEE by sending a request to pubs-permissions@ieee.org}
\thanks{This paper was presented in part at the 54th Annual Allerton Conference on Communication, Control, and Computing (Allerton), 2016.}
}
\maketitle
\IEEEpeerreviewmaketitle
\begin{abstract}
We consider the estimation of a signal from the knowledge of its noisy linear random Gaussian projections. A few examples where this problem is relevant are compressed sensing, sparse superposition codes, and code division multiple access. There has been a number of works considering the mutual information for this problem using the replica method from statistical physics. Here we put these considerations on a firm rigorous basis. First, we show, using a Guerra-Toninelli type interpolation, that the replica formula yields an upper bound to the exact mutual information. Secondly, for many relevant practical cases, we present a converse lower bound via a method that uses spatial coupling, state evolution analysis and the I-MMSE theorem. This yields a single letter formula for the mutual information and the minimal-mean-square error for random Gaussian linear estimation of all discrete bounded signals. In addition, we prove that the low complexity approximate message-passing algorithm is optimal outside of the so-called hard phase, in the sense that it asymptotically reaches the minimal-mean-square error. 

In this work spatial coupling is used primarily as a proof technique. However our results also prove two important features of spatially coupled noisy linear random Gaussian estimation. First there is no algorithmically hard phase. This means that for such systems approximate message-passing always reaches the minimal-mean-square error. Secondly, in the limit of infinitely long coupled chain, the mutual information associated to spatially coupled systems is the same as the one of uncoupled linear random Gaussian estimation.
\end{abstract}
\tableofcontents
\section{Introduction}
%
Random linear projections and random matrices are ubiquitous in
computer science and play an important role in machine learning, statistics and communications. In particular, the task of estimating a signal from  linear random projections has a myriad of
applications such as compressed sensing (CS)~\cite{candes2006near}, code
division multiple access (CDMA) in
communications~\cite{verdu1999spectral}, error correction via sparse superposition codes~\cite{barron2010sparse}, or Boolean group testing~\cite{atia2012boolean}. It is thus natural to ask what are the
 information theoretic limits for the estimation of a
signal from the knowledge of its noisy random linear
projections.

A particularly influential approach to this question has been through
the use of the replica method of statistical physics~\cite{mezard1990spin}, which allows to compute non rigorously the
mutual information (MI) and the associated theoretically achievable minimal-mean-square error (MMSE). The replica method typically predicts the optimal performance through the solution 
of non-linear equations
which interestingly coincide, for a range of parameters, with the predictions for the performance of a message-passing
 algorithm. 
In this context the algorithm is usually called {\it approximate 
message-passing} (AMP)~\cite{donoho2009message,krzakala2012statistical,krzakala2012probabilistic}. 

In this contribution we prove rigorously that the replica formula for the MI is 
asymptotically exact in the case of random Gaussian linear projections. For example, our results put on a firm 
rigorous basis the Tanaka formula for CDMA~\cite{tanaka2002statistical} and allow to rigorously obtain the Bayesian MMSE in CS. 
In addition, we carefully discuss algorithmic consequences of our analysis, and prove that AMP reaches the MMSE for a large 
class of such problems except for a region called the \emph{hard phase}. While AMP is an efficient low complexity algorithm, in the hard phase there is no known polynomial complexity local algorithm that allows to reach the MMSE, and it is believed that no such algorithms exist (hence the name ``hard phase'').

Plenty of papers about structured linear problems make use of the replica method. In statistical physics, these date back to the late 80's with the study of the perceptron and neural networks~\cite{gardner1988space,gardner1988optimal,mezard1989space}. Of particular influence has been the work of Tanaka on CDMA~\cite{tanaka2002statistical} which has opened the way to a large set of contributions in information theory~\cite{guo2003multiuser,guo2005randomly}. In particular, the MI (or the free energy) in CS has been considered in a number of publications, e.g.~\cite{guo2009single,rangan2009asymptotic,kabashima2009typical,ganguli2010statistical,wu2012optimal,krzakala2012statistical,tulino2013support,krzakala2012probabilistic}.

In a very interesting line of work, the replica formula has emerged following the study of the AMP algorithm. Again, the story of this algorithm is deeply rooted in statistical physics, with the work of Thouless, Anderson and Palmer~\cite{thouless1977solution} (thus the name ``TAP'' sometimes given to this approach). The earlier version, to the best of our knowledge, appeared in the late 80's in the context of the perceptron problem~\cite{mezard1989space}. For linear estimation, it was again developed initially in the context of CDMA~\cite{kabashima2003cdma}. It is, however, only after the application of this approach to CS~\cite{donoho2009message} that the method has gained its current popularity. Of particular importance has been the development of the rigorous proof of {\it state evolution} (SE), an iterative equation that allows to track the performance of AMP, using techniques developed by~\cite{bayati2011dynamics} and~\cite{bayati2015universality}. Such techniques have their roots in the analysis of iterative forms of the TAP equations by Bolthausen \cite{Bolthausen2014}. 
 Interestingly, the SE fixed points correspond to the extrema of the replica symmetric (RS) potential computed using the replica method, strongly hinting that AMP achieves the MMSE for many problems 
 where it reaches the global minimum.

While our proof technique uses AMP and SE, it is based on two important additional ingredients. The first is the Guerra-Toninelli  interpolation method~\cite{guerra2005introduction,korada2010}, that allows 
in particular to show that the RS potential yields an upper bound to the MI. This was already done for the CDMA problem in~\cite{korada2010}
(for binary signals) and here we extend this work to any discrete signal distribution. The converse requires more work and uses the second ingredient, namely a {\it spatially coupled} version of the model. In the context of CS such spatial coupling (SC) constructions were introduced in 
~\cite{kudekar2010effect,donoho2013information,krzakala2012statistical,krzakala2012probabilistic}. It was observed 
in \cite{krzakala2012statistical,krzakala2012probabilistic} and already proved in \cite{donoho2013information} that there is no hard phase for the AMP algorithm in the asymptotic regime of an infinite spatially coupled system. 
In the present paper spatial coupling is first used, not so much as an engineering construction, but as a mean to analyze the underlying uncoupled original system. Additionally, we also extend the results of \cite{donoho2013information} by proving that, not only does spatial coupling remove the existence of the hard phase at infinitesimal noise, but also provably leads to Bayes optimal results for {\it any} values of the signal-to-noise ratio and undersampling ratio.
We use methods developed in the recent analysis of capacity-achieving 
spatially coupled low-density parity-check codes~\cite{kudekar2011threshold,hassani2010coupled,pfister2012,kumar2014} and sparse superposition codes~\cite{barbier2014replica,barbier2015approximate,barbier2016proof}. 

We have recently applied a similar strategy to the factorization of low rank matrices~\cite{krzakala2016mutual,XXT}. This, we believe, shows that the techniques and results developed 
in this paper are not only relevant for random linear estimation, but also in a broader context, and opens the way to prove many other results on estimation problems previously obtained with the heuristic replica method.

The techniques developed in this paper work well for the Bayes-optimal setting. One may wonder to what extent they can be applied away from this setting e.g., the LASSO or even L0 estimators. In the Bayes-optimal setting,
the Bayes law implies simple but
remarquable identities, often called Nishimori identities (see Appendix \ref{app:Nishimori}), that prevent the phenomenon of replica symmetry breaking, and as a consequence the replica symmetric formula for the
mutual information holds 
irrespective of the regime of parameters. In the absence of Nishimori identities our analysis breaks down. Another important feature of our analysis is the ``full threshold saturation'' of the message passing threshold to 
the information-theoretic one. 
For non-Bayesian problems
threshold saturation may exist, but may only be partial (see e.g., \cite{HassaniMacris2013}). A more in-depth study of these problems for random linear estimation away from the Bayesian-optimal setting is beyond the scope of this paper.

A summary of the present work has already appeared in \cite{BarbierDMK16}. The recent 
work~\cite{private,DBLP:journals/corr/ReevesP16} also proves the replica formula for the MI in random linear estimation 
using a very different approach. In the present paper the replica formula is obtained under sligthly stronger hypothesis which 
could be improved to exactly match those of 
\cite{DBLP:journals/corr/ReevesP16} at the expense of a few more technicalities. This is explained at the appropriate places.
%
\section{Setting}
\label{sec:partI}
We use capital letters for random variables and small letters for particular realizations. Bold letters are vectors or matrices.
\subsection{Gaussian random linear estimation}
\label{sec:settings}
In Gaussian random linear estimation one is interested in reconstructing a signal $\bs=(s_i)_{i=1}^N\in \mathbb{R}^N$ from few noisy
measurements $\by= (y_\mu)_{\mu=1}^M\!\in \!\mathbb{R}^M$ obtained from the projection of $\bs$ by a random i.i.d Gaussian \emph{measurement matrix} $\bm{\phi}=(\phi_{\mu i})_{\mu=1, i=1}^{M, N}\!\in\! \mathbb{R}^{M \times N}$. We study the problem in a probabilistic (Bayesian) setting, and therefore assume that the signal $\bs$ is a particular realization of the random variable $\bS$ (and similarly for the noise, measurement matrix etc). We consider i.i.d additive white Gaussian noise (AWGN) of known variance $\Delta$. Let
the standardized noise components be $Z_\mu \sim \mathcal{N}(0,1)$, $\mu\in\{1,\dots, M\}$. Then the measurement model is
$\by = \bm{\phi}\tbf{s} + \tbf{z} \sqrt{\Delta}$, or equivalently
\be \label{eq:CSmodel}
y_{\mu} = \sum_{i=1}^N\phi_{\mu i} s_i + z_\mu \sqrt{\Delta}.
\ee
The signal $\bs$ may be structured in the sense that it is made of $L$ i.i.d $B$-dimensional \emph{sections} $\bs_l \in \mathbb{R}^B$, $l\in\{1,\ldots,L\}$, distributed according to a discrete prior $P_0(\bs_l) = \sum_{k=1}^K p_k \delta(\bs_l - \tbf{a}_k)$ with a finite number $K$ of terms and all $\tbf{a}_k$'s with bounded components
$\max_{k, j} \vert a_{kj}\vert \leq s_{\rm max}$ (here $1\leq k\leq K$, $1\leq j\leq B$). By a slight abuse of notation we denote $\bS\sim P_0$ to actually mean that each of the sections of $\bS$ are drawn i.i.d from $P_0$, i.e. $\bS\sim P_0^{\otimes L}$. 
It is useful to keep in mind that  $\bs= (s_i)_{i=1}^N= (\bs_l)_{l=1}^L$ where $\bs_l\in \mathbb{R}^B$ and the total number of signal components is $N=LB$. The case $B=1$ corresponds to a structureless signal with purely scalar i.i.d components.
We stress that 
$K$, $s_{\rm max}$ and $B$ are independent of $N, M, L$. We will refer to such priors simply as {\it discrete priors}, and to avoid extra technical complications 
we present all results and proofs in this context. 
We can extend {\it some} of our results to the case 
 of priors that are mixtures of discrete and 
absolutely continuous parts and we briefly indicate how at appropriate places. 

The matrix $\bm{\phi}$ has i.i.d Gaussian entries $\phi_{\mu i}\sim\mathcal{N}(0,1/L)$ (this scaling of the variance implies that the measurements have $\mathcal{O}(1)$ fluctuations). The \emph{measurement rate} is $\alpha\defeq M/N$. 

We borrow concepts from statistical mechanics and we often find it convenient to call the asymptotic large system size limit, where $N, M, L\to \infty$ with 
$\alpha$ and $B$ fixed, the ``thermodynamic limit''. This regime where $\alpha$ is fixed is also sometimes referred to as the ``high dimensional'' regime in statistics.

The above setting is referred to as the \emph{CS model}, and despite being more general than compressed sensing, we employ the vocabulary of this field.

The joint distribution of the signal and the measurements is equal to the AWGN channel transition probability $P^{\rm cs}(\by|\bs,\bm{\phi})$ times the prior 
\begin{align}
P^{\rm cs}(\bs,\by|\bm{\phi}) = (2\pi\Delta)^{-M/2}\exp\Big(-\frac{1}{2\Delta}\sum\limits_{\mu=1}^M ([\bm{\phi} \bs]_\mu - y_\mu)^2\Big)
\prod_{l=1}^LP_0(\bs_l)\,.
\end{align}
The information-theoretical optimal way of estimating the signal $\bs$ follows from its posterior distribution. By the Bayes formula the posterior probability that the signal takes a value $\bx$ given the knowledge of $\by$ and $\bm{\phi}$ is 
\begin{align}
P^{\rm cs}(\bs=\bx|\by,\bm{\phi}) =P^{\rm cs}(\bx|\by,\bm{\phi})&= \frac{1}{\mathcal{Z}^{\rm cs}(\by,\bm{\phi})}\exp\Big(-\frac{1}{2\Delta}\sum\limits_{\mu=1}^M ([\bm{\phi} \bx]_\mu - y_\mu)^2\Big)
\prod_{l=1}^LP_0(\bx_l). \label{eq:posteriorCS}
\end{align}
From now on we always denote by $\bx$ a sample drawn according to the posterior distribution \eqref{eq:posteriorCS}, $\bX$ the associated random variable, and keep the notation $\bs$ for the ground-truth signal to be inferred. The posterior normalization
\begin{align} \label{patFunc}
\mathcal{Z}^{\rm cs}(\by,\bm{\phi}) =  \int d\bx \exp\Big(-\frac{1}{2\Delta}
\sum\limits_{\mu=1}^M ([\bm{\phi} \bx]_\mu - y_\mu)^2\Big)
\prod_{l=1}^LP_0(\bx_l)
\end{align}
is also called the \emph{partition function}. 
Note that it is related to the distribution of measurements through
$$P^{\rm cs}(\by|\bm{\phi})\! =\! \mathcal{Z}^{\rm cs}(\by,\bm{\phi})(2\pi\Delta)^{-M/2}.$$ The MI (per section) $i^{\rm cs}\defeq I(\bS;\bY|\bm{\Phi})/L$ is by definition
\begin{align} \label{eq:true_mutual_info}
i^{\rm cs} &= \frac{1}{L}\mathbb{E}_{\bm{\Phi}, \bS,\bY}\Big[\!\ln\!\Big( \frac {P^{\rm cs}(\bS,\bY|\bm{\Phi})}{P_0(\bS)P^{\rm cs}(\bY|\bm{\Phi})}\Big)\Big] = \frac{1}{L}\mathbb{E}_{\bm{\Phi}, \bS,\bY}\Big[\ln\Big( \frac {P^{\rm cs}(\bY|\bS,\bm{\Phi})(2\pi\Delta)^{M/2}}{{\mathcal{Z}}^{\rm cs}(\bY,\bm{\Phi})}\Big)\Big]
\nonumber \\
&~=-\frac{1}{L}h(\bY|\bS,\bm{\Phi}) + \frac{M}{2L}\ln(2\pi\Delta) - \frac{1}{L}\mathbb{E}_{\bm{\Phi}, \bY}[\ln (\mathcal{Z}^{\rm cs}(\bY,\bm{\Phi})) ]
\nonumber \\
&~
 = -\frac{\alpha B}2 - \frac{1}{L}\mathbb{E}_{\bm{\Phi}, \bY}[\ln(\mathcal{Z}^{\rm cs}(\bY,\bm{\Phi}))],
\end{align}
where $\bS\sim P_0$. The last equality is obtained noticing that the conditional entropy of $P^{\rm cs}(\by|\bs,\bm{\phi})$
is simply the differential entropy of i.i.d Gaussian random variables of variance $\Delta$, that is
$h(\bY|\bS,\bm{\Phi}) = (M/2)\big(\ln(2\pi\Delta)+1\big)$.
Note that up to an additive constant the MI is equal to 
\begin{align}
f^{\rm cs}\defeq- \frac{1}{L}\mathbb{E}_{\bm{\Phi}, \bY}[\ln(\mathcal{Z}^{\rm cs}(\bY,\bm{\Phi}))]
\end{align}
which is the average \emph{free energy} (per section) 
in statistical physics. We will consider at some point the free energy for a 
given measurement and measurment matrix realization defined as $f^{\rm cs}(\by,\bm{\phi}) \defeq- \ln(\mathcal{Z}^{\rm cs}(\by,\bm{\phi}))/L$ and show that it concentrates. 

The usual MMSE estimator which minimises the mean-square error is
$\mathbb{E}[ \bX\vert \by,\bm{\phi}] $, which is the mean of the posterior \eqref{eq:posteriorCS} (recall that we use $\bs$ for the ground-truth value of the signal, $\bS\sim P_0$ the associated random variable, and $\bX\sim P^{\rm cs}(\cdot|\by,\bm{\phi})$ for a random variable drawn according to the posterior), and the MMSE per section is
\begin{align} \label{MSEdef}
{\rm mmse} \defeq  \frac{1}{L}\mathbb{E}_{\bm{\Phi}, \bS, \bY}[\|\bS\! -\! \mathbb{E}[ \bX\vert \bY,\bm{\Phi}]\|^2]=\frac{1}{L}\mathbb{E}_{\bm{\Phi}, \bS, \bZ}[\|\bS\! -\! \mathbb{E}[ \bX\vert \bm{\Phi} \bS +\bZ\sqrt{\Delta},\bm{\Phi}]\|^2].
\end{align}
Unfortunately, 
this quantity is rather difficult to access directly from the 
MI. For this reason, it is more convenient to consider 
the {\it measurement MMSE} defined as 
\begin{align} \label{defymmse}
{\rm ymmse} \defeq \frac{1}{M}\mathbb{E}_{\bm{\Phi}, \bS, \bZ}[\|\bm{\Phi}(\bS \!-\! \mathbb{E}[ \bX\vert \bm{\Phi} \bS +\bZ\sqrt{\Delta},\bm{\Phi}])\|^2]
\end{align}
which is directly related to the MI by an I-MMSE relation~\cite{GuoShamaiVerdu_IMMSE}:
\begin{align}\label{y-immse}
\frac{di^{\rm cs}}{d\Delta^{-1}} = \frac{\alpha B}{2}{\rm ymmse}.
\end{align}
We verify this relation for the present setting by explicit algebra in appendix~\ref{app:immse}. 

We will prove in sec.~\ref{secMMSErelation} the following non-trivial relation between the MMSE's for almost every (a.e.) $\Delta$:
\begin{align}\label{xymmse_noTh}
{\rm ymmse} = \frac{{\rm mmse}}{1+ {\rm mmse}/\Delta} + \smallO_L(1).
\end{align} 
In this paper we always denote by $\smallO_L(1)$ a quantity such that $\lim_{L\to\infty}\smallO_L(1)=0$ and that might depend on the various fixed problem parameters such as $(P_0, B, \Delta,\alpha)$ etc. 

Thus, if we can compute the MI, we can compute the measurement MMSE and conversely. Moreover from the 
measurement MMSE we get the usual MMSE and conversely. 

\subsection{Replica symmetric formula}\label{repliform}
\label{sec:RSformula}
Define $v \defeq \mathbb{E}[\Vert\bS\Vert^2]/L = \sum_{k=1}^K p_k \|\tbf{a}_k\|^2$ and for $0\leq E \leq v$,
\begin{align}
\Sigma(E;\Delta)^{-2} &\defeq \frac{\alpha B}{\Delta+ E},\label{eq:defSigma2} \\
\psi(E;\Delta)&\defeq \frac{1}{2}\Big(\alpha B\ln(1+ E/\Delta)- \frac{E}{\Sigma(E;\Delta)^2}\Big).\label{eq:psi}
\end{align}
Let $i(\widetilde \bS;\widetilde \bY)$ be the MI for a $B$-dimensional \emph{denoising model}
$\widetilde\by =\widetilde\bs + \widetilde\bz\, \Sigma$ with 
$\widetilde \bs$, $\widetilde \bz$, $\widetilde \by$ all in $\mathbb{R}^B$ and $\widetilde \bS \sim P_0$, $\widetilde \bZ \sim \mathcal{N}(0,\mathbf{I}_B)$, $\mathbf{I}_B$ the $B$-dimensional identity matrix (recall capital letters are for random variables, small ones for realizations). A straightforward exercise leads
to 
\begin{align}
i(\widetilde \bS;\widetilde \bY) &\defeq -\mathbb{E}_{\widetilde\bS,\widetilde\bZ}\Big[\ln\Big(\mathbb{E}_{\widetilde\bX} \Big[\exp\Big(- \sum_{i=1}^B\frac{(\widetilde X_i - (\widetilde S_i + \widetilde Z_i \Sigma) )^2}{2\Sigma^2}\Big)\Big]\Big)\Big]-\frac{B}{2}, \label{eq:i_denoising}
\end{align}
where $\widetilde\bX \sim P_0$. 

The replica method yields the \emph{replica symmetric (RS) formula} for the MI of model \eqref{eq:CSmodel},
\begin{align}\label{rsformula}
\lim_{L\to \infty} i^{\rm cs} = \min_{E\in [0, v]} i^{\rm RS}(E; \Delta) ,
\end{align}
where the \emph{RS potential} is 
\begin{align}
i^{\rm RS}(E;\Delta) \defeq \psi(E;\Delta) + i(\widetilde\bS;\widetilde \bY). 
\label{eq:rs_mutual_info}
\end{align}
This formula was first derived by Tanaka \cite{tanaka2002statistical} for binary signals using the heuristic replica method, and later on for different and more general settings in
\cite{guo2009single,krzakala2012statistical,
krzakala2012probabilistic,tulino2013support,phdBarbier,kabashima2009typical,rangan2009asymptotic} still using the replica method. See also \cite{reeves2012sampling,reeves2012compressed} for comparisons between rigorous bounds and replica predictions. One of our main contribution is to prove that the replica predictions is indeed correct, at least in some settings that will precisely describe in sec.~\ref{sec:results}.

In the following we will denote 
\begin{align}
\widetilde E(\Delta) \defeq \underset{E\in [0, v]}{\rm argmin} ~i^{\rm RS}(E; \Delta) \label{eq:defeTildeE}
\end{align}
when it is unique (this is the case except at isolated first order phase transition points). In order to alleviate the subsequent notations we often do not explicitly write the $\Delta$ dependence of $\widetilde E$ when the context is clear.

We consider models with a discrete prior $P_0$ such that (s.t)~\eqref{eq:rs_mutual_info}~has at most three stationary points (see the discussion in 
sec.~\ref{sec:scenarios}). This implies that $i^{\rm RS}(\widetilde E(\Delta);\Delta)$ has \emph{at most one non-analyticity point} denoted $\Delta_{\rm RS}$ which is precisely the point where the ${\rm argmin}$ in \eqref{eq:defeTildeE} is non-unique. A complete proof of this fact is a lengthy but elementary exercise in real analysis and we give the main thrust of the argument in Appendix \ref{appendix_AnalyticityAndDeltaRS}. When $i^{\rm RS}(\widetilde E(\Delta);\Delta)$ is analytic over $\mathbb{R}^+$ we simply set $\Delta_{\rm RS} \!=\!\infty$. 

The most common non-analyticity in this context is a non-differentiability point of 
$i^{\rm RS}(\widetilde E(\Delta);\Delta)$. By virtue 
of \eqref{y-immse} and \eqref{xymmse_noTh} this corresponds to a jump discontinuity of the MMSE's, and one speaks of a
\emph{first order phase transition}. Another possibility is a discontinuity in higher derivatives of the 
MI, in which case the MMSE's are continuous but non differentiable and one speaks of \emph{higher order phase transitions}. 

For the reader interested in visualizing the potential shape in various settings of $(P_0, \alpha,\Delta,B)$ we refer to \cite{krzakala2012statistical,krzakala2012probabilistic} (for $B=1$) or \cite{barbier2015approximate,barbierSchulkeKrzakala} (for $B\ge 2$) where many examples can be found.

%

\subsection{Approximate message-passing and state evolution}

\subsubsection[Approximate message-passing algorithm]{\bf Approximate message-passing algorithm}

Define the following rescaled variables $\bm{\phi}_0 \defeq \bm{\phi} / \sqrt{\alpha B}$, $\by_0 \defeq \by / \sqrt{\alpha B}$. These definitions are useful in order to be coherent with the definitions of \cite{donoho2009message,bayati2011dynamics} in order to apply directly their theorems. The AMP algorithm constructs a sequence of estimates $\widehat{\bs}\!\!~^{(t)} \in \mathbb{R}^N$ and ``residuals'' $\bz\!\!~^{(t)} \in \mathbb{R}^M$ (these play the role of an effective noise not to be confused with $\bz$) according to the following iterations
\begin{align}
\bz\!\!~^{(t)} &= \by_0 - \bm{\phi}_0\bs\!\!~^{(t)} + \bz\!\!~^{(t-1)} \frac{1}{N\alpha}\sum_{i=1}^N [\eta^\prime(\bm{\phi}_0\bz\!\!~^{(t-1)} + \widehat{\bs}\!\!~^{(t-1)};\tau_{t-1}^2)]_i, \label{eq:AMP1}\\
\widehat{\bs}\!\!~^{(t+1)} &= \eta(\bm{\phi}_0 \bz\!\!~^{(t)} + \widehat{\bs}\!\!~^{(t)};\tau_{t}^2),
\end{align}
with initialization $\widehat{\bs}\!\!~^{(0)}=0$ (any quantity with negative time index is also set to the zero vector). In the Bayesian optimal setting, the section wise \emph{denoiser} $\eta(\by;\Sigma^2)$ (which returns a vector with same dimension as its first argument) is the MMSE estimator associated to an effective AWGN channel $\by = \bx + \bz\Sigma$, $\bx, \by, \bz\in \mathbb{R}^N$, $\bZ\sim \mathcal{N}(0,\textbf{I}_N)$. The $l$-th section of the $N$-dimensional vector
$\eta(\by;\Sigma^2)$
is a $B$-dimensional vector given by
\begin{align} \label{denoiser}
[\eta(\by;\Sigma^2)]_l  = \frac{\int d\bx_l\, \bx_l \, P_0(\bx_l) \exp\Big(- \frac{\Vert\by_l - \bx_l\Vert^2}{2\Sigma^2}\Big)}{\int d\bx_l\, P_0(\bx_l) \exp\Big(- \frac{\Vert\by_l - \bx_l\Vert^2}{2\Sigma^2}\Big)} 
= \frac{\sum_{k=1}^K \tbf{a}_k p_k  \exp\Big(- \frac{\Vert\by_l - \tbf{a}_k\Vert^2}{2\Sigma^2}\Big)}{\sum_{k=1}^K p_k \exp\Big(- \frac{\Vert\by_l - \tbf{a}_k\Vert^2}{2\Sigma^2}\Big)}.
\end{align}
The last form is obtained using the explicit form of the discrete prior. 
In \eqref{eq:AMP1} $[\eta^\prime(\by;\Sigma^2)]_i$ denotes the $i$-th scalar component of the gradient of $\eta$ w.r.t its first argument. In order to define $\tau_{t}$ we need the following function. Define the ${\rm mmse}$ function associated to the $B$-dimensional denoising model $\widetilde\by =\widetilde\bs + \widetilde\bz\, \Sigma$ introduced previously as
\begin{align}\label{defmmsefunction}
{\rm mmse}(\Sigma^{-2}) \defeq \mathbb{E}_{\widetilde\bS, \widetilde\bZ}\big[\|\widetilde\bS\!-\! \mathbb{E}[\widetilde\bS\vert \widetilde\bS \!+\! \widetilde\bZ\Sigma ]\|^2\big] = \mathbb{E}_{\widetilde\bS, \widetilde\bZ}\big[\|\widetilde\bS\!-\! \eta(\widetilde\bS \!+\! \widetilde\bZ\Sigma;\Sigma^2) \|^2\big].
\end{align}
Then $\tau_{t}$ is a sequence of effective AWGN variances precomputed by the following recursion:
\be
\tau_{t+1}^2 = \frac{\Delta + {\rm mmse}(\tau_{t}^{-2})}{\alpha B}, \ t\ge 0 \quad \text{with} \quad \tau_{0}^2 = \frac{\Delta + v}{\alpha B}.
\ee 

\subsubsection[State evolution]{\bf State evolution}

The asymptotic performance of AMP for the CS model can be rigorously tracked by {\it state evolution} in the scalar $B\!=\!1$ case~\cite{bayati2011dynamics,donoho2013information}. The vectorial $B\!\ge\!2$ case requires extending the rigorous analysis of SE, which at the moment has not been done to the best of our knowledge. Nevertheless, we conjecture that SE (see \eqref{recursion-uncoupled-SE} below) tracks AMP for any $B$. This is numerically confirmed in~\cite{barbier2015approximate} and proven for power allocated sparse superposition codes~\cite{rush2015capacity} (these correspond to a special vectorial case with $B\ge 2$).

Denote the asymptotic MSE per section obtained by AMP at iteration $t$ as
\begin{align} \label{MSEamp}
E^{(t)} \defeq \lim_{L\to\infty} \frac{1}{L} \|\bs \!- \! {\widehat{\bs}}\!\!~^{(t)}\|^2.
\end{align}
The SE recursion tracking the performance of AMP is
\begin{align}\label{recursion-uncoupled-SE}
E^{(t+1)} = {\rm mmse}(\Sigma(E^{(t)}; \Delta)^{-2}),
\end{align}
with initialization $E^{(0)} = v$, that is without any knowledge about the signal other than its prior distribution. Monotonicity properties of the mmse function \eqref{defmmsefunction} imply that $E^{(t)}$ is a decreasing sequence s.t $\lim_{t\to \infty}E^{(t)} \!=\! E^{(\infty)}$ exists, see \cite{barbier2016proof} for the proof of this fact.
%
%
%

\subsubsection[Algorithmic threshold and link with the potential]{\bf Algorithmic threshold and link with the potential}

Let us give a natural definition for the AMP threshold.

\begin{definition}[AMP algorithmic threshold]\label{algo-thresh-def} 
$\Delta_{\rm AMP}$ is the supremum of all $\Delta$ s.t the SE fixed point equation 
$E = {\rm mmse}(\Sigma(E ; \Delta)^{-2})$
has a \emph{unique} solution for all noise values in $[0, \Delta]$.
\end{definition}
 
\begin{remark}[SE and $i^{\rm RS}$ link] \label{remark:extrema_irs_fpSE}
It is easy to prove by simple algebra that if $E$ is a fixed point of the SE recursion \eqref{recursion-uncoupled-SE}, then it is an extremum of \eqref{eq:rs_mutual_info}, that is the fixed point equation $E = {\rm mmse}(\Sigma(E ; \Delta)^{-2})$ implies 
$\partial i^{\rm RS}(E;\Delta)/\partial E = 0$. 
Therefore $\Delta_{\rm AMP}$ is also the smallest solution of 
$\partial i^{\rm RS}/ \partial E\!=\! \partial^2 i^{\rm RS}/\partial E^2 \!=\! 0$; in other words it is 
the ``first'' horizontal inflexion point appearing in $i^{\rm RS}(E;\Delta)$ when $\Delta$ increases. 
In particular $\Delta_{\rm AMP} \leq \Delta_{\rm RS}$ because $\Delta_{\rm RS}$ is precisely the point where the argmin 
in \eqref{eq:defeTildeE} is not unique. 
\end{remark}
\section{Main Results}\label{sec:results}

\subsection{Mutual information, MMSE and optimality of AMP}
\subsubsection{\bf Mutual information and information theoretic threshold}
Our first result states that the minimum of the RS potential \eqref{eq:rs_mutual_info} upper bounds the asymptotic MI.
\begin{thm}[Tight upper bound] \label{th:upperbound}
For model \eqref{eq:CSmodel} with any $B$ and discrete prior $P_0$,
\begin{equation}\label{equIneqGuerra}
\lim_{L\to \infty} i^{\rm cs}  \le \min_{E\in [0, v]} i^{\rm RS} (E;\Delta).
\end{equation}
\end{thm}

This result generalizes the one already obtained for CDMA in~\cite{korada2010}.
One of our main result is the equality in the scalar case $B=1$, namely the proof of the RS formula derived previously in \cite{tanaka2002statistical,guo2009single,krzakala2012statistical,
krzakala2012probabilistic,phdBarbier} based on the heuristic replica method from statistical physics.
\begin{thm}[RS formula for $i^{\rm cs}$] \label{th:replica}
Take $B=1$ and assume $P_0$ is a discrete
prior s.t the RS potential $i^{\rm RS}(E;\Delta)$ in \eqref{eq:rs_mutual_info} has at most three stationary points (as a function of $E$). Then for any $\Delta$ the RS formula is exact, that is
\begin{align}\label{eq:replicaformula}
\lim_{L\to \infty} i^{\rm cs} = \min_{E\in [0, v]} i^{\rm RS}(E;\Delta).\end{align}
\end{thm}

It is conceptually useful to define the following threshold.
\begin{definition}[Information theoretic (or optimal) threshold]\label{def-delta-opt} 
Define $\Delta_{\rm Opt} \defeq\sup\{\Delta \ \text{s.t}\ \lim_{L\to \infty}i^{\rm cs}$ is analytic in $]0, \Delta[\}$.
\end{definition}

This is also one of the most fundamental definitions of a (static) phase transition threshold and also plays an important role in our analysis.
Theorem~\ref{th:replica} gives us an explicit formula to {\it compute} the information theoretic threshold, namely $\Delta_{\rm Opt} = \Delta_{\rm RS}$. 
Notice that we have not assumed anything on the number of non-analyticity points (phase transitions) of $\lim_{L\to \infty} i^{\rm cs}$. 
\subsubsection{\bf Minimal mean-square errors}
An important result is the relation between the measurement MMSE and usual MMSE (proven in sec.~\ref{secMMSErelation}). The next
theorem is independent of the previous ones. In particular the proof does not rely on SE which allows to relax the constraint $B=1$ (for which SE is known to rigorously track AMP).
\begin{thm}[MMSE relation]\label{thmMMSE}
For model \eqref{eq:CSmodel} with any (finite) $B$, any discrete prior $P_0$ and for a.e. $\Delta$, the usual and measurement MMSE's given by \eqref{MSEdef} and \eqref{defymmse} are related by
\begin{align}\label{xymmse}
{\rm ymmse} = \frac{{\rm mmse}}{1+ {\rm mmse}/\Delta} + \smallO_L(1).
\end{align} 
\end{thm}
\begin{corollary}[MMSE and measurement MMSE]\label{cor:MMSE}
Under the same assumptions as in Theorem~\ref{th:replica} and for any $\Delta\!\neq\! \Delta_{\rm RS}$, the usual and measurement MMSE given by \eqref{MSEdef} and \eqref{defymmse} satisfy
\begin{align}
\lim_{L\to \infty}{\rm mmse}&=\widetilde E(\Delta), \label{MMMSE_Etilde}\\
\lim_{L\to \infty}{\rm ymmse}&=\frac{\widetilde E(\Delta)}{1+\widetilde E(\Delta)/\Delta},
\label{yMMMSE_Etilde}
\end{align}
where $\widetilde E(\Delta)$ is the unique global minimum of $i^{\rm RS}(E;\Delta)$
for $\Delta\neq \Delta_{\rm RS}$. 
\end{corollary}

\begin{IEEEproof}
The proof follows from standard arguments that we immediately sketch here.
We first remark that the sequence $i^{\rm cs}$ (w.r.t $L$) is concave in $\Delta^{-1}$ and the thermodynamic limit 
$\lim_{L\to \infty} i^{\rm cs}$ exists. 
Concavity is intuitively clear from the I-MMSE relation \eqref{y-immse} which implies that
$d i^{\rm cs}/d\Delta^{-1}$ is non-decreasing in $\Delta^{-1}$. A detailed proof of concavity that applies in the present context is found in \cite{5165186}. The proof of existence of the limit in \cite{korada2010}
for a binary distribution $P_0$ directly extends to the more general setting here.
Alternatively an interpolation method similar to that of sec.~\ref{sec:mutualInfoSCmodel} 
shows that the sequence is super-additive which implies existence of the thermodynamic limit (by Fekete's lemma). 
By a standard theorem of real analysis
the limit of a sequence of concave functions is: $i)$ concave and continuous on any compact subset, $ii)$ differentiable almost everywhere, $iii)$ the limit and derivative can be exchanged at every differentiability point. Here we know
from Theorem \ref{th:replica} that the only 
non-differentiability point of $\lim_{L\to \infty} i^{\rm cs}$ is $\Delta_{\rm RS}$.
Therefore 
thanks to the I-MMSE relation \eqref{y-immse} we deduce that for $\Delta\neq \Delta_{\rm RS}$,
\begin{align}
\lim_{L\to \infty}{\rm ymmse} = \frac{2}{\alpha B}
\frac{d}{d\Delta^{-1}}\lim_{L\to \infty} i^{\rm cs} = \frac{2}{\alpha B}\frac{d}{d\Delta^{-1}} i^{\rm RS}(\widetilde E(\Delta);\Delta)\,.
\end{align}
The explicit calculation of the total derivative is found in appendix~\ref{app:Differentiation} and yields relation \eqref{yMMMSE_Etilde}. 
Finally \eqref{MMMSE_Etilde}
follows from \eqref{xymmse} in Theorem \ref{thmMMSE} and \eqref{yMMMSE_Etilde}. 
\end{IEEEproof}

The proofs of Theorems~\ref{th:upperbound} and \ref{th:replica} are presented in sec.~\ref{sec:partII} and \ref{sec:partIII}. We conjecture that 
Theorem~\ref{th:replica} and Corollary~\ref{cor:MMSE} hold for any $B$. Their proofs require a control of AMP by SE, a result that to the best 
of our knowledge is currently available in the literature only for $B=1$. Proving SE for all $B$ would imply that all the results of this 
paper are valid for the structured case, and we believe that this is not out of reach\footnote{Since completion of the first version of this manuscript some works related to the extension of the state evolution analysis to more general settings, including structured priors, have appeared \cite{DBLP:journals/corr/MaRB17,DBLP:journals/corr/abs-1708-03950}. It is probable that these combined with the present analysis allow to generalize our results to the $B\ge 2$ case. We tend to believe that it is the case but leave this question open for future work.}.

\begin{remark}[More general priors and phase transition scenarios]\label{remark-comparing-with-GH}
In \cite{DBLP:journals/corr/ReevesP16} priors are more general in two respects. First they can have mixtures of discrete and continuous parts. Second
the RS potential $i^{\rm RS}(E;\Delta)$ may have more than three stationary points but with at most two that become global minima. This second point means 
that the single first order phase transition scenario discussed below (see paragraph \ref{sec:scenarios}) is the same. All 
our proofs and statements also hold in this case
but we have formulated them in a slightly more restrictive way just for clarity of the general picture. Concerning the first point one can presumably
directly extend Theorems ~\ref{th:upperbound} and \ref{th:replica} (and hence 
identity \eqref{yMMMSE_Etilde}) to priors with continuous and discrete parts. The idea is to use limiting arguments, approximating
a general prior by a sequence of discrete ones, and taking the limit of \eqref{equIneqGuerra} and \eqref{eq:replicaformula}. In this process one must make sure that
one stays within the class of priors such that the RS potential $i^{\rm RS}(E;\Delta)$ has at most three stationary points 
(or more but with at most two global minima) and it is
possible to do so as long as the limiting prior also satisfies this condition.
\end{remark}

\subsubsection{\bf Optimality of approximate message-passing in Gaussian random linear estimation}
Let us now give the results related to the optimality of AMP for the CS model \eqref{eq:CSmodel}. These are proven in sec.~\ref{sec:AMP}. Let the measurement MSE of AMP be 
\begin{align} 
{\rm ymse}_{\rm AMP}^{(t)} &\defeq \lim_{L\to\infty} \frac{1}{M} \|\bm{\phi}(\bs -  {\widehat{\bs}}\!\!~^{(t)})\Vert^2.\label{yMSEAMP}
\end{align}
Moreover, recall the definition of the asymptotic MSE of AMP $E^{(t)}$ given in \eqref{MSEamp}.
\begin{thm}[Optimality of AMP] \label{thm:optAMP}
Under the same assumptions as in Theorem~\ref{th:replica} and if $\Delta < \Delta_{\rm AMP}$ or $\Delta > \Delta_{\rm RS}$, then AMP is almost surely optimal in the following sense:
\begin{align}
\lim_{t\to \infty}E^{(t)} &= \lim_{L\to\infty} {\rm mmse},\label{thmOptAMP}\\
\lim_{t\to \infty}{\rm ymse}_{\rm AMP}^{(t)} &= \lim_{L\to\infty} {\rm ymmse}, \label{thmOptAMPy}
\end{align}
where ${\rm mmse}$ and ${\rm ymmse}$ are defined by \eqref{MSEdef} and \eqref{defymmse}.
\end{thm}
\subsubsection{\bf Results on spatially coupled Gaussian random linear estimation}
Another fundamental result, that is key in our proof of the previous theorems but is in itself of independent interest, is an equivalence 
theorem (see Theorem \ref{lemma:modelsEquiv}). It roughly states that in the limit of infinit spatially coupled chain, the MI of the CS model \eqref{eq:CSmodel} and the 
spatially coupled CS models are equal. Informally speaking: denote $i_{\Gamma, w}^{\rm seed}$ and $i_{\Gamma, w}^{\rm per}$ the MI for two different spatially coupled models, where the former is associated to an open coupled chain, the latter to a closed periodic chain. Then the following theorem states that these coupled systems have same MI as the plain model when the number of sections $L$ diverges, and, for the open chain model, when then the number of sub-systems $\Gamma$ in the coupled chain also diverges afterwards. We already state this result and an important corollary here, for the convenience of the reader, but
refer to sec.~\ref{sec:deltaopt_deltars} for the precise definitions of the spatially coupled CS models. 
\begin{thm}[Invariance of the MI] \label{lemma:modelsEquiv}
The following MI limits exist and are equal 
for any (odd) $\Gamma$ and $w\!\in\!\{0,\ldots,(\Gamma\!-\!1)/2\}$: 
\be \label{eq:firstPart_lemmaInvMI}
\lim_{L\to \infty} i_{\Gamma, w}^{\rm per} = \lim_{L\to \infty} i^{\rm cs}.
\ee
Moreover for any fixed $w$, we also have 
\be\label{eq:secondPart_lemmaInvMI}
\lim_{\Gamma\to \infty}\lim_{L\to \infty} i_{\Gamma, w}^{\rm seed} = \lim_{L\to \infty} i^{\rm cs}.
\ee 
\end{thm}

Let us mention that the proof of this result is based on a crucial new interpolation technique developed specifically for this purpose, and called {\it sub-extensive interpolation}, which we believe is of independent interest\footnote{Since completion of the first version of this manuscript the sub-extensive interpolation method has been further developed \cite{Barbier_IMMSE_subExt}. It then evolved into another type of interpolation technique called ``adaptive interpolation method'', see \cite{barbier_stoInt,2019arXiv190106516B,barbier2017phase}.}. The high-level idea is to compare two statistical models by interpolating one onto the other. But, as oposed to the canonical Guerra-Toninelli interpolation \cite{guerra2002thermodynamic} where this is done ``all at once'', in the sub-extensive interpolation method this is done in small increments, which permits a more refined control along the interpolation path. We refer to sec.~\ref{sec:mutualInfoSCmodel} for a precise explanation of the method.

In addition, as a corollary of the exactness of state evolution for the SC models \cite{bayati2011dynamics,donoho2013information}, 
the {\it threshold saturation} Lemma~\ref{lemma:threshSat} and of Corollary~\ref{cor:MMSE} we obtain the following optimality result of AMP for spatially coupled Gaussian random linear estimation. This result states that the mean square-error of AMP in reconstructing the signal in the spatially coupled system, $E^{c,(t)}$, as well as its mean square-error ${\rm ymse}_{\rm AMP}^{{\rm c},(t)}$ in reconstructing the linear projections $\bm{\phi} \bs$ both tend to the optimal values as the number of iterations diverges, and when the noise does not exceed the information theoretic threshold.
\begin{thm}[Optimality of AMP for spatially coupled estimation]\label{optamp}
Under the same assumptions as in Theorem~\ref{th:replica} and if $\Delta < \Delta_{\rm Opt}=\Delta_{\rm RS}$, then AMP is almost surely optimal for spatially coupled Gaussian random linear estimation in the following sense:
\begin{align}
\lim_{t\to \infty}E^{{\rm c},(t)} &= \lim_{L\to\infty} {\rm mmse},\label{thmOptAMP}\\
\lim_{t\to \infty}{\rm ymse}_{\rm AMP}^{{\rm c},(t)} &= \lim_{L\to\infty} {\rm ymmse}, \label{thmOptAMPy}
\end{align}
where the asymptotic MSE $E^{{\rm c},(t)}$ and measurement MSE ${\rm ymse}_{\rm AMP}^{{\rm c},(t)}$ of AMP for 
the spatially coupled models are defined similarly as \eqref{MSEamp} and \eqref{yMSEAMP}, respectively, but with a spatially 
coupled measurement matrix $\bm{\phi}$ and where ${\rm mmse}$ and ${\rm ymmse}$ are defined by \eqref{MSEdef} and \eqref{defymmse}.
\end{thm}

Note that this corollary is not explicitly used in this paper and comes as a bonus. In \cite{donoho2013information}, a similar 
argument was used to prove that, with vanishing noise, spatial coupling allowed for an optimal reconstruction as long as the 
undersampling rate (or measurement rate) exceeds the upper R\'enyi information dimension of the signal. This result allows for a wide set of signal prior distributions
involving discrete as well as continuous parts. Our analysis is restricted to the case of discrete priors, but in this case
theorem \ref{optamp} generalizes the result of \cite{donoho2013information}. Informally, it simply states that spatial coupling always allows to reach Bayes optimal performances.
\subsection{The single first order phase transition scenario} \label{sec:scenarios}
In this contribution, we assume that $P_0$ is discrete and s.t \eqref{eq:rs_mutual_info} has \emph{at most} three stationary points. 
Let us briefly discuss what this hypothesis entails. 

Three scenarios are possible: $\Delta_{\rm AMP}\! <\!\Delta_{\rm RS}$ (one \emph{first order} phase transition); 
$\Delta_{\rm AMP}\!=\!\Delta_{\rm RS}\! <\!\infty$ (one \emph{higher order} phase transition); 
$\Delta_{\rm AMP}\!=\!\Delta_{\rm RS}\!=\! \infty$ (no phase transition). Here we will consider the 
most interesting (and challenging) first order phase transition case where a gap between the algorithmic AMP and 
information theoretic performance appears.
The cases of no or higher order phase transition, which present no algorithmic gap, follow as special cases from
our proof. It should be noted that in these two cases spatial coupling is not really needed and the proof 
may be achieved by an ``area theorem'' as already argued in \cite{MontanariTse06}.

Recall the notation $\widetilde E(\Delta) \!=\! {\rm argmin}_{E\in[0,v]} i^{\rm RS}(E;\Delta)$. At $\Delta_{\rm RS}$, when the ${\rm argmin}$ is a set with two elements, one can think of $\widetilde E(\Delta)$ as a discontinuous function.

The picture for the stationary points of \eqref{eq:rs_mutual_info} is as follows. 
For $\Delta \!<\!\Delta_{\rm AMP}$ there is a unique stationary point 
which is a global minimum $\widetilde E(\Delta)$ and we have $\widetilde E(\Delta) \!=\! E^{(\infty)}$, the fixed point of SE \eqref{recursion-uncoupled-SE}. At 
$\Delta_{\rm AMP}$ the function $i^{\rm RS}$ develops
a horizontal inflexion point, and for $\Delta_{\rm AMP}\!<\!\Delta\! <\! \Delta_{\rm RS}$ there are three stationary points: a local 
minimum corresponding to $E^{(\infty)}$, a local maximum, and the global minimum $\widetilde E(\Delta)$. It is not 
difficult to argue that 
$\widetilde E(\Delta) \!<\! E^{(\infty)}$ in the interval $\Delta_{\rm AMP}\!<\!\Delta \!<\! \Delta_{\rm RS}$. At $\Delta_{\rm RS}$ 
the local and global minima switch roles, so at this point the global minimum
$\widetilde E(\Delta)$ has a jump discontinuity. For all $\Delta\!>\!\Delta_{\rm RS}$ there is at least one stationary point which 
is the global minimum $\widetilde E(\Delta)$ and $\widetilde E(\Delta)\!=\! E^{(\infty)}$ (the other stationary points can merge and annihilate each other 
as $\Delta$ increases). 

Finally we note that with the help of the implicit function theorem for real analytic functions, we can show that $\widetilde E(\Delta)$ is an analytic function of $\Delta$ except at $\Delta_{\rm RS}$. Therefore $i^{\rm RS}(\widetilde E(\Delta), \Delta)$ is analytic in $\Delta$ except at $\Delta_{\rm RS}$. 

Let us mention that references \cite{private,DBLP:journals/corr/ReevesP16} require a weaker ``single-crossing'' assumption that includes the scenarios covered by our proof\footnote{One example of prior where our assumption of at most three stationary points is not verified is a uniform $P_0$ supported on $\{0, 1, 10\}$ (for a measurement rate $\alpha$ low enough). We thank a reviewer for pointing out this explicit example.}. We believe that our proof technique may be extended to include additional scenarios with more generic priors at the cost of additional technicalities.

\section{Strategy of proof}\label{sec:partII}

Let us start with a word about subsequent notations used. It is useful to distinguish two types of expectations.
The first one are expectations w.r.t posterior distributions, e.g., the expectation $\mathbb{E}_{\bX|\by,\bm{\phi}}$ w.r.t \eqref{eq:posteriorCS}, which we will most of the time denote as 
Gibbs averages $\langle\! -\! \rangle$. The second one are the expectations w.r.t all so-called {\it quenched} variables, e.g., $\bm{\Phi}$, $\bS$, $\bZ$, which will be denoted by $\mathbb{E}$. Subscripts in these expectations will be explicitly written down only when necessary to avoid confusions. 
For example the MMSE estimator becomes with these notations $\mathbb{E}[ \bX\vert \by,\bm{\phi}] = \langle \bX \rangle$ and the mmse \eqref{MSEdef} is simply ${\rm mmse} = \mathbb{E}[\|\bS\! -\! \langle \bX \rangle]\|^2]/L$.
%
%
For 
the measurement MMSE \eqref{defymmse} we have ${\rm ymmse} = \mathbb{E}[\|\bm{\Phi}(\bS \!-\! \langle \bX \rangle)\|^2]/M$.
%
%

\subsection{A general interpolation} \label{sec:generalInterpolation}
We have already seen in sec.~\ref{repliform} that the RS potential \eqref{eq:rs_mutual_info} involves the MI of a denoising model. One of the main tools that we use is an interpolation between this denoising model and the original CS model \eqref{eq:CSmodel} (the denoising model here is the same up to its dimensionality, thus we use the same notation).
Consider a set of observations $[\by, \widetilde\by]$ from the following channels
\begin{align}  \label{eq:defChannels}
\begin{cases}
\by = \bm{\phi}\bs + \bz\frac{1}{\sqrt{\gamma(t)}}, \\ 
\widetilde{\by} = \bs + \widetilde{\bz}\frac{1}{\sqrt{\lambda(t)}} ,
\end{cases}
\end{align}
where $\bZ\sim\mathcal{N}(0,\mathbf{I}_M)$, $\widetilde \bZ\sim\mathcal{N}(0,\mathbf{I}_N)$, $t\in[0, 1]$ is the interpolating parameter and the ``signal-to-noise functions'' $\gamma(t)$ and $\lambda(t)$ satisfy the constraint
\begin{align} 
 \frac{\alpha B}{\gamma(t)^{-1} + E} + \lambda(t) = \frac{\alpha B}{\Delta +E} = \Sigma(E;\Delta)^{-2},
 \label{snrconstraint}
\end{align}
with the following boundary conditions
\begin{align}
\begin{cases}
\gamma(0) = 0, \quad \gamma(1) = 1/\Delta, \\
\lambda(0) = \Sigma(E;\Delta)^{-2}, \quad \lambda(1) = 0.
\end{cases}
\label{eq:boundariesSNR}
\end{align}
We also require $\gamma(t)$ to be strictly increasing. Notice that \eqref{snrconstraint} implies 
\be 
\frac{d\lambda(t)}{dt} = - \frac{d\gamma(t)}{dt}\frac{\alpha B}{(1+\gamma(t)E)^{2}}, \label{eq:relation_gamma_lambda}
\ee
so that $\lambda(t)$ is strictly decreasing with $t$. 

In order to prove concentration properties that are needed in our proofs, we will actually work with a more complicated {\it perturbed interpolated model} where we add a set of extra observations 
that come from another ``side channel'' denoising model 
\begin{align}
\widehat \by = \bs + \widehat \bz\frac{1}{\sqrt h}, \label{eq:side_channel}
\end{align}
$\widehat\bZ \sim \mathcal{N}(0,\mathbf{I}_N)$. Here the snr $h$ is ``small'' and one should keep in mind that it will be removed in the process of the proof, i.e. $h\!\to\! 0_+$.

Define $\mathring{\by}\!\defeq\! [\by, \widetilde \by,\widehat \by,\bm{\phi}]$ as the concatanation of all observations and the measurement matrix. Moreover, it will be useful to use the following notation: $\bar \bx \defeq \bx-\bs$. In particular note
$[\bm{\phi} \bar \bx]_\mu = \sum_{i=1}^N \phi_{\mu i}(x_i-s_i)$. Our central object of study is 
the posterior of the general perturbed interpolated model:
\begin{align}
P_{t,h}(\bs=\bx|\mathring \by)=P_{t,h}(\bx|\mathring \by) = \frac{1}{\mathcal{Z}_{t,h}(\mathring{\by})}\exp\big(-\mathcal{H}_{t,h}(\bx|\mathring{\by})\big)
\prod_{l=1}^L P_0(\bx_l),
\label{gibbs-general}
\end{align}
where the \emph{Hamiltonian} is
\begin{align} 
&\mathcal{H}_{t,h}(\bx|\mathring{\by}) \defeq  \frac{\gamma(t)}{2}\sum_{\mu=1}^M\! \Big([\bm{\phi}\bar \bx]_\mu\!\!-\! 
\frac{z_{\mu}}{\sqrt{\gamma(t)}}\Big)^2\!\! +\! \frac{\lambda(t)}{2}\sum_{i=1}^N \!\Big(\bar x_i\!-\!\frac{\widetilde z_i}{\sqrt{\lambda(t)}}\Big)^2
+ \frac{h}{2}\sum_{i=1}^N \!\Big(\bar x_i \!-\! \frac{\widehat z_i}{\sqrt{h}} \Big)^2 + \sqrt{h} s_{\rm max}\sum_{i=1}^N |\widehat z_i |, 
\label{eq:int_hamiltonian}
\end{align}
and the partition function 
\begin{align}
\mathcal{Z}_{t,h}(\mathring{\by}) = \int d\bx \exp\big(-\mathcal{H}_{t,h}(\bx|\mathring{\by})\big) \prod_{l=1}^L P_0(\bx_l).
\end{align}
We replaced $\by$, $\widetilde\by$, $\widehat\by$ by their expressions \eqref{eq:defChannels} and \eqref{eq:side_channel}. 
 Expectations w.r.t the Gibbs measure \eqref{gibbs-general} are denoted $\langle \!-\!\rangle_{t,h}$, namely 
\begin{align*}
\langle g(\bX)\rangle_{t,h}	\defeq\int d\bx P_{t,h}(\bx|\mathring{\by})    g(\bx) 
\end{align*}
  and expectations w.r.t all 
quenched random variables $\bm{\phi}$, $\bs$, $\by$, $\widetilde \by$, $\widehat \by$ (or equivalently the independent $\bm{\phi}$, $\bs$, $\bz$, $\widetilde\bz$, $\widehat\bz$) by $\mathbb{E}=\mathbb{E}_{\bm{\Phi}}\mathbb{E}_{\bS}\mathbb{E}_{\bY|\bm{\Phi},\bS}\mathbb{E}_{\widetilde \bY|\bS}\mathbb{E}_{\widehat\bY|\bS}=\mathbb{E}_{\bm{\Phi}}\mathbb{E}_{\bS}\mathbb{E}_{\bZ}\mathbb{E}_{\widetilde \bZ}\mathbb{E}_{\widehat\bZ}$. 
The last term appearing in the Hamiltonian does not depend on $\bx$ and cancels in the posterior. The reason for adding this term is purely technical: it makes an additive contribution to the 
MI and free energy, which makes them concave in $h$. 

The MI for the perturbed interpolated model is defined 
similarly as \eqref{eq:true_mutual_info} and one obtains 
\begin{equation} \label{ith}
i_{t,h} = -B\Big(\frac{\alpha}{2} + 1\Big) - \frac{1}{L}\mathbb{E}[\ln(\mathcal{Z}_{t,h}(\mathring{\bY}))].
\end{equation}
Note that $i_{1,0}=i^{\rm cs}$ given by \eqref{eq:true_mutual_info}. It is also useful to define 
the free energy  per section for a given realisation of quenched variables $$f_{t,h}(\mathring{\by})\defeq- \frac1L\ln(\mathcal{Z}_{t,h}(\mathring{\by})).$$

We immediately prove an easy but useful lemma.
\begin{lemma}[Concavity in $h$ of the mutual information] \label{lemma:f_concav_h}
The MI for the perturbed interpolated model $i_{t,h}$ is concave in $h$ for all $t$. The same is true for the 
free energy $f_{t, h}(\mathring{\by})$.
\end{lemma}
\begin{IEEEproof}
Let 
\begin{align}
\mathcal{L} &\defeq\frac{1}{L} \sum_{i=1}^N\Big( \frac{x_i^2}{2} - x_i s_i - \frac{x_i \widehat z_i}{2\sqrt{h}} \Big)\,. \label{eq:def_L}
\end{align}
One can compute the first derivative of $i_{t,h}$ and finds
\begin{align}
\frac{di_{t,h}}{dh} &= -\frac1L\EE\Big[\frac{d\mathcal{Z}_{t,h}(\mathring{\bY})}{dh}\frac{1}{\mathcal{Z}_{t,h}(\mathring{\bY})}\Big]=\frac1L\EE\Big[ \Big\langle\frac{d\mathcal{H}_{t,h}(\bX|\mathring{\bY})}{dh}\Big\rangle_{t,h}\Big] \nonumber\\
& = \EE\Big[ \langle \mathcal{L}\rangle_{t,h} + \frac{1}{2L} \sum_{i=1}^L S_i^2 + \frac{s_{\rm max}}{2\sqrt{h}L} \sum_{i=1}^N |\widehat Z_i|\Big] =  \EE[ \langle \mathcal{L}\rangle_{t,h}] + \frac{v}{2} + \frac{s_{\rm max}B }{\sqrt{2\pi h}}, \label{eq:first_der_h_fInt}
\end{align}
because, recalling the definition of the Hamiltonian \eqref{eq:int_hamiltonian}, we have
\begin{align*}
 \frac1L\frac{d\mathcal{H}_{t,h}(\bx|\mathring{\by})}{dh}= \mathcal{L} + \frac{1}{2L} \sum_{i=1}^L \Big(s_i^2+\frac{s_i\widehat z_i}{\sqrt{h}}\Big) + \frac{s_{\rm max}}{2\sqrt{h}L} \sum_{i=1}^N |\widehat z_i|	
\end{align*}
and $S_i$ is independent of the zero mean Gaussian variable $\widehat Z_i$. We now compute the second derivative:
\begin{align}
\frac{d^2i_{t,h}}{dh^2} &= -L\EE\big[ \langle \mathcal{L}^2\rangle_{t,h} - \langle \mathcal{L}\rangle_{t,h}^2\big] - \frac{1}{4h^{3/2}L} \sum_{i=1}^N\EE\big[s_{\rm max} |\widehat Z_i|-\langle X_i\rangle_{t,h} \widehat Z_i\big]. \label{eq:seconde_der_h_fInt}
\end{align}
In order to obtain \eqref{eq:seconde_der_h_fInt}, note that 
\begin{align*}
\frac{d}{dh}\EE[ \langle \mathcal{L}\rangle_{t,h}]&=\EE \frac{d}{dh}\int d\bx {\cal L} \frac{e^{-\mathcal{H}_{t,h}(\bx|\mathring{\bY})}}{\mathcal{Z}_{t,h}(\mathring{\bY})} \prod_{l=1}^L P_0(\bx_l)\nonumber\\
&=\EE \int d\bx {\cal L} \Big( \frac{d\mathcal{L}}{dh}-\frac{d\mathcal{H}_{t,h}(\bx|\mathring{\bY})}{dh}-\frac{d\mathcal{Z}_{t,h}(\mathring{\bY})}{dh}\frac{1}{\mathcal{Z}_{t,h}(\mathring{\bY})}\Big)\frac{e^{-\mathcal{H}_{t,h}(\bx|\mathring{\bY})}}{\mathcal{Z}_{t,h}(\mathring{\bY})} \prod_{l=1}^L P_0(\bx_l).
\end{align*}
We also have 
\begin{align*}
\frac{d\mathcal{Z}_{t,h}(\mathring{\bY})}{dh}\frac{1}{\mathcal{Z}_{t,h}(\mathring{\bY})}=-\Big\langle\frac{d\mathcal{H}_{t,h}(\bX|\mathring{\bY})}{dh}\Big\rangle_{t,h}.
\end{align*}
Therefore
\begin{align*}
-\EE \int d\bx {\cal L} \Big(\frac{d\mathcal{H}_{t,h}(\bx|\mathring{\bY})}{dh}+\frac{d\mathcal{Z}_{t,h}(\mathring{\bY})}{dh}\frac{1}{\mathcal{Z}_{t,h}(\mathring{\bY})}\Big)\frac{e^{-\mathcal{H}_{t,h}(\bx|\mathring{\bY})}}{\mathcal{Z}_{t,h}(\mathring{\bY})} \prod_{l=1}^L P_0(\bx_l)=-L\EE\big[ \langle \mathcal{L}^2\rangle_{t,h} - \langle \mathcal{L}\rangle_{t,h}^2\big].
\end{align*}
Combining everything gives \eqref{eq:seconde_der_h_fInt}.
We observe that the second derivative is non-positive since 
$|\langle X_i\rangle_{t,h}|\!\le\! s_{\rm max}$ and therefore $i_{t,h}$ is concave in $h$. 
The second derivative of the free 
energy $d^2f_{t, h}(\mathring{\by})/dh^2$ is given by the right hand side of \eqref{eq:seconde_der_h_fInt} but without the expectation $\mathbb{E}$.
Therefore $f_{t, h}(\mathring{\by})$ is also concave in $h$. 
\end{IEEEproof}
\begin{remark}[Thermodynamic limit]\label{thermolimit}
The interpolation methods used in this paper imply super-additivity of the mutual information and thus (by Fekete's lemma) the existence of the thermodynamic limit $\lim_{L\to \infty} i_{t,h}$.
The reader is refered to paragraph \ref{thermo} in Sec. \ref{sec:mutualInfoSCmodel} for a discussion of this point.
Concavity in $h$ thus implies that the convergence of the sequence $i_{t,h}$ is uniform on all $h$-compact subsets and therefore 
$\lim_{h\to 0}\lim_{L\to \infty} i_{t,h} = \lim_{L\to \infty} \lim_{h\to 0} i_{t,h}$ (note that the MI is bounded for any $h$, so $h=0$ can be included in the compact subset). This property will be used later on.  
\end{remark}
%
 %
 \begin{remark}[Interpretation of \eqref{snrconstraint}]\label{remark:interpretation}
 Constraint \eqref{snrconstraint}, or \emph{snr conservation}, is essential. It expresses that as $t$ decreases from $1$ to $0$, we slowly decrease the snr of the CS measurements and 
make up for it in the denoising model. When $t\!=\!0$ the snr vanishes for the CS model, and no information is available about $\bs$ from the compressed measurements, information comes only from the denoising model. Instead at $t\!=\!1$ the noise is infinite in the denoising model and letting also $h\!\to\!0$ we recover the CS model \eqref{eq:CSmodel}. Let us further interpret \eqref{snrconstraint}. Given a CS model of snr $\Delta^{-1}$, by remark~\ref{remark:extrema_irs_fpSE} and \eqref{recursion-uncoupled-SE}, the global minimum of \eqref{eq:rs_mutual_info} is the MMSE of an ``effective'' denoising model of snr $\Sigma(E;\Delta)^{-2}$. Therefore, 
the interpolated model \eqref{eq:int_hamiltonian} (at $h\!=\!0$) is asymptotically equivalent (in the sense that it has the same MMSE)
to two independent denoising models: an ``effective'' one of snr $\Sigma(E;\gamma(t)^{-1})^{-2}$ associated to the CS model, 
and another one with snr $\lambda(t)$. Proving Theorem~\ref{th:replica} requires 
the interpolated model to be designed s.t its MMSE equals the MMSE of the CS model \eqref{eq:CSmodel} for almost all $t$. Knowing that the estimation of $\bs$ in the interpolated model comes 
from \emph{independent} channels, this MMSE constraint induces \eqref{snrconstraint}.
\end{remark}
\begin{remark}[Nishimori identity]\label{remark:Nishi}
We place ourselves in the \emph{Bayes optimal 
setting} which means that $P_0,\Delta, \gamma(t)$, $\lambda(t)$ and $h$ are known. The perturbed interpolated model is carefully designed so that each of the three $\bx$-dependent terms in \eqref{eq:int_hamiltonian} corresponds to a ``physical'' channel model with a properly defined transition probability. As a consequence the \emph{Nishimori identity} holds. This remarkable and general identity that follows from the Bayes formula plays an important role in our calculations.
For any (integrable) function $g(\bx, \bs)$ where $\bs$ is the signal, we have
\begin{align}
&\mathbb{E}[\langle g(\bX, \bS)\rangle_{t,h}] = \mathbb{E}[\langle g(\bX, \bX')\rangle_{t,h}], \label{eq:NishimoriId}
\end{align}
where $\bX$ and $\bX'$ are i.i.d vectors, called ``replicas'', distributed according to the {\it product measure} associated to \eqref{gibbs-general}, namely the measure
$P_{t,h}(\bx|\mathring{\by})P_{t,h}(\bx^\prime|\mathring{\by})$. We slightly abuse notation here by 
denoting the expectation operator w.r.t the posterior measure for $\bX$ and this product measure for $\bX, \bX^\prime$ with the 
same bracket $\langle\! -\! \rangle_{t, h}$:
\begin{align*}
\langle g(\bX,\bX') \rangle_{t, h}\defeq \int d\bx d\bx' P_{t,h}(\bx|\mathring{\by})P_{t,h}(\bx^\prime|\mathring{\by}) g(\bx,\bx').
\end{align*}
See appendix~\ref{app:Nishimori} for a derivation of the basic identity \eqref{eq:NishimoriId} as well as many other useful consequences.  
\end{remark}

\subsection{Various MMSE's}
We will need the following I-MMSE lemma that straightforwardly extends to the perturbed interpolated model the usual I-MMSE relation \eqref{y-immse} for the AWGN channel. The proof, for the simpler CS model, is found in appendix~\ref{app:immse} but it remains valid here as well because the Nishimori identity \eqref{eq:NishimoriId} holds for the perturbed interpolated model. Let 
\begin{align} \label{eq:ymmseth_}
{\rm ymmse}_{t,h} \defeq \frac{1}{M}\mathbb{E}[\|\bm{\Phi}(\bS\!-\!\langle \bX\rangle_{t,h})\|^2]. 
\end{align}
\begin{lemma}[I-MMSE] \label{lemma:Immse}
The perturbed interpolated model at $t=1$ verifies the following I-MMSE relation:
\begin{align}
\frac{di_{1,h}}{d\Delta^{-1}} = \frac{\alpha B}{2}{\rm ymmse}_{1,h}.
\end{align}
\end{lemma}

Let us give a useful link between ${\rm ymmse}_{t,h}$ and the usual MMSE for this model,
\begin{align}\label{eq:Eeth_}
E_{t,h} \defeq \frac{1}{L}\mathbb{E}[\|\bS\!-\! \langle\bX\rangle_{t,h}\|^2].
\end{align} 
For the perturbed interpolated model the following holds (see sec.~\ref{app:T} for the proof).

\begin{lemma}[MMSE relation] \label{lemma:MSEequivalence}
For a.e. $h$,
\begin{align}
{\rm ymmse}_{t,h} = \frac{E_{t,h}}{1 + \gamma(t)E_{t,h}}+\smallO_L(1). 
\label{mmse-hh}
\end{align}
\end{lemma}

In this lemma $\lim_{L\to \infty}\smallO_L(1)=0$ but in our proof $\smallO_L(1)$ is not uniform in $h$ and diverges like $h^{-1/2}$ as $h\!\to\! 0$. For this reason we cannot interchange the limits $L\!\to\!\infty$ and $h\!\to\!0$
in \eqref{mmse-hh}. This is not only a technicality, because in the presence of a first order phase transition one has to somehow deal with the discontinuity in the MMSE. 
\subsection{The integration argument} \label{sec:integrationArgument}
\subsubsection[Sub-optimality inequality]{\bf Sub-optimality inequality}
The AMP algorithm is sub-optimal and this can be expressed as a useful inequality. When used for inference over the CS model \eqref{eq:CSmodel} (i.e. over the perturbed interpolated model with $t\!=\!1, h\!=\!0$)
one gets $\limsup_{L\to\infty}E_{1,0} \leq E^{(\infty)}$. 
Adding new measurements coming from a side channel
can only improve optimal inference thus $E_{1, h}\leq E_{1,0}$ so that
$\limsup_{L\to\infty}E_{1,h} \leq E^{(\infty)}$.
Combining this with Lemma~\ref{lemma:MSEequivalence} and using that $E(1
+E/\Delta)^{-1}$ is an increasing function of $E$, one gets for a.e. $h$
\begin{align}
 \limsup_{L\to\infty} {\rm ymmse}_{1,h} \leq \frac{E^{(\infty)}}{1+ E^{(\infty)}/\Delta}.
 \label{ymmseinequality}
\end{align}
We note that we could use a version of Theorem~\ref{thmMMSE} valid for a.e. $\Delta$ to get the same inequality for $h=0$ and a.e. $\Delta$. This does not make any major difference nor simplification in the subsequent argument so we prefer to use the weaker Lemma~\ref{lemma:MSEequivalence} at this point.
\subsubsection[Below the algorithmic threshold $\Delta \leq \Delta_{\rm AMP}$]{\bf Below the algorithmic threshold
 $\Delta < \Delta_{\rm AMP}$}
In this noise regime $E^{(\infty)}\!=\!\widetilde{E}$ the global minimum 
of $i^{\rm RS}$ (recall \eqref{eq:defeTildeE} and remark~\ref{remark:extrema_irs_fpSE}) so we replace $E^{(\infty)}$ by $\widetilde E$ in the r.h.s of \eqref{ymmseinequality}. An elementary calculation done in appendix~\ref{app:Differentiation} shows that 
\begin{align}
\frac{di^{\rm RS}(\widetilde E;\Delta)}{d\Delta^{-1}} = \frac{\alpha B}{2} \frac{\widetilde E}{1+ \widetilde E/\Delta}. \label{eq:dirsddelta}
\end{align}
Using this identity and Lemma~\ref{lemma:Immse}, the inequality \eqref{ymmseinequality} becomes
\begin{align}
\limsup_{L\to\infty} \frac{di_{1,h}}{d\Delta^{-1}} 
\leq \frac{di^{\rm RS}(\widetilde E;\Delta)}{d\Delta^{-1}}
\end{align}
or equivalently
\begin{align}
\liminf_{L\to\infty} \frac{di_{1,h}}{d\Delta} \geq \frac{di^{\rm RS}(\widetilde E;\Delta)}{d\Delta}.
\end{align}
Integrating the last inequality over $[0, \Delta]\subset [0, \Delta_{\rm AMP}]$ and using Fatou's lemma we get for a.e. $h$
\begin{align}
i^{\rm RS}(\widetilde E;\Delta) -i^{\rm RS}(\widetilde E;0) \leq 
\liminf_{L\to\infty} \big(i_{1,h}|_{\Delta} -  i_{1,h}|_{\Delta=0}\big).\label{eq:beforeSimplification_0}
\end{align}
By remark \ref{thermolimit} we can replace $\liminf$ by $\lim$ in \eqref{eq:beforeSimplification_0}, take the limit $h\!\to\! 0$ and permute the limits. This yields
\begin{align}
i^{\rm RS}(\widetilde E;\Delta) -i^{\rm RS}(\widetilde E;0) \leq 
\lim_{L\to\infty} i^{\rm cs}|_{\Delta} -  \lim_{L\to\infty} i^{\rm cs}|_{\Delta=0}.\label{eq:beforeSimplification_2}
\end{align}
In the noiseless case $\Delta\!=\!0$ we have from the replica potential $i^{\rm RS}(\widetilde E; 0)=H(\bS)$
and also $\lim_{L\to \infty}i^{\rm cs}\vert_{\Delta=0} = H(\bS)$, with $H(\bS)$ the Shannon entropy of $\bS\!\sim\!P_0$. 
We stress that these two statements are true irrespective of $\alpha$ and $\rho$ because the alphabet is discrete. 
A justification is found in appendix~\ref{app:zeronoise}. 
For a mixture of continuous and discrete alphabet we still have $i^{\rm RS}(\widetilde E; 0)= \lim_{L\to \infty}i^{\rm cs}\vert_{\Delta=0}$ but the proof is non trivial. Thus we obtain from \eqref{eq:beforeSimplification_2} that the RS potential evaluated at its minimum $\widetilde E$ is a lower bound to the true asymptotic MI when $\Delta\! <\! \Delta_{\rm AMP}$,
\be
i^{\rm RS}(\widetilde E;\Delta) \le \lim_{L\to\infty} i^{\rm cs}\,. 
\label{eq:lowerBOund}
\ee
Combined with Theorem~\ref{th:upperbound}, this yields Theorem~\ref{th:replica} for all $\Delta\! \in\! [0,\Delta_{\rm AMP}]$. 
\subsubsection[The hard phase $\Delta_{\rm AMP}\leq \Delta\leq\Delta_{\rm RS}$]{\bf The hard phase $\Delta\in [\Delta_{\rm AMP}, \Delta_{\rm RS}]$}
Notice first that $\Delta_{\rm AMP} \!\leq\! \Delta_{\rm Opt}$. Indeed since $\Delta_{\rm RS} \!\geq \!\Delta_{\rm AMP}$ (by their definitions) and both functions $i_{\rm RS}(E;\Delta)$ and $\lim_{L\to\infty}i^{\rm cs}$ are equal up to $\Delta_{\rm AMP}$, knowing that $i_{\rm RS}(E;\Delta)$ is analytic until $\Delta_{\rm RS}$ implies directly $\Delta_{\rm AMP} \!\leq\! \Delta_{\rm Opt}$.


\emph{Assume for a moment that} $\Delta_{\rm Opt} \!=\! \Delta_{\rm RS}$. Thus both $\lim_{L\to\infty}i^{\rm cs}$ and $i^{\rm RS}(\widetilde E; \Delta)$ are analytic on $]0, \Delta_{\rm Opt}[$
which, since they are equal on $[0, \Delta_{\rm AMP}]\!\subset\! [0, \Delta_{\rm RS}]$,
implies (by unicity of the analytic continuation) that they must be equal for all $\Delta\!<\!\Delta_{\rm RS}$. 
Concavity in $\Delta$ implies continuity of $\lim_{L\to \infty}i^{\rm cs}$ which allows to conclude that Theorem 
\ref{th:replica} holds at $\Delta_{\rm RS}$ too. 
%
%
\subsubsection[Above the static phase transition $\Delta \geq \Delta_{\rm RS}$]{\bf Above the static phase transition $\Delta \geq \Delta_{\rm RS}$}
For this noise regime, we have again $E^{(\infty)}\!=\!\widetilde E$ the global 
minimum of $i^{\rm RS}$. We can start again from \eqref{ymmseinequality} with $E^{(\infty)}$ replaced by $\widetilde E$
and apply a similar integration argument with the integral now running from $\Delta_{\rm RS}$ to $\Delta$ from which we get, after taking the limit $h\to 0$ (thanks to remark \ref{thermolimit})
\begin{align}
i^{\rm RS}(\widetilde E;\Delta) -i^{\rm RS}(\widetilde E;\Delta_{\rm RS}) \leq 
\lim_{L\to\infty} i^{\rm cs}|_{\Delta} -  \lim_{L\to\infty}i^{\rm cs}|_{\Delta_{\rm RS}}\,.
\label{eq:beforeSimplification_1}
\end{align}
The validity of the replica formula at $\Delta_{\rm RS}$ (just proved above under the assumption $\Delta_{\rm Opt} \!=\! \Delta_{\rm RS}$) is crucial to complete this argument. It allows to cancel $i^{\rm RS}(\widetilde E;\Delta_{\rm RS})$ and $\lim_{L\to\infty} i^{\rm cs}|_{\Delta_{\rm RS}}$ which implies the inequality \eqref{eq:lowerBOund} for $\Delta \ge \Delta_{\rm RS}$. In view of
Theorem~\ref{th:upperbound} this completes the proof of Theorem~\ref{th:replica}.

{\it It remains to show} $\Delta_{\rm Opt} \!=\! \Delta_{\rm RS}$. This is where spatial coupling and threshold saturation come as new crucial ingredients. 
\subsection{Proof of $\Delta_{\rm Opt}\!=\!\Delta_{\rm RS}$ using spatial coupling} \label{sec:deltaopt_deltars}
%
%
\begin{figure}[!t]
\centering
\includegraphics[draft=false, width=.35\textwidth]{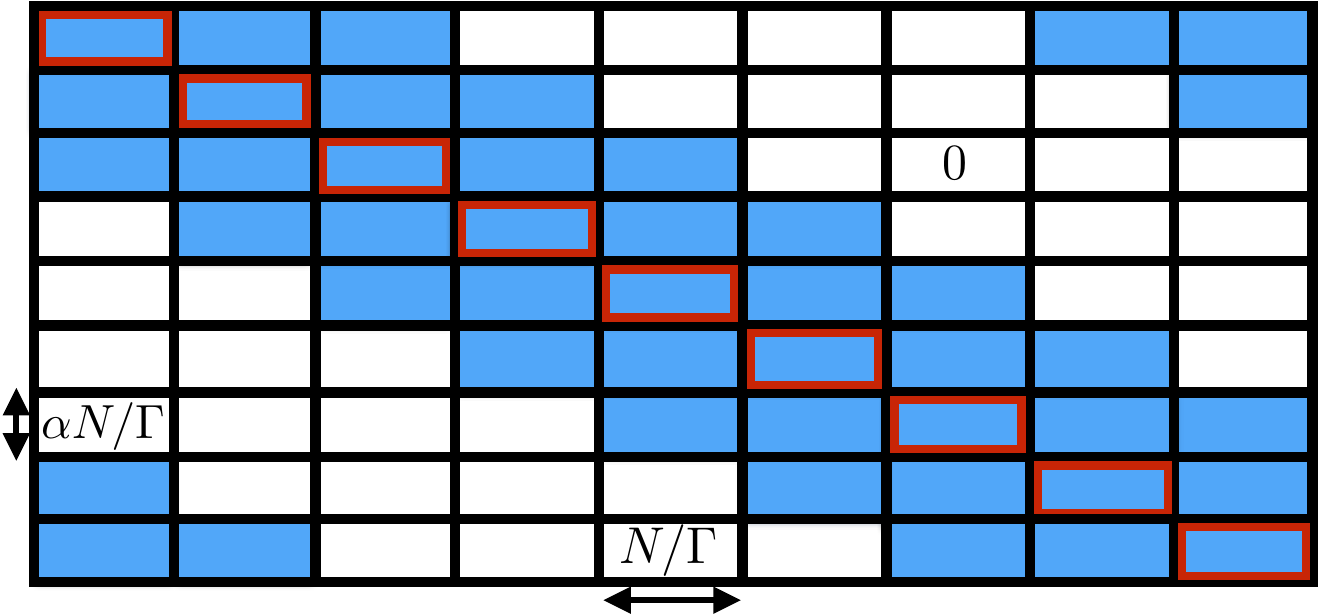}
\includegraphics[draft=false, width=.353\textwidth]{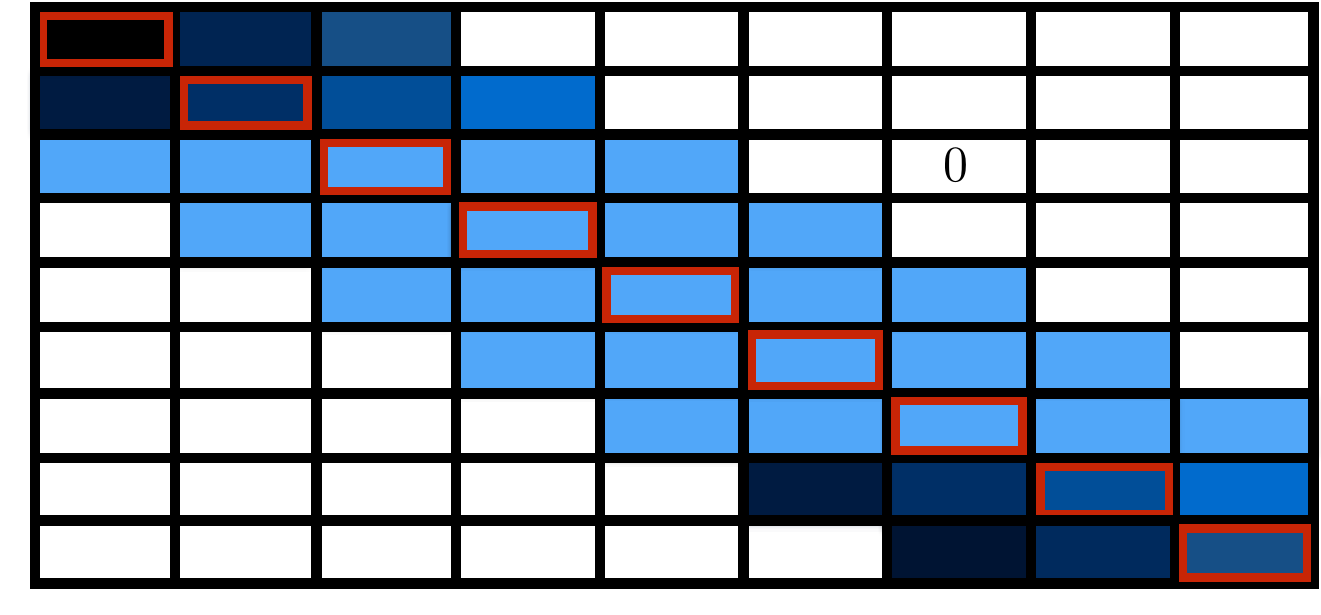}
\caption{Spatially coupled measurement matrices $\in\!\mathbb{R}^{M\times N}$ with a ``band diagonal'' structure. They are made of $\Gamma\!\times\! \Gamma$ \emph{blocks} indexed by $(r,c)$ ($\Gamma$ odd, here $\Gamma\!=\!9$), each with $N/\Gamma$ columns and $M/\Gamma$ rows. Index $r$ corresponds to ``block-rows'' and index $c$ to ``block-columns''. The i.i.d entries in block $(r,c)$ are $\mathcal{N}(0,J_{r,c}/L)$. The \emph{coupling strength} is controlled by the variance matrix $\tbf{J} \in \mathbb{R}^{\Gamma\times\Gamma}$. We consider two slightly different constructions. The \tbf{periodic model} (left): it has $w$ forward and $w$ backward \emph{coupling blocks} (here $w\!=\!2$) with $J_{r,c}\!=\!\Gamma/(2w\!+\!1)$ if $|r\!-\!c|\!\le\!w$ $(\!\!\!\!\mod \Gamma$), $0$ else (white blocks with only zeros). The \tbf{open model} (right): the \emph{coupling window} $w$ remains unchanged except at the boundaries where the periodicity is broken. Moreover the coupling strength is stronger at the boundaries (darker color).}
\label{fig:opSpCoupling}
\end{figure}
\subsubsection[Spatial coupling]{\bf Spatial coupling}
In order to show the equality of the thresholds, we need the introduction of two closely related \emph{spatially coupled CS models}. Their construction is described by Fig.~\ref{fig:opSpCoupling} which shows two measurement matrices replacing the homogeneous one of the CS model \eqref{eq:CSmodel}. When a SC measurement matrix is used, it naturally induces a decomposition of the signal $\bs=(s_i)_{i=1}^N$ into $\Gamma$ ``blocks'' $c=1,\dots, \Gamma$ (with $\Gamma$ odd). Block $c$ of the signal corresponds to its $N/\Gamma$ components on which act the matrix entries inside the $c$-th ``block-column''. It also induces a block decomposition of the measurement vector $\by=(y_\mu)_{\mu=1}^M$ into $\Gamma$ blocks $r=1, \dots, \Gamma$. Block $r$ corresponds to its $M/\Gamma$ components obtained from the product between the measurement matrix and the signal where the former is restricted to its $r$-th ``block-row''.

The matrix on the left of Fig.~\ref{fig:opSpCoupling} corresponds to taking periodic boundary conditions. This is called the 
{\it periodic} SC system (or model). On the right of Fig.~\ref{fig:opSpCoupling} the SC system is ``open". This open system is also called {\it seeded} SC system because we assume that the signal components belonging to the boundary blocks $\in\!\mathcal{B}$ are known and fixed. The number of boundary blocks is of the order of the coupling window $w$. The same construction was used for spatially coupled sparse superposition codes~\cite{barbier2016proof,barbier2016threshold} and we refer to these papers for more details. Introducing this ``information seed'' by fixing the boundaries is essential when proving the \emph{threshold saturation} phenomenon described in the next subsection. The stronger variance at the boundaries of the open model help this information seed trigger a ``reconstruction wave'' that propagates the reconstructed signal inwards. This phenomenon (intimately related to crystal growth by nucleation) is what allows SC to reach such good practical performance, namely reconstruction of the signal by AMP at low measurement rate $\alpha$.

Here, the seeded SC model is not introduced for practical purposes. Instead, we introduce it to prove properties for the non coupled original CS model. Indeed, we are able to show that the seeded and original CS models have identical mutual informations in the thermodynamic limit. Proving this fact directly for the seeded model is rather cumbersome, and this is why the periodic model is first introduced as an intermediate step in the proof. Moreover because of threshold saturation, the algorithmic transition blocking AMP ``is removed'' for seeded SC models, allowing us to probe the ``hard phase'' (which is not hard anymore for the seeded SC system) by analysing the AMP algorithm.

Like for the CS model with \eqref{eq:int_hamiltonian}, the SC model is associated to an interpolated Hamiltonian. In order to prove necessary concentrations, we also associate to each block of the signal an independent AWGN side channel. The Hamiltonian of the perturbed interpolated SC model is thus
\begin{align} 
\mathcal{H}_{t,h}(\bx|\mathring{\by}) \defeq 
&\frac{\gamma(t)}{2}\sum_{\mu=1}^M\! \Big([\bm{\phi}\bar \bx]_\mu\!\!-\! 
\frac{z_{\mu}}{\sqrt{\gamma(t)}}\Big)^2 + \frac{\lambda(t)}{2}\sum_{i=1}^N \!\Big(\bar x_i\!-\!\frac{\widetilde z_i}{\sqrt{\lambda(t)}}\Big)^2 \nonumber \\
& + \sum_{c=1}^\Gamma \frac{h_c}{2}\sum_{i_c=1}^{N/\Gamma} \!\Big(\bar x_{i_c} \!-\! \frac{\widehat z_{i_c}}{\sqrt{h_c}} \Big)^2
+ \sum_{c=1}^\Gamma \sqrt{h_c} s_{\rm max}\sum_{i_c=1}^{N/\Gamma} |\widehat z_{i_c}|, 
\label{eq:int_hamiltonian_SC}
\end{align}
where $\{h_c\}$ are (small) snr values per block, and $\bm{\phi}$ is the SC measurement random matrix (see Fig.~\ref{fig:opSpCoupling} for the structure of the variance of the matrix elements). 
As before, $\mathring{\by}=[{\by}, {\widetilde{\by}}, {\widehat{\by}},\bm{\phi}]$. Taking $\Gamma\!=\!1$ in this model, one recovers the (homogeneous) perturbed interpolated model \eqref{eq:int_hamiltonian}. With slight abuse of notation, we will denote by $\langle\!-\!\rangle_{t,h}$ the Gibbs averages associated to this Hamiltonian (i.e. w.r.t the posterior \eqref{gibbs-general} but with the Hamiltonian \eqref{eq:int_hamiltonian_SC} replacing \eqref{eq:int_hamiltonian}). This is the same notation as for the Gibbs averages corresponding to \eqref{eq:int_hamiltonian}, however the difference will be clear from the context.
\subsubsection[Threshold saturation]{\bf Threshold saturation}
The performance of AMP for the CS model, when SC matrices are used, is tracked by an MSE \emph{profile} $\tbf{E}^{(t)}$, a vector $\in\![0,v]^\Gamma$ whose components are local MSE's describing the quality of the reconstructed signal. More precisely, define $\bs_c$ as the vector made of the $L/\Gamma$ sections belonging to the $c$-th block of the signal, ${\widehat{\bs}}_c\!\!\!\!~^{(t)}$ its AMP estimate at iteration $t$. Then the $r$-th component of the profile is
\be
E_r^{(t)} = \lim_{L\to\infty}\frac{1}{L} \sum_{c=1}^\Gamma J_{r,c} \|\bs_c - {\widehat{\bs}}_c\!\!\!\!~^{(t)} \|^2,\qquad r\in\{1, \ldots, \Gamma\}.
\ee

Consider the seeded SC system. The MSE profile $\tbf{E}^{(t)}$ can be computed by SE. The precense of the seed is reflected by $E_r^{(t)} = 0$ for all $t$ if $r\!\in\!\mathcal{B}$. Apart from the boundary blocks we have
\begin{align}
&E_r^{(t+1)} = \frac{1}{\Gamma}\sum_{c=1}^\Gamma \!J_{r,c} \, {\rm mmse}(\Sigma_c(\tbf{E}^{(t)}; \Delta)^{-2}), \qquad r\notin \mathcal{B}, \\ 
&\Sigma_c(\tbf{E}; \Delta)^{-2} \defeq \frac{\alpha B}{\Gamma}\sum_{r=1}^{\Gamma} \frac{J_{r,c}}{\Delta +E_r},\label{eq:SEcoupled2}
\end{align}
with initialization $E_r^{(0)} = v$ for all $r\!\notin\mathcal{B}$, as required by AMP. Denote $\tbf{E}^{(\infty)}$ the fixed point profile of this SE recursion. Furthermore, denote $E_{\rm good}(\Delta)$ the smallest solution of the fixed point equation associated to the uncoupled SE recursion \eqref{recursion-uncoupled-SE}, or equivalently the smallest value of $E$ corresponding to an extrememum of $i^{\rm RS}(E;\Delta)$ (see remark \ref{remark:extrema_irs_fpSE}).
\begin{definition}[AMP algorithmic threshold of the seeded SC model]
The AMP algorithmic threshold for the seeded SC model is
\be
\Delta_{\rm AMP}^{\rm c}\!\defeq\! \liminf_{w\to\infty}\liminf_{\Gamma\to\infty}
\sup\{\Delta\!>\!0  \mid E_r^{(\infty)}\!\le\!E_{\rm good}(\Delta)\, \forall\, r\}.
\ee
The order of the limits is essential.
\end{definition}

It is proved in \cite{barbier2016proof} by three of us that when AMP is used for seeded SC systems, \emph{threshold saturation} occurs.
\begin{lemma}[Threshold saturation]\label{lemma:threshSat}
The AMP algorithmic threshold of the seeded SC system saturates to the RS threshold $\Delta_{\rm RS}$, that is
\be
\Delta_{\rm AMP}^{\rm c}\geq \Delta_{\rm RS}.
\ee
\end{lemma}

In fact we can prove the equality holds, but we shall not need it here. 
\subsubsection[Invariance of the optimal threshold]{\bf Invariance of the optimal threshold}
Call the MI per section for the periodic and seeded SC systems, respectively, $i^{\rm per}_{\Gamma,w}$ and $i_{\Gamma, w}^{\rm seed}$. Using an interpolation method we will 
show in sec.~\ref{sec:mutualInfoSCmodel} the following asymptotic equivalence property between the coupled and original CS models. This non-trivial key result says that despite the 1-dimensional ``chain'' structure introduced in coupled models, the MI is preserved.
\begin{thm}[Invariance of the MI] \label{lemma:modelsEquiv}
The following MI limits exist and are equal 
for any (odd) $\Gamma$ and $w\!\in\!\{0,\ldots,(\Gamma\!-\!1)/2\}$: 
\be \label{eq:firstPart_lemmaInvMI}
\lim_{L\to \infty} i_{\Gamma, w}^{\rm per} = \lim_{L\to \infty} i^{\rm cs}.
\ee
Moreover for any fixed $w$, we also have 
\be\label{eq:secondPart_lemmaInvMI}
\lim_{\Gamma\to \infty}\lim_{L\to \infty} i_{\Gamma, w}^{\rm seed} = \lim_{L\to \infty} i^{\rm cs}.
\ee 
\end{thm}

This implies straightforwardly that the information theoretic (or optimal) threshold $\Delta_{\rm Opt}^{\rm c}$ of the seeded SC model, defined as the first non-analyticity point, as $\Delta$ increases, of its asymptotic MI (i.e., the l.h.s of \eqref{eq:secondPart_lemmaInvMI}), is the same as the one of the original uncoupled CS model (i.e., the r.h.s of \eqref{eq:secondPart_lemmaInvMI}):
\be
\Delta_{\rm Opt}^{\rm c}= \Delta_{\rm Opt}. \label{eq:eqOptcOpt}
\ee 
This equality means that the phase transition occurs at the same threshold for the seeded SC and uncoupled CS models. This will be essential later on. 
\subsubsection[The inequality chain]{\bf The inequality chain}
We claim the following:
\begin{align}
\Delta_{\rm RS} \le \Delta_{\rm AMP}^{\rm c} \le \Delta_{\rm Opt}^{\rm c} = \Delta_{\rm Opt} \le \Delta_{\rm RS},
\end{align}
and therefore we obtain the desired result
\be
\Delta_{\rm Opt} = \Delta_{\rm RS}.
\ee 
The first inequality is Lemma~\ref{lemma:threshSat}. The second inequality follows from sub-optimality of AMP for the seeded SC system. The equality is \eqref{eq:eqOptcOpt} which follows from Theorem~\ref{lemma:modelsEquiv}. The last inequality requires a final argument that we now explain.  

Recall that $\Delta_{\rm Opt}\! <\! \Delta_{\rm AMP}$ is not possible. Let us show that $\Delta_{\rm RS}\!\in \,]\Delta_{\rm AMP}, \Delta_{\rm Opt}[$ is also impossible. Since $\Delta_{\rm RS} \geq \Delta_{\rm AMP}$ this will imply
$\Delta_{\rm RS}\geq \Delta_{\rm Opt}$. We proceed by contradiction so we suppose this is true. Then each side of \eqref{eq:replicaformula} is analytic on $]0, \Delta_{\rm RS}[$ and 
since they are equal for $]0, \Delta_{\rm AMP}[ \subset ]0, \Delta_{\rm RS}[$, they must 
be equal on the whole range $]0, \Delta_{\rm RS}[$ by unicity of analytic continuation and also at $\Delta_{\rm RS}$ by continuity. 
For $\Delta\!> \!\Delta_{\rm RS}$ the fixed point of SE 
is $E^{(\infty)} \!=\! \widetilde{E}$ the global minimum of $i^{\rm RS}(E;\Delta)$, hence, the integration argument can be used once more on an interval $[\Delta_{\rm RS}, \Delta]$ which implies that \eqref{eq:replicaformula} holds for all $\Delta$. But then $i^{\rm RS}(\widetilde E;\Delta)$ is analytic at $\Delta_{\rm RS}\!\in \,]\Delta_{\rm AMP}, \Delta_{\rm Opt}[$ which is a \emph{contradiction}. 

The proof of the main Theorem~\ref{th:replica} is now complete. The rest of the paper contains the proofs of the other theorems and of the various intermediate lemmas.
\section{Guerra's interpolation method: Proof of Theorem~\ref{th:upperbound}}\label{sec:partIII}
The goal of this section is to prove Theorem~\ref{th:upperbound}. The ideas are to a fair extent identical to \cite{korada2010}. In this section, the Gibbs averages are taken considering the perturbed interpolated model \eqref{eq:int_hamiltonian}. 

First note that the interpolation \eqref{eq:int_hamiltonian} has been designed specifically so that, using \eqref{ith}, $i_{0,0}=i(\widetilde{\bS};\widetilde{\bY})$ given by \eqref{eq:i_denoising} where $\widetilde{\by}$ comes from the denoising model discussed above \eqref{eq:i_denoising} with noise variance equal to $\Sigma(E;\Delta)^2$. It implies from \eqref{eq:rs_mutual_info} that
\begin{align}
i^{\rm RS}(E;\Delta) =  i_{0,0} + \psi(E;\Delta), 
\label{eq:relation_fden_frs}
\end{align}
%
%
By the fundamental theorem of calculus, we have
\be
i_{1,h} = i_{0,h} + \int_0^1 dt \,\frac{di_{t,h}}{dt}. \label{eq:fundThCalc}
\ee
Using \eqref{eq:relation_fden_frs} this is equivalent to 
\begin{align}
i_{1,h} &= i^{\rm RS}(E;\Delta)+ (i_{0,h} -i_{0,0}) + \int_0^1 dt R_{t, h} \, ,
\label{eq:fcs_fdedn_plus_remainder}\\
R_{t, h} &= \frac{di_{t,h}}{dt} - \psi(E;\Delta).
\label{eq:remainder}
\end{align}
We derive a useful expression for the remainder $R_{t,h}$ which shows that it is negative up to a negligible term. For this purpose, it is useful to re-write $\psi(E;\Delta)$ given by \eqref{eq:psi} in a more convenient form. Using \eqref{eq:boundariesSNR} and \eqref{eq:relation_gamma_lambda} one gets
\begin{align}
\frac{\alpha B}{2} \ln(1+E/\Delta)&=\frac{\alpha B}{2}\int_0^1 dt \frac{d\gamma(t)}{dt} \frac{E}{1+\gamma(t)E}, \label{eq:lastStepOK_0}\\
\frac{E}{2\Sigma(E;\Delta)^2}&= - \frac{1}{2}\int_0^1 dt\frac{d\lambda(t)}{dt}E = \frac{\alpha B}{2}\int_0^1 dt \frac{d\gamma(t)}{dt} \frac{E}{(1+\gamma(t)E)^2}, \label{eq:lastStepOK}
\end{align}
Using these two identities we obtain
\begin{align}\label{intermediate-psi-equ}
\psi(E;\Delta) = \frac{\alpha B}{2}\int_0^1 dt\, \frac{d\gamma(t)}{dt} \Big(\frac{E}{1+\gamma(t) E} - \frac{E}{(1+\gamma(t) E)^2}\Big).
\end{align}
Let us now deal with the derivative term in the remainder. Straightforward differentiation gives 
\begin{align}
\frac{d i_{t,h}}{dt} &= \frac{1}{2L}({\cal A} + {\cal B}), \label{eq:ab}\\
{\cal A} &\defeq \frac{d\gamma(t)}{dt}\sum_{\mu=1}^M \mathbb{E}\big[\big\langle [\bm{\Phi}\bar \bX]_\mu^2 - \gamma(t)^{-1/2}[\bm{\Phi}\bar \bX]_\mu Z_{\mu} \big\rangle_{t,h}\big], \quad {\cal B} \defeq \frac{d\lambda(t)}{dt} \sum_{i=1}^N \mathbb{E}\big[\big\langle \bar X_i^2 - \lambda(t)^{-1/2}\bar X_i \widetilde Z_i\big\rangle_{t,h}\big]. \label{eq:b}
\end{align}
These two quantities can be simplified using 
Gaussian integration by parts. For example, integrating by parts w.r.t $Z_\mu$,
\begin{align}
\gamma(t)^{-1/2}\mathbb{E}_{\bZ}[ \langle [\bm{\phi}\bar \bX]_\mu \rangle_{t,h}Z_\mu] &= \mathbb{E}_{\bZ}[\langle [\bm{\phi}\bar \bX]_\mu^2 \rangle_{t,h} - \langle [\bm{\phi}\bar \bX]_\mu \rangle_{t,h}^2], \label{eq:ippNoise}
\end{align}
which allows to simplify $\mathcal{A}$, 
\begin{align}
{\cal A} &= \frac{d\gamma(t)}{dt} \sum_{\mu=1}^M \mathbb{E}[\langle [\bm{\Phi}\bar \bX]_\mu \rangle_{t,h}^2]\,. \label{eq:was71}
\end{align}
For $\mathcal{B}$ we proceed similarly with an integration by parts w.r.t $\widetilde Z_i$, and find
\begin{align}
{\cal B} &=\frac{d\lambda(t)}{dt} \sum_{i=1}^N \mathbb{E}[\langle \bar X_i \rangle_{t,h}^2 ]. 
\label{eq:was71_2}
\end{align}
Now, recalling the definitions \eqref{eq:ymmseth_} and \eqref{eq:Eeth_} of ${\rm ymmse}_{t,h}$ and $E_{t,h}$,
using Lemma~\ref{lemma:MSEequivalence} and \eqref{eq:relation_gamma_lambda}, we obtain that for a.e $h$,
\begin{align}
\frac{{\cal A}}{2L} &= \frac{d\gamma(t)}{dt} \frac{\alpha B}{2} {\rm ymmse}_{t,h} = \frac{d\gamma(t)}{dt}\frac{\alpha B}{2}\frac{E_{t,h}}{1\!+\!\gamma(t) E_{t,h}} + \smallO_L(1), 
\label{eq:secondTerm_fint_A}\\
\frac{\mathcal{B}}{2L} &= \frac{d\lambda(t)}{dt} \frac{E_{t,h}}{2} = - \frac{d\gamma(t)}{dt}\frac{\alpha B}{2}\frac{ E_{t,h}}{(1+\gamma(t) E)^{2}}. 
\label{eq:secondTerm_fint}
\end{align}
Finally, combining \eqref{eq:ab}, \eqref{intermediate-psi-equ} and \eqref{eq:remainder} we get
for a.e $h$
\begin{align} 
\int_0^1 dtR_{t,h} &= \int_0^1 dt\,\frac{d\gamma(t)}{dt}\frac{\alpha B}{2}\Big(\frac{E_{t,h}}{1+\gamma(t) E_{t,h}} -\frac{E_{t,h}}{(1+\gamma(t) E)^2}- \frac{E}{1+\gamma(t) E}
+ \frac{E}{(1+\gamma(t) E)^2} \Big) + \smallO_L(1)\label{eq:remainder_long}\\
&= - \int_0^1 dt\frac{d\gamma(t)}{dt}\frac{\alpha B}{2}\frac{\gamma(t)(E-E_{t,h})^2}{(1+\gamma(t) E)^2 (1+\gamma(t) E_{t,h})} + \smallO_L(1).\label{eq:R_negative}
\end{align}
Since $\gamma(t)$ is an increasing function we see that, quite remarquably, $R_{t,h}$ is negative up to a vanishing term for a.e $h$. Since the limit $\lim_{L\to \infty} i_{1, h}$ exists we obtain 
from \eqref{eq:fcs_fdedn_plus_remainder} that for a.e $h$
\be
\lim_{L\to \infty}i_{1,h} \leq i^{\rm RS}(E;\Delta) + (i_{0,h} - i_{0,0}).
\ee
Note from \eqref{eq:int_hamiltonian} that $i_{0,h}$ is independent of $L$ for all $h$. Moreover $\lim_{L\to \infty}i_{t,h}$ is concave (see Lemma~\ref{lemma:f_concav_h}) and thus continuous in any compact set containing $h=0$. Therefore this inequality is in fact valid for all 
$h$ in a compact set containing $h=0$. Now let $h\to 0$ and since $\lim_{h\to 0}\lim_{L\to \infty}i_{t,h} = \lim_{L\to \infty}\lim_{h\to 0}i_{t,h}$ (recall remark \ref{thermolimit}) we get that for any trial $E\in[0,v]$,
\be
\lim_{L\to\infty}i_{1,0} \leq i^{\rm RS}(E;\Delta),
\ee
which is the statement of Theorem~\ref{th:upperbound} recalling that $i_{1,0}=i^{\rm cs}$.
\section{Linking the measurement and standard MMSE: Proof of Lemma~\ref{lemma:MSEequivalence}} \label{app:T}
In this section we prove Lemma \ref{lemma:generalMMSE} below, which is a generalisation of Lemma 
\ref{lemma:MSEequivalence}.  We place ourselves in the general setting where a SC measurement matrix is used and consider the perturbed interpolated SC model \eqref{eq:int_hamiltonian_SC}. 
\begin{lemma}[General MMSE relation] \label{lemma:generalMMSE}
Consider model \eqref{eq:int_hamiltonian_SC}. Take a set $\mathcal{S}_r\subset \{1,\dots, M\}$ of rows of the measurement matrix where all the rows belong to a common block-row with index $r$ and where $\vert \mathcal{S}_r\vert = (M/\Gamma)^u$ with $0< u \leq 1$. Let ${\rm ymmse}_{{\cal S}_r,t,h}$ the measurement MMSE associated to
$\mathcal{S}_r$:
\begin{align}
{\rm ymmse}_{\mathcal{S}_r,t,h} \defeq \frac{\Gamma^u}{M^u}\sum_{\mu_r\in \mathcal{S}_r}\mathbb{E}[\langle [\bm{\Phi}\bar \bX]_{\mu_r} \rangle_{t,h}^2] \,.
\end{align}
Then it verifies for a.e. $h$
\be
{\rm ymmse}_{{\cal S}_r,t,h}=  \frac{\sum_{c\in r^w} \frac{E_{c,t,h}}{2w+1}}{1+ \gamma(t)\sum_{c\in r^w} \frac{E_{c,t,h}}{2w+1}} +\smallO_L(1), \label{eq:ymmserth_Ecth}
\ee
where $r^w=\{r-w,\cdots,r+w\}$ and 
\begin{align}
E_{c,t,h}\defeq \frac{\Gamma}{L}\mathbb{E}\Big[\sum_{i_c=1}^{N/\Gamma} (\langle X_{i_c} \rangle_{t,h} - S_{i_c})^2\Big]. \label{eq:Ecth_0}
\end{align}
is the MMSE associated to the $c$-th block of the signal for the perturbed interpolated SC model.
\end{lemma}
\begin{remark}[Homogeneous case]
Lemma~\ref{lemma:MSEequivalence} is a special case, recovered from this more general lemma obtained for the SC model, by taking the homogeneous 
measurement matrix case, that is $\Gamma \!=\!1$, $w\!=\!0$ and ${\cal S}_{r=1} = \{1,\ldots,M\}$ (with $u=1$).
\end{remark}
\begin{remark}[Asymptotics]
It is useful to keep in mind that $M=\alpha N =\alpha  BL$ so $(M/\Gamma)^u = \mathcal{O}(L^u)$. 
\end{remark}
\begin{remark}[Other models] The proof shows that such generalised MMSE relations hold as long as the Hamiltonians 
are constituted of terms corresponding to AWGN channels and are carefully designed such that the Nishimori identities hold.
\end{remark}


\begin{IEEEproof}
The Nishimori identity \eqref{eq:Nish1} in appendix \ref{app:Nishimori} is valid for the perturbed interpolated model \eqref{eq:int_hamiltonian_SC},  so
\begin{align}
2\EE[\langle[\bm{\Phi}\bar\bX]_\mu\rangle_{t,h}^2]=\EE[\langle[\bm{\Phi}\bar\bX]_\mu^2\rangle_{t,h}]\,. \label{eq:Nish_squarreInOUt}
\end{align}
%
Thus the per-block measurement MMSE verifies
\begin{align}
{\rm ymmse}_{\mathcal{S}_r,t,h} &\defeq \frac{\Gamma^u}{M^u}\!\sum_{\mu_r\in \mathcal{S}_r}\!\mathbb{E}[\langle [\bm{\Phi}\bar \bX]_{\mu_r} \rangle_{t,h}^2] = \frac{\Gamma^u}{M^u}\! \sum_{\mu_r\in \mathcal{S}_r}\! \mathbb{E}\Big[\Big\langle [\bm{\Phi}\bar \bX]_{\mu_r}^2 - \frac{[\bm{\Phi}\bar \bX]_{\mu_r} Z_{\mu_r}}{\sqrt{\gamma(t)}} \Big\rangle_{t,h}\Big] \nonumber \\
&= \frac{\Gamma^{u}}{2M^u}\! \sum_{\mu_r\in \mathcal{S}_r}\! \mathbb{E}[\langle [\bm{\Phi}\bar \bX]_{\mu_r}^2 \rangle_{t,h}]. \label{eq:toManipulate}
\end{align}
The second equality is obtained using an integration by part w.r.t the noise  similarly to the steps \eqref{eq:ippNoise}--\eqref{eq:was71}. The last equality is obtained using \eqref{eq:Nish_squarreInOUt}.
We also notice that \eqref{eq:toManipulate} is equivalent to
\begin{align}
&{\rm ymmse}_{{\cal S}_r,t,h} = \frac{\Gamma^{u}}{M^u\sqrt{\gamma(t)}}\sum_{\mu_r\in {\cal S}_r}\EE[Z_{\mu_r}\langle [\bm{\Phi}\bar \bX]_{\mu_r}\rangle_{t,h}]. \label{eq:ident_ymmse}
\end{align}
Indeed the equality between the second and third forms of ${\rm ymmse}_{\mathcal{S}_r,t,h}$ in \eqref{eq:toManipulate} implies 
$$-\frac{\Gamma^{u}}{M^u\sqrt{\gamma(t)}}\sum_{\mu_r\in {\cal S}_r}\EE[Z_{\mu_r}\langle [\bm{\Phi}\bar \bX]_{\mu_r}\rangle_{t,h}] =-\frac{\Gamma^{u}}{2M^u}\! \sum_{\mu_r\in \mathcal{S}_r}\! \mathbb{E}[\langle [\bm{\Phi}\bar \bX]_{\mu_r}^2 \rangle_{t,h}]$$
and the right hand side is equal to $-{\rm ymmse}_{{\cal S}_r,t,h}$ by the last equality in \eqref{eq:toManipulate}. This proves \eqref{eq:ident_ymmse}.
%
%
Define $U_{\mu_r} \!\defeq\! \sqrt{\gamma(t)} \,[\bm{\Phi} \bar \bX]_{\mu_r} \!-\! Z_{\mu_r}$. Recalling that all the rows in $\mathcal{S}_r$ belong to block-row $r$ and using an integration by parts of \eqref{eq:ident_ymmse} w.r.t $\phi_{\mu i}\!\sim\!\mathcal{N}(0,J_{r, c_i}/L)$ ($c_i$ being the block-column to which index $i$ belongs) leads to
\begin{align}
{\rm ymmse}_{{\cal S}_r,t,h} &=\frac{\Gamma^u}{M^u L}\sum_{\mu_r\in{\cal S}_r} \sum_{i=1}^N J_{r,c_i}\EE[Z_{{\mu_r}} \langle U_{\mu_r} {\bar  X}_i\rangle_{t,h} \langle {\bar  X}_i\rangle_{t,h}-Z_{{\mu_r}}\langle U_{\mu_r} {\bar X}^2_i\rangle_{t,h}] \nonumber\\
&=\frac{\Gamma^{u+1}}{M^u L} \sum_{\mu_r\in{\cal S}_r}\frac{1}{2w+1} \sum_{c\in r^w} \sum_{i_c=1}^{N/\Gamma} \EE[Z_{{\mu_r}} \langle U_{\mu_r} {\bar  X}_{i_c}\rangle_{t,h} \langle {\bar  X}_{i_c}\rangle_{t,h}-Z_{{\mu_r}}\langle U_{\mu_r} {\bar X}^2_{i_c}\rangle_{t,h}] \nonumber\\
&=\frac{\Gamma^{u+1}}{M^u L} \sum_{\mu_r\in {\cal S}_r}\frac{1}{2w+1} \sum_{c\in r^w} \sum_{i_c=1}^{N/\Gamma} \EE[ Z_{{\mu_r}}^2 S_{i_c}\langle \bar X_{i_c}\rangle - \sqrt{\gamma(t)} Z_{{\mu_r}}S_{i_c}\langle [\bm{\Phi}\bar \bX]_{\mu_r} \bar X_{i_c}\rangle - Z_{{\mu_r}}\langle U_{\mu_r} {\bar X}^2_{i_c}\rangle_{t,h}],
\end{align}
where the second equality comes from the construction of $\textbf{J}$ (see the caption of Fig.~\ref{fig:opSpCoupling}) and the third equality comes from the identity \eqref{eq:niceId} proved in appendix \ref{app:Nishimori}
combined with the Nishimori identity \eqref{eq:nishCond}.
%
Replacing $U_{\mu_r}$ by its expression,
\begin{align}
{\rm ymmse}_{{\cal S}_r,t,h} &= \frac{\Gamma^{u+1}}{M^u L}\sum_{\mu_r\in{\cal S}}\frac{1}{2w+1} \sum_{c\in r^w} \sum_{i_c=1}^{N/\Gamma}\EE[ Z_{{\mu_r}}^2 S_{i_c}\langle \bar X_{i_c}\rangle - \sqrt{\gamma(t)} Z_{{\mu_r}}S_{i_c}\langle [\bm{\Phi}\bar \bX]_{\mu_r} \bar X_{i_c}\rangle \nonumber \\
&\ \ \ \ - \sqrt{\gamma(t)} Z_{{\mu_r}}\langle [\bm{\Phi}\bar \bX]_{\mu_r} {\bar X}_{i_c}(X_{i_c}-S_{i_c})\rangle_{t,h} + Z_{{\mu_r}}^2\langle {\bar X}_{i_c}(X_{i_c}-S_{i_c})\rangle_{t,h}] \nonumber\\
&=\frac{\Gamma^{u+1}}{M^u L}\sum_{\mu_r\in{\cal S}_r}\frac{1}{2w+1} \sum_{c\in r^w} \sum_{i_c=1}^{N/\Gamma}\EE[Z_{{\mu_r}}^2\langle {\bar X}_{i_c}X_{i_c}\rangle_{t,h} - \sqrt{\gamma(t)} Z_{{\mu_r}}\langle [\bm{\Phi}\bar \bX]_{\mu_r} {\bar X}_{i_c}X_{i_c}\rangle_{t,h} ] \nonumber \\
&= {\cal Y}_{{\cal S}_r,1} - {\cal Y}_{{\cal S}_r,2}, \label{eq:allpieces} 
\end{align}
where we have defined
\begin{align}
{\cal Y}_{{\cal S}_r,1} &\defeq \EE\Big[\frac{\Gamma^u}{M^u}\sum_{\mu_r\in{\cal S}_r} Z_{\mu_r}^2 \frac{1}{2w+1} \sum_{c\in r^w}  \langle{\cal E}_c\rangle_{t,h}\Big], \label{eq:Y1}\\
{\cal Y}_{{\cal S}_r,2} &\defeq \sqrt{\gamma(t)}\, \EE\Big[\frac{\Gamma^u}{M^u}\sum_{\mu_r\in{\cal S}_r} Z_{\mu_r}\Big\langle [\bm{\Phi} \bar \bX]_{\mu_r}  \frac{1}{2w+1} \sum_{c\in r^w} {\cal E}_c\Big\rangle_{t,h}\Big], \label{eq:Y2beforeDecoup}
\end{align}
together with $\mathcal{E}_c \defeq (\Gamma/L)\sum_{i_c=1}^{N/\Gamma} X_{i_c} \bar X_{i_c}$. By the law of large numbers $(\Gamma/M)^{u}\sum_{\mu_r\in{\cal S}_r} z_{\mu_r}^2 = 1+\smallO_L(1)$ almost surely as $L\to \infty$ so that using the Nishimori identity $\EE[\langle {\cal E}_c\rangle_{t,h}]=E_{c,t,h}$, we reach 
\be
{\cal Y}_{{\cal S}_r,1}= \frac{1}{2w+1}\sum_{c\in r^w} E_{c,t,h}+\smallO_L(1). 
\label{eq:73}
\ee
%
%
For the other term ${\cal Y}_{{\cal S}_r,2}$ we will show below that for a.e $h$,
\be
{\cal Y}_{{\cal S}_r,2} = \gamma(t)\,{\rm ymmse}_{{\cal S}_r,t,h}\frac{1}{2w+1}\sum_{c\in r^w} E_{c,t,h} +\smallO_L(1).
\label{y2}
\ee
From \eqref{y2}, \eqref{eq:73} and \eqref{eq:allpieces} we get that for a.e $h$,
\begin{align}
&{\rm ymmse}_{{\cal S}_r,t,h}=\frac{1}{2w+1}\sum_{c\in r^w} E_{c,t,h} - \gamma(t)\,{\rm ymmse}_{{\cal S}_r,t,h}\frac{1}{2w+1}\sum_{c\in r^w} E_{c,t,h} +\smallO_L(1),
\end{align}
which is equivalent to \eqref{eq:ymmserth_Ecth}.

{\it It remains to prove} \eqref{y2}. We prove in sec.~\ref{app:decoupling} that $\mathcal{E}_c$ satisfies a concentration property, namely Proposition~\ref{lemma:concentration}. This proposition 
implies that for a.e $h$\footnote{
This is actually the point where the perturbation of the interpolated model is really needed. Ideally we would like to obtain this result for $h=0$. However continuity of this quantity may fail if one happens to be at a phase transition point and it is therefore difficult get a control of the limit $h\to 0$.},
\begin{align}
\lim_{L\to\infty}\mathbb{E}\big[\big\langle (\mathcal{E}_c - \mathbb{E}[\langle \mathcal{E}_c\rangle_{t,h}])^2\big\rangle_{t,h}\big] =0.
\label{concen}
\end{align}
 
Set $a= \mathcal{E}_c $ and 
$b = Z_{\mu_r} [\bm{\Phi} \bar \bX]_{\mu_r}$.
By 
Cauchy-Schwarz
\begin{align}
\EE[\langle ab\rangle_{t,h}] = \EE[\langle a\rangle_{t,h}]\EE[\langle b\rangle_{t,h}] +\mathcal{O}\Big(\sqrt{\EE[\langle (a - \EE[\langle a\rangle_{t,h}])^2 \rangle_{t,h}]\EE[\langle b^2\rangle_{t,h}]}\Big). \label{useConc}
\end{align}
Because of \eqref{concen} the variance of $a$ appearing under the square-root is $\smallO_L(1)$ for a.e $h$. We now show the second moment of $b$ is bounded so that the $\mathcal{O}(\cdot)$ in 
\eqref{useConc} tends to $0$ for a.e $h$. From 
Cauchy-Schwarz
\be
\EE[\langle b^2\rangle_{t,h}] \le \sqrt{\EE[Z_{\mu_r}^4] \EE[\langle [\bm{\Phi} \bar \bX]_{\mu_r}^4\rangle_{t,h}]} = \sqrt{3\EE[\langle [\bm{\Phi} \bar\bX]_{\mu_r}^4\rangle_{t,h}]}.
\ee
Recall $\bar \bx = \bx -\bs$. Expanding the fourth power in $[\bm{\Phi} \bar\bX]_{\mu_r}^4$ the only terms that appear are of the form
\be
\sqrt{\EE[\langle [\bm{\Phi} \bX]_{\mu_r}^m\rangle_{t,h}[\bm{\Phi} \bS]_{\mu_r}^n]} \label{eq79}
\ee
for some finite $m,n \ge 0$ (and in this case $\le 4$). To bound such terms we use again Cauchy-Schwarz once more
and then the Nishimori identity \eqref{eq:NishId_0} in appendix \ref{app:Nishimori} to obtain
\be
\EE[\langle [\bm{\Phi} \bX]_{\mu_r}^m\rangle_{t,h} [\bm{\Phi} \bS]_{\mu_r}^n] \le \sqrt{\EE[\langle [\bm{\Phi} \bX]_{\mu_r}^{2m}\rangle_{t,h}]\EE[[\bm{\Phi} \bS]_{\mu_r}^{2n}]} = \sqrt{\EE[[\bm{\Phi} \bS]_{\mu_r}^{2m}]\EE[[\bm{\Phi} \bS]_{\mu_r}^{2n}]}. \label{eq78}
\ee
Now since $\bs$ has i.i.d bounded components, $\bm \phi$ as well, and moreover are independent, the central limit theorem implies that $[\bm{\phi} \bs]_{\mu_r}$ converges in distribution to a random Gaussian variable with finite variance. Thus the moments in \eqref{eq78} are bounded and thus $\EE[\langle b^2\rangle_{t,h}]$ as well. 
With our choice of $a$ and $b$ and these observations \eqref{useConc} implies from \eqref{eq:Y2beforeDecoup} that for a.e $h$,
\begin{align}
{\cal Y}_{{\cal S}_r,2}&= \sqrt{\gamma(t)} \frac{\Gamma^u}{M^u}\sum_{\mu_r\in {\cal S}_r}\EE[Z_{\mu_r}\langle [\bm{\Phi} \bar \bX]_{\mu_r}\rangle_{t,h}]\frac{1}{2w+1}\sum_{c\in r^w} E_{c,t,h}  +\smallO_L(1), \label{eq:firstEq_}
\end{align}
using again $\EE[\langle {\cal E}_c\rangle_{t,h}]=E_{c,t,h}$. 
%
Finally from \eqref{eq:firstEq_} and \eqref{eq:ident_ymmse} we recognize that this relation is nothing else than 
\eqref{y2}.
\end{IEEEproof}

\section{Invariance of the mutual information: Proof of Theorem~\ref{lemma:modelsEquiv}} \label{sec:mutualInfoSCmodel}
%
%
The interpolation method originates in the work of Guerra and 
Toninelli \cite{guerra2002thermodynamic,guerra2005introduction} on the Sherrington-Kirkpatrick spin glass 
model and that of \cite{FranzLeone} for spin systems on sparse graphs. There are by now many variants of these methods, 
see for example \cite{korada2009exact,korada2010,PanchenkoTalagrand2004,Montanari2005,Macris2007,MacrisKudekar2009,bayati2013,HassaniMacris2013,6284230,MacrisGiurgiuUrbanke2016}. 
In order to prove Theorem~\ref{lemma:modelsEquiv} we introduce a 
new type of interpolation that we call {\it sub-extensive interpolation method}. This method borrows ideas from the Guerra-Toninelli 
interpolation \cite{guerra2002thermodynamic} for dense systems and from the 
combinatorial approach of \cite{bayati2013} suitable for sparse systems. In the
Guerra-Toninelli approach one interpolates from one system to another by a global smooth change of the interactions. In contrast, in the combinatorial approach, one interpolates in discrete steps by changing 
one interaction (or constraint) at a time. Here we combine these two ideas: We will smoothly modify one step at a time a large but {\it sub-extensive} number of constraints along the interpolation path. 
%
%
%
\subsection{Invariance of the mutual information between the CS and periodic SC models}\label{sec:mutualInfoSCmodel_1}
In this subsection we prove the first equality of Theorem~\ref{lemma:modelsEquiv}. For that purpose, we will compare three models: The \emph{decoupled} $w\!=\!0$ model, the SC $0\!<\!w\!<\!(\Gamma\!-\!1)/2$ model and the \emph{homogeneous} $w\!=\!(\Gamma\!-\!1)/2$ model. In all cases, a periodic matrix (Fig.~\ref{fig:opSpCoupling}, left) is considered.
%
%
%
\subsubsection[The $\rho$-ensembles]{\bf The $\rho$-ensembles}
Recall that the periodic SC matrices are decomposed in $\Gamma\times\Gamma$ blocks (see Fig.~\ref{fig:opSpCoupling}). Focus only on the {\it block-row} decomposition. Block-rows are indexed by 
$r\in \{1,\dots, \Gamma\}$ and there are $M/\Gamma$ rows in each block-row. We consider a ``virtual'' {\it thinner} decomposition into {\it sub-block-rows}: each of the $\Gamma$ block-rows is decomposed in 
$(M/\Gamma)^{1-u}$ sub-block-rows with $(M/\Gamma)^u$ rows in each sub-block-row (here $0<u\ll 1$). The total number of such sub-block-rows is thus $\tau\! \defeq\! \Gamma (M/\Gamma)^{1-u}$. The number of rows belonging to one sub-block-row scales sub-extensively like ${\cal O}(L^u)$ and their total number $\tau$ scales like ${\cal O}(L^{1-u})$. 
Let $\rho\!\in\! \{0,\dots,\tau\}$ and define a periodic SC matrix $\bm{\phi}_\rho$ as follows: $\tau \!-\! \rho$ of its sub-block-rows have a coupling window $w_0$ and the remaining $\rho$ ones have a coupling $w_\tau$. 
This defines the $\rho$-ensemble of measurement matrices (here the ordering of the sub-block-rows is irrelevant because a permutation amounts to order measurements differently and does not affect the MI).

For $w_0=0$ and $w_\tau=w$ the $\rho\in\{ 0, \dots, \tau\}$ ensembles interpolate between the decoupled $w=0$ ensemble corresponding
to $\rho=0$ and the SC ensemble with coupling window $w$ corresponding to $\rho=\tau$. Similarly, for  
$w_0=w$ and $w_\tau=(\Gamma\!-\!1)/2$ the $\rho\in\{ 0, \dots, \tau\}$ ensembles interpolate between the SC ensemble with window $w$ corresponding to $\rho=0$ and the homogeneous ensemble with $w\!=\!(\Gamma\!-\!1)/2$ corresponding to $\rho=\tau$.
\subsubsection[An intermediate basis ensemble]{\bf An intermediate basis ensemble}
Our goal is to compare the MI of the $\rho$ and $(\rho \!+\! 1)$-ensembles. We first construct a
{\it basis matrix} $\bm{\phi}_\rho^*$ as follows. Consider $\bm{\phi}_\rho$ and select uniformly at random a block-row index
$r\in \{0,\dots, \Gamma\}$. Inside this block-row, select uniformly at random a 
sub-block-row among the ones that have a coupling window $w_0$. Denote by $\mathcal{S}_r$ this sub-block-row. Then remove $\mathcal{S}_r$ from $\bm{\phi}_\rho$ (this process can be repeated until a sub-block-row with proper $w_0$ is found). 
This gives an intermediate {\it basis matrix} $\bm{\phi}_\rho^*$ with $\tau \!-\! \rho\! -\! 1$ sub-block-rows with coupling $w_0$ and $\rho$ sub-block-rows with coupling $w_\tau$. 
Now, if we insert in place of $\mathcal{S}_r$ the 
sub-block-row with same index of a random SC matrix $\bm{\phi}_0$ with coupling window $w_0$, we get back a matrix 
from the $\rho$-ensemble. Instead, if we insert in place of $\mathcal{S}_r$ the sub-block-row of a 
random SC matrix $\bm{\phi}_\tau$ with coupling window $w_\tau$, we get a matrix from the $(\rho\!+\!1)$-ensemble. Matrices $\bm{\phi}_0$ and $\bm{\phi}_\tau$ will be denoted $\bm{\phi}_q$, $q\in \{0, \tau\}$. 
%
%
\subsubsection[Comparing $\rho$ and $(\rho\!+\!1)$-ensembles]{\bf Comparing $\rho$ and $(\rho\!+\!1)$-ensembles}
We estimate the variation of MI when going from the basis system with matrix $\bm{\phi}_\rho^*$ to a system 
from the $\rho$-ensemble or $(\rho\!+\!1)$-ensemble. To do so we use a ``smooth'' and ``global'' interpolation. Define
\begin{align} \label{eq:int_hamiltonian_coupled}
\mathcal{H}_{q,t, r, \mathcal{S}_r}(\bx|\mathring{\by}) =~\! &\frac{1}{2\Delta}\sum_{\mu\notin \mathcal{S}_r} \Big([\bm{\phi}_\rho^*\bar \bx]_{\mu}- z_{\mu} \sqrt{\Delta}\Big)^2 + \frac{t}{2\Delta}\sum_{\nu\in\mathcal{S}_r} \Big([\bm{\phi}_q\bar \bx]_{\nu}- z_{\nu} \sqrt{\frac{\Delta}{t}}\Big)^2 \nonumber \\
+ &\sum_{c=1}^\Gamma \frac{h_c}{2}\sum_{i_c=1}^{N/\Gamma} \!\Big(\bar x_{i_c} \!-\! \frac{\widehat z_{i_c}}{\sqrt{h_c}} \Big)^2 + \sum_{c=1}^\Gamma \sqrt{h_c} s_{\rm max}\sum_{i_c=1}^{N/\Gamma} |\widehat z_{i_c}|.
\end{align}
The last two terms are needed to use concentration properties similar to those of sec.~\ref{app:decoupling}. The Hamiltonian is conditioned on the choice of the random block-row index $r$ and sub-block-row $\mathcal{S}_r$, and on all other usual quenched variables $\bs$, $\mathring \by$ and $\bm{\phi}$.
The interpolation parameters are $t\in[0,1]$ and $q\in\{0,\tau\}$. When $t\!=\!0$ there is no dependence on $q$ and this is the Hamiltonian of the basis system with matrix 
$\bm{\phi}^*_\rho$. Instead at $(t = 1, q=0)$ this is the Hamiltonian of the $\rho$-ensemble, and at $(t=1, q=1)$ this is the Hamiltonian of the $(\rho\!+\!1)$-ensemble. Keep in mind that this Hamiltonian depends on $\{h_c\}$ but we leave this dependence implicit, and for the other quantities that we will introduce in this section as well, for the sake of readability.

Denote $i^{\rm per}_{q,t, r, \mathcal{S}_r}$ the MI associated to this general interpolated model when $r$ and $\mathcal {S}_r$ are fixed. For $t=0$ as noted above there is no dependence on $q$ so 
$i^{\rm per}_{q=0,t=0, r,\mathcal{S}_r} = i^{\rm per}_{q=\tau,t=0, r,\mathcal{S}_r}$ (and if we further average over $\mathcal{S}_r$ and $r$ we get the MI of the basis system). 
For $t=1$, if we average over $\mathcal{S}_r$ and $r$ we get $\mathbb{E}_{r, \mathcal{S}_r}[i^{\rm per}_{q=0,t=1, r, \mathcal{S}_r}] = i_\rho$ the MI of the $\rho$-ensemble and 
$\mathbb{E}_{r, \mathcal{S}_r}[i^{\rm per}_{q=\tau,t=1, r, \mathcal{S}_r}] = i_{\rho+1}$ the MI of the $(\rho\!+\!1)$-ensemble.

From the fundamental theorem of calculus
\be
i^{\rm per}_{q,t=1, r,\mathcal{S}_r} = i^{\rm per}_{q,t=0, r,\mathcal{S}_r} + \int_{0}^1 dt \,\frac{d}{dt} i^{\rm per}_{q,t, r,\mathcal{S}_r}
\label{eq:ipert1t0}
\ee
so subtracting the $q=0$ and $q=\tau$ cases we obtain
\begin{align}
i_\rho - i_{\rho+1} = \int_{0}^1 dt \, \mathbb{E}_{r, \mathcal{S}_r}\Big[\frac{d}{dt} i^{\rm per}_{q=0,t, r, \mathcal{S}_r} - \frac{d}{dt} i^{\rm per}_{q=\tau,t, r, \mathcal{S}_r}\Big]. 
\label{eq:diffiper}
\end{align}
We now follow similar steps as in sec.~\ref{sec:partIII} to compute the derivative w.r.t $t$. First one has
\begin{align}
\frac{d}{dt} i^{\rm per}_{q,t, r,\mathcal{S}_r} &= \frac{1}{2\Delta L} \sum_{\nu\in {\cal S}_r} \mathbb{E}\Big[\Big\langle [\bm{\Phi}_q\bar \bX]_{\nu}^2 - \frac{[\bm{\Phi}_q\bar \bX]_\nu Z_{\nu}}{\sqrt{t/\Delta}} \Big\rangle_{q,t,r, \mathcal{S}_r}\Big], \label{eq:A_twoterms_coup}
\end{align}
where $\langle-\rangle_{q,t,r, \mathcal{S}_r}$ is the Gibbs average associated to the Hamiltonian \eqref{eq:int_hamiltonian_coupled}. Define the measurement MMSE associated to the subset $\mathcal{S}_r$ and the normalized MMSE of a block $c$, as
\begin{align}
{\rm ymmse}_{q,t,r, {\cal S}_r}\defeq \frac{\Gamma^u}{M^u} \sum_{\mu\in{\cal S}_r}\EE[\langle [\bm{\Phi}_q \bar \bX]_{\mu} \rangle_{q,t,r,
\mathcal{S}_r}^2 ], \label{eq:ymmserth}\\
E_{c,q, t,r, \mathcal{S}_r}\defeq \frac{\Gamma}{L}\mathbb{E}\Big[\sum_{i_c=1}^{N/\Gamma} (\langle X_{i_c} \rangle_{q,t,r, \mathcal{S}_r} - S_{i_c})^2\Big], 
\label{eq:Ecth}
\end{align}
where recall $\{i_c\}$ are the components belonging to block $c$. Integrating by parts w.r.t the noise variables and using Lemma~\ref{lemma:generalMMSE}, a derivation similar to the one of equations \eqref{eq:was71} and \eqref{eq:secondTerm_fint_A}, transforms \eqref{eq:A_twoterms_coup} to
\begin{align}
\frac{d}{dt}i^{\rm per}_{q,t, r, \mathcal{S}_r}
&=  \frac{M^{u}}{2\Delta\Gamma^u L} {\rm ymmse}_{q, t, r,\mathcal{S}_r} = \frac{M^u}{2\Delta\Gamma^u L}\frac{\sum_{c\in r^{w_q}} \frac{E_{c,q, t,r, \mathcal{S}_r}}{2w_q+1}}{1+ (t/\Delta)\sum_{c\in r^{w_q}} \frac{E_{c,q, t,r, \mathcal{S}_r}}{2w_q+1}} + \smallO(L^{u-1}), 
\label{eq:ar} 
\end{align}
for a.e. $h$, where recall $r^w\!\defeq\!\{r\!-\!w,\ldots,r\!+\!w\}$, $0<u\ll 1$, and $\lim_{L\to\infty}\smallO(L^{u-1}) = 0$. 
We note that Lemma \ref{lemma:generalMMSE} applies here because Hamiltonian \eqref{eq:int_hamiltonian_coupled} is constructed so that all terms correspond to AWGN channels and Nishimori identities hold as well as the concentration property in Proposition~\ref{lemma:concentration}. Replacing \eqref{eq:ar} in \eqref{eq:diffiper} one gets for a.e. $h$
\be
i_\rho - i_{\rho+1} = \frac{M^u}{2\Gamma^u L}
 \int_{0}^1 dt \, \mathbb{E}_{r, \mathcal{S}_r}\Big[\frac{\sum_{c\in r^{w_0}}\frac{E_{c,q=0, t,r, \mathcal{S}_r}}{2w_0+1}}{\Delta+ t\sum_{c\in r^{w_0}} \frac{E_{c,q=0, t,r, \mathcal{S}_r}}{2w_0+1}}  - \frac{\sum_{c\in r^{w_\tau}}\frac{E_{c,q=\tau, t,r, \mathcal{S}_r}}{2w_\tau+1}}{\Delta+ t\sum_{c\in r^{w_\tau}} \frac{E_{c,q=\tau, t,r, \mathcal{S}_r}}{2w_\tau+1}} \Big]
 +\smallO(L^{u-1}). 
\label{eq:diffiper_1}
\ee

The MMSE profile $E_{c, q, t, r, \mathcal{S}_r}$ depends only very weakly on $q$, $t$, $r$, $\mathcal{S}_r$ because $\mathcal{S}_r$ is sub-extensive, and this is actually the reason we chose it in such a way. More precisely, 
define $E_{c, \rho} = (\Gamma/L)\mathbb{E}\big[\sum_{i_c=1}^{N/\Gamma} (\langle X_{i_c} \rangle_{\rho} - S_{i_c})^2\big]$ where $\langle\! -\! \rangle_\rho$ is the average corresponding to the Hamiltonian of the $\rho$-ensemble (obtained for $t=1$ and $q=0$). We prove in sec.~\ref{proofLemma6.1} that for a.e. $h$ (see Corollary~\ref{independencecorollary}),
\begin{align}\label{independence}
E_{c, q, t, r, \mathcal{S}_r} = E_{c, \rho} + \smallO_{L}(1).
\end{align}
Since $M^u/(2\Gamma^u L) = {\cal O}(L^{u-1})$, and performing explicitly the expectation $\EE_{r,{\cal S}_r}$ (note that $E_{c,\rho}$ does not depend on ${\cal S}_r$) we then get for $u$ small enough and a.e $h$,
\begin{align}
i_\rho - i_{\rho+1} & = \frac{M^u}{2\Gamma^u L}
 \int_{0}^1 dt \, \frac{1}{\Gamma}\sum_{r=1}^\Gamma\Big[\frac{\sum_{c\in r^{w_0}}\frac{E_{c,\rho}}{2w_0+1}}{\Delta+ t\sum_{c\in r^{w_0}} \frac{E_{c,\rho}}{2w_0+1}}  - \frac{\sum_{c\in r^{w_\tau}}\frac{E_{c,\rho}}{2w_\tau+1}}{\Delta+ t\sum_{c\in r^{w_\tau}} \frac{E_{c,\rho}}{2w_\tau+1}} \Big]
 +\smallO(L^{u-1}) 
 \nonumber \\ &
 =
 \frac{M^u}{2\Gamma^u L}\, \frac{1}{\Gamma}\sum_{r=1}^\Gamma\Big[
 \ln\Big(\Delta+ \sum_{c\in r^{w_0}}\frac{E_{c,\rho}}{2w_0+1}\Big)  
  - 
  \ln\Big(\Delta+ \sum_{c\in r^{w_\tau}}\frac{E_{c,\rho}}{2w_\tau+1}\Big) \Big]
 +\smallO(L^{u-1}). 
\label{eq:diffiper_2}
\end{align}
Thanks to this identity we can easily compare the MI of the decoupled, coupled and homogeneous models.
%
%
%
%
%
\subsubsection[Comparison of homogeneous and coupled models]{\bf Comparison of homogeneous and coupled models}

We consider the $\rho\in\{0,\dots,\tau\}$ ensembles for the choice $w_0 \!=\! w$ (with $0\!<\!w\!<\!(\Gamma-1)/2$) and $w_\tau\! =\! (\Gamma-1)/2$. With this choice and because of the periodicity of the model 
\begin{align}
\frac{1}{2w_\tau +1}\sum_{c\in r^{w_\tau}} E_{c,\rho} = \frac{1}{\Gamma} \sum_{c=1}^\Gamma E_{c,\rho}
=
\frac{1}{\Gamma}\sum_{r=1}^\Gamma\frac{1}{2w+1}\sum_{c\in r^w} E_{c,\rho}\,.
\end{align}
Therefore \eqref{eq:diffiper_2} and concavity of the logarithm immediately imply
$i_\rho \le i_{\rho+1} + \smallO(L^{u-1})$.
Now since $i_{\rho=0}= i_{\Gamma, w}^{\rm per}$, the MI of the periodic SC system with coupling $w$, and $i_{\rho=\tau}= i_{\Gamma, (\Gamma -1)/2}^{\rm per}$ we obtain
$i_{\Gamma, w}^{\rm per} \le i_{\Gamma, (\Gamma -1)/2}^{\rm per} + \tau \,\smallO(L^{u-1})$ and since $\tau = {\cal O}(L^{1-u})$ we get (for a.e. $h$)
\be
i^{\rm per}_{\Gamma, w}\le i^{\rm per}_{\Gamma, (\Gamma-1)/2} +\smallO_L(1). 
\label{eq118}
\ee 
%

\subsubsection[Comparison of decoupled and coupled models]{\bf Comparison of decoupled and coupled models}

We now consider the $\rho\in\{0,\dots,\tau\}$ ensembles for the choice $w_0 \!=\! w$ (with $0\!<\!w\!<\!(\Gamma\!-\!1)/2$) and $w_\tau \!=\! 0$. Because of the periodicity of the model and with this choice
\begin{align}
\frac{1}{\Gamma}\sum_{r=1}^\Gamma \ln\Big(\Delta+ \sum_{c\in r^{w_\tau}}\frac{E_{c,\rho}}{2w_\tau +1} \Big)
& = \frac{1}{\Gamma} \sum_{r=1}^\Gamma \ln(\Delta+ E_{r,\rho})
=
\frac{1}{\Gamma}\sum_{r=1}^\Gamma \frac{1}{2w+1}\sum_{c\in r^w} \ln(\Delta+  E_{c,\rho})\,.
\end{align}
Convavity of the logarithm now implies $i_\rho \ge i_{\rho+1} + \smallO(L^{u-1})$ for a.e $h$, which leads to 
\be
 i_{\Gamma, w}^{\rm per} \geq  i_{\Gamma, 0}^{\rm per} + \smallO_L(1). 
\label{eq121}
\ee
\subsubsection[Super-additivity and existence of thermodynamic limit]{\bf Superadditivity and existence of thermodynamic limit}\label{thermo}
We note that the last inequality \eqref{eq121} is an instance of the superadditivity property of the mutual information. To understand this point consider the $L$ dependence of the mutual information 
$i^{\rm cs}_L$
of the basic system defined by the parameters $(L, N=LB, \alpha, B)$ (here $\alpha, B$ are held fixed). Consider three systems with sizes, $L$ even, $L_1=L/2$ and $L_2=L/2$. 
We interpret $L \, i^{\rm cs}_L$ as the mutual information of a coupled system with two blocks in the measurement matrix, i.e., $\Gamma=2$ and 
``complete coupling window'' between these two blocks, and we interpret
the sum $L_1 \,i^{\rm cs}_{L_1} + L_2 \,i^{\rm cs}_{L_2}$ as the total mutual information of two decoupled systems each of size $L/2 = (N/\Gamma)B$. Therefore \eqref{eq121} is nothing else than 
the superadditivity statement $i^{\rm cs}_L \geq \frac{L_1}{L} i^{\rm cs}_{L_1} + \frac{L_2}{L} i^{\rm cs}_{L_2} + \smallO_L(1)$. 
By the {\it same} proof as above this inequality can easily be generalized to any three systems of sizes $L_1$, $L_2$, $L$ satisfying $L= L_1+L_2$. In fact with a bit more work (see Appendix \ref{appendix_thermolimit}) we show that the terms 
$\smallO_L(1)$ can be improved to $\mathcal{O}(L^{-\eta})$. Namely we have
$i^{\rm cs}_L \geq \frac{L_1}{L}i^{\rm cs}_{L_1} +  \frac{L_2}{L} i^{\rm cs}_{L_2} + \mathcal{O}(L^{-\eta})$
for some small $\eta >0$. Improved, but standard, versions of Fekete's Lemma on super-additive sequences (see e.g., \cite{Ruelle} or Appendix B of \cite{bayati2013}, and references therein for even stronger results) then imply the existence 
of the limit $\lim_{L\to +\infty} i^{\rm cs}(L)$. 

We remark that in fact the inequalities \eqref{eq118} and \eqref{eq121} can also be improved to hold with an $\mathcal{O}(L^{-\eta})$ instead of $\smallO_L(1)$. Indeed these results ultimately rest on Proposition \ref{lemma:concentration}. However we do not directly need
this improvement to conclude the proof in the next paragraph. 
\subsubsection[Combining everything]{\bf Combining everything and conclusion of proof}
We now prove relation \eqref{eq:firstPart_lemmaInvMI} in Theorem~\ref{lemma:modelsEquiv}. 
First note that by construction the homogeneous model with $w=(\Gamma -1)/2$ is the model \eqref{eq:CSmodel} and 
the decoupled model with $w=0$ is a union of $\Gamma$ independent models of size $(M/\Gamma)\times (N/\Gamma)$. Existence of the thermodynamic limit therefore implies 
\be
\lim_{L\to\infty}i^{\rm per}_{\Gamma, (\Gamma-1)/2} = \lim_{L\to \infty} i_{\Gamma, 0}^{\rm per} =
\lim_{L\to\infty}i^{\rm cs}.
\label{eq:limitsEqual}
\ee
On the other hand
from \eqref{eq118} and \eqref{eq121} we have
\be
\lim_{L\to\infty} i^{\rm per}_{\Gamma, (\Gamma-1)/2}  \geq \limsup_{L\to\infty} i^{\rm per}_{\Gamma, w}\geq \liminf_{L\to\infty} i^{\rm per}_{\Gamma, w}\geq \lim_{L\to\infty} i^{\rm per}_{\Gamma,0}.
\label{ouf}
\ee
The existence of $\lim_{L\to +\infty} i^{\rm per}_{\Gamma, w}$ as well as
relation \eqref{eq:firstPart_lemmaInvMI} then follows from \eqref{eq:limitsEqual} and \eqref{ouf}.

The careful reader will have noticed that a priori we proved this result for a.e. $h$. However by the usual 
concavity in $h$ arguments we know that these limits are uniform in $h$ and thus the limits $h\to 0$ and $L\to \infty$ can be exchanged. The result is thus valid for $h=0$. 
This concludes the proof of \eqref{eq:firstPart_lemmaInvMI}. We defer the proof of \eqref{eq:secondPart_lemmaInvMI} in sec.~\ref{subsec:secondpartlemma39}. 
\subsection{Variation of MMSE profile}\label{proofLemma6.1}
In this section we prove \eqref{independence}. This is done through a global interpolation. 
%
\begin{lemma}[MMSE variation 1]  \label{lemma:smallEdiff}
For a.e $h$
we have $E_{c,q,t,r,\mathcal{S}_r} =E_{c,q,t =0,r,\mathcal{S}_r} + \smallO_L(1)$ for all $c\in\{1,\ldots,\Gamma\}$. Observe from \eqref{eq:int_hamiltonian_coupled} that $E_{c,q,t =0,r,\mathcal{S}_r}$ is 
independent of $q\in \{0, \tau\}$.
\end{lemma}
\begin{IEEEproof}
Note that thanks to the Nishimori identity \eqref{eq:NishimoriId}
\be
E_{c,q, t, r, \mathcal{S}_r} = \EE \Big[\frac{\Gamma}{L} \sum_{i_c=1}^{N/\Gamma} (\langle X_{i_c}\rangle_{q, t, r, \mathcal{S}_r} - S_{i_c})^2 \Big]= \EE[\langle \mathcal{E}_c\rangle_{q, t, r, \mathcal{S}_r}],
\ee
where we recall $\mathcal{E}_c=(\Gamma/L)\sum_{i_c=1}^{N/\Gamma} x_{i_c}(x_{i_c} - s_{i_c})$.
Thus from the fundamental theorem of calculus
\begin{align}
\int_{\epsilon}^a dh |E_{c, q, t, r, \mathcal{S}_r}-E_{c, q, 0,r, \mathcal{S}_r}| & =
 \int_{\epsilon}^a dh\Big|\int_{0}^t ds \EE\Big[ \frac{d }{ds}\langle\mathcal{E}_c \rangle_{q, s, r, \mathcal{S}_r} \Big]\Big|
\nonumber \\ &
\le \frac{1}{2\Delta}\sum_{\nu \in \mathcal{S}_r} \int_{0}^t ds \int_{\epsilon}^a dh\big|\EE[\langle\mathcal{E}_c G_\nu^{(q)} \rangle_{q,s,r, \mathcal{S}_r} -\langle\mathcal{E}_c \rangle_{q,s,r, \mathcal{S}_r}\langle G_\nu^{(q)}\rangle_{q,s,r, \mathcal{S}_r} ]\big|,  \label{eq:G0G1used}
\end{align}
where $G_\nu^{(q)} \defeq  [\bm{\phi}_q\bar \bx]_\nu^2 - [\bm{\phi}_q\bar \bx]_\nu z_\nu \sqrt{\Delta/s}$ and we used the Fubini theorem to exchange the order of the 
integrals (the integrand can be shown to be bounded and the integral is over a finite interval).
Concentration of $\mathcal{E}_c$ as in Proposition \ref{lemma:concentration} is valid for the present model. Indeed 
all is needed in the proofs of sec.~\ref{app:decoupling} are that all terms in the Hamiltonian can be interpreted as AWGN channels and Nishimori identities. Using that the $h$-integral of $\mathcal{E}_c$ concentrates we can check the integrand is $\smallO_L(1)$ for a.e 
$h$ by arguments already used at the end of sec.~\ref{app:T}. We give the main steps here for completeness. 

Set $\delta\mathcal{E}_c\! \defeq\!\mathcal{E}_c \!-\! \EE[\langle \mathcal{E}_c\rangle]$ where we drop the 
subscripts $q, s, r, \mathcal{S}_r$ in the Gibbs average. Using Cauchy-Schwarz
\begin{align}
\int_\epsilon^a dh \big|\EE[\langle \mathcal{E}_c G_\nu^{(q)} \rangle - \langle\mathcal{E}_c \rangle\langle G_\nu^{(q)}\rangle ] \big| &= \int_\epsilon^a dh \big|\EE[\langle \delta\mathcal{E}_c  G_\nu^{(q)} \rangle-\langle \delta\mathcal{E}_c \rangle  \langle G_\nu^{(q)} \rangle ]\big| \nonumber \\
&\le \int_\epsilon^a dh \EE[\langle |\delta\mathcal{E}_c G_\nu^{(q)} |\rangle] + \int_\epsilon^a dh \EE[\langle|\langle \delta\mathcal{E}_c\rangle  G_\nu^{(q)} |\rangle]
\nonumber \\ 
&\le \sqrt{\int_\epsilon^a dh\EE[\langle G_\nu^{(q) 2} \rangle ]} \Big(\sqrt{\int_\epsilon^a dh\EE[\langle \delta\mathcal{E}_c^2 \rangle ] } + \sqrt{\int_\epsilon^a dh\EE[\langle \delta\mathcal{E}_c\rangle^2 ]}\Big) \nonumber \\
&\le  2\sqrt{\int_\epsilon^a dh\EE[\langle G_\nu^{(q) 2} \rangle ] \int_\epsilon^a dh\EE[\langle \delta\mathcal{E}_c^2 \rangle ]}\,. 
\label{eq:last_integrand}
\end{align}
We check that $\mathbb{E}[\langle G_\nu^{(q)})^2\rangle]$ is bounded.
To do so we expand the square and use Gaussian integration by parts over the noise variables so that the only 
remaining terms are of the form $\mathbb{E}[[\bm{\Phi}_q\bS]_\nu^n\langle[\bm{\Phi}_q\bX]_\mu^m \rangle]$.
Finally we proceed exactly as in \eqref{eq78} using Cauchy-Schwarz again and Nishimori identities to reduce such terms to estimates of quantities $\mathbb{E}[[\bm{\Phi}_q \bS]_\lambda^\ell]$ which are $\mathcal{O}(1)$. 
Now since by Proposition \ref{lemma:concentration} $\int_\epsilon^a dh\EE[\langle \delta\mathcal{E}_c^2\rangle] \!=\! \mathcal{O}(L^{-1/10})$ 
for a.e $h$, we obtain that \eqref{eq:last_integrand} is $\mathcal{O}(L^{-1/20})$. Using this and recalling that $|\mathcal{S}_r| = \mathcal{O}(L^u)$, one reaches that \eqref{eq:G0G1used} is ${\cal O}(L^{u-1/20})$. To conclude the proof of the lemma choose $u$ small enough and then use the same arguments (found below \eqref{conclemma}) that led to the identity \eqref{concen} (the MMSE's are bounded uniformly in $L$).
\end{IEEEproof}
\begin{corollary}[MMSE variation 2]\label{independencecorollary}
For a.e $h$
we have $E_{c,q,t,r,\mathcal{S}_r} =E_{c,\rho} + \smallO_L(1)$ for all $c\in\{1,\ldots,\Gamma\}$.
\end{corollary}
\begin{IEEEproof}
Recall that $E_{c, \rho}$ was defined as the MMSE of block $c$ for the $\rho$-ensemble, i.e. the Hamiltonian \eqref{eq:int_hamiltonian_coupled} corresponding to $t=1$ and $q=0$. Thus 
$E_{c, \rho} = E_{c,q=0,t=1,r,\mathcal{S}_r}$ and from Lemma \ref{lemma:smallEdiff} $E_{c,q=0,t=1,r,\mathcal{S}_r}
= E_{c,q=0,t,r,\mathcal{S}_r} +\smallO_L(1)$. Thus $E_{c, \rho} = E_{c,q=0,t,r,\mathcal{S}_r} +\smallO_L(1)$ and this proves the corollary for $q=0$. For the case $q=\tau$ we note that Lemma 
\ref{lemma:smallEdiff} also implies $E_{c,q=0,t=1,r,\mathcal{S}_r} = E_{c,q=0,t=0,r,\mathcal{S}_r} +\smallO_L(1)$ and since the $t=0$ model is independent of $q$ we have $E_{c,q=0,t=0,r,\mathcal{S}_r} = E_{c,q=\tau,t=0,r,\mathcal{S}_r}$, so
$E_{c, \rho} = E_{c,q=\tau,t=0,r,\mathcal{S}_r} + \smallO_L(1)$. Applying Lemma 
\ref{lemma:smallEdiff} once more we have $E_{c,q=\tau,t=0,r,\mathcal{S}_r} = E_{c,q=\tau,t,r,\mathcal{S}_r} + \smallO_L(1)$ and we conclude 
$E_{c, \rho} = E_{c,q=\tau,t,r,\mathcal{S}_r} + \smallO_L(1)$.
\end{IEEEproof}
\subsection{Invariance of the mutual information between the periodic and seeded SC systems} \label{subsec:secondpartlemma39}
We conclude this section by proving that in a proper limit, the seeded and periodic SC models have identical MI. We show that the difference between MI of these models is $\mathcal{O}(w/\Gamma)$ and thus vanishes when $\Gamma\!\to\!\infty$ for a fixed coupling window. As a consequence from \eqref{eq:firstPart_lemmaInvMI} this proves equation \eqref{eq:secondPart_lemmaInvMI} in the second part of Theorem~\ref{lemma:modelsEquiv}.

The arguments below are essentially the same as those developed in \cite{barbier2016proof,barbier2016threshold}.
The only difference between the periodic and seeded SC systems is the boundary condition: The signal components belonging to the $8w$ boundary blocks $\in \mathcal{B} \!=\! \{1:4w\}\cup\{\Gamma-4w+1:\Gamma\}$ are known for the seeded system. Thus the Hamiltonians of the periodic and seeded SC systems, $\mathcal{H}^{\rm per}$ and $\mathcal{H}^{\rm seed}$, satisfy the identity 
\begin{align}
\mathcal{H}^{\rm seed}(\bx) &= \mathcal{H}^{\rm per}(\bx) - \delta \mathcal{H}(\bx), \label{eq:linkHamilOpSeeded}\\
\delta \mathcal{H}(\bx) &= \frac{1}{2\Delta}\sum_{r \in \mathcal{B}}\sum_{\mu_r=1}^{M/\Gamma} \Big([\bm{\phi}\bar \bx]_{\mu_r}- z_{\mu_r}\sqrt{\Delta}\Big)^2 \label{deltaH} ,
\end{align}
recalling that $\{\mu_r\}$ is the set of ``measurement indices'' belonging to the block-row $r$ in the block decomposition of Fig.~\ref{fig:opSpCoupling}. The dependence of the Hamiltonians w.r.t the quenched random variables is implicit. Let $\mathcal{Z}^{\rm seed}$ and $\langle\! -\!\rangle_{\rm seed}$ be the partition function and posterior mean, respectively, associated to $\mathcal{H}^{\rm seed}(\bx)$:
\begin{align}
\langle A(\bX)\rangle_{\rm seed} = \frac{1}{\mathcal{Z}^{\rm seed}} \int d\bx\, A(\bx)  
e^{-\mathcal{H}^{\rm seed}(\bx)} \prod_{l=1}^L P_0(\bx_l),\quad 
\mathcal{Z}^{\rm seed} = \int d\bx\, e^{-\mathcal{H}^{\rm seed}(\bx)} \prod_{l=1}^L P_0(\bx_l),
\end{align}
and similarly with $\mathcal{Z}^{\rm per}$, $\langle -\rangle_{\rm per}$ for $\mathcal{H}^{\rm per}(\bx)$. One obtains from \eqref{eq:linkHamilOpSeeded} the following identities
\begin{align}
i^{\rm per}_{\Gamma,w} & = -\frac{\alpha B}{2} - \frac{1}{L}\EE[\ln (\mathcal{Z}^{\rm per})] =  -\frac{\alpha B}{2} - \frac{1}{L}\EE[\ln( \mathcal{Z}^{\rm seed} \langle e^{-\delta \mathcal{H}} \rangle_{\rm seed} )] = i^{\rm seed}_{\Gamma,w} - \frac{1}{L}\EE[\ln (\langle e^{-\delta \mathcal{H}} \rangle_{\rm seed}) ],\label{eq:Z_openVSclosed} \\
i^{\rm seed}_{\Gamma,w} & = -\frac{\alpha B}{2} - \frac{1}{L}\EE[\ln (\mathcal{Z}^{\rm seed})] = -\frac{\alpha B}{2} - \frac{1}{L}\EE[\ln( \mathcal{Z}^{\rm per} \langle e^{\delta \mathcal{H}} \rangle_{\rm per} )]=i^{\rm per}_{\Gamma,w} - \frac{1}{L}\EE[\ln (\langle e^{\delta \mathcal{H}} \rangle_{\rm per}) ].\label{eq:Z_openVSclosed_2}
\end{align}
Using the convexity of the exponential, we get
\begin{equation}\label{eq:openVSclosed_sandwich}
i_{\Gamma,w}^{\rm{seed}} + \frac{1}{L}\mathbb{E}[ \langle \delta \mathcal{H} \rangle_{\rm per}] \le i_{\Gamma,w}^{\rm per} \le i_{\Gamma,w}^{\rm{seed}} + \frac{1}{L}\mathbb{E}[ \langle \delta \mathcal{H} \rangle_{\rm seed}].
\end{equation}
Due to the knowledge of the signal components at the $8w$ boundary blocks for the seeded system, one gets straightforwardly (set $\bar{\bx}=0$ in \eqref{deltaH})
\be\label{eq:openVSclosed_order_0}
\frac{1}{L}\mathbb{E}[ \langle \delta \mathcal{H} \rangle_{\rm seed} ] = \frac{4w \alpha B}{\Gamma} = \mathcal{O}(w/\Gamma).
\ee
Let us now study the lower bound. Using a Gaussian integration by part (as done in \eqref{eq:ippNoise}) and the Nishimori identity \eqref{eq:Nish_squarreInOUt}, we obtain 
\begin{align}\label{eq:openVSclosed_order}
\frac{1}{L}|\mathbb{E}[ \langle \delta \mathcal{H} \rangle_{{\rm per}} ]| \le \frac{4w\alpha B}{\Gamma}\mathcal{O}(1) = \mathcal{O}(w/\Gamma).
\end{align}
%
Therefore, both bounds tighten as $\Gamma\to \infty$ for any fixed $w$. Taking 
$L\!\to\! \infty$ and then $\Gamma \!\to\! \infty$ we get from \eqref{eq:openVSclosed_sandwich}, \eqref{eq:openVSclosed_order_0}, \eqref{eq:openVSclosed_order} that
\be
\lim_{\Gamma\to\infty}\lim_{L\to\infty} i_{\Gamma, w}^{\rm per} = \lim_{\Gamma\to\infty}\lim_{L\to\infty} i_{\Gamma, w}^{\rm seed}.
\ee
Combining this with \eqref{eq:firstPart_lemmaInvMI} yields \eqref{eq:secondPart_lemmaInvMI} and proves the second part of Theorem~\ref{lemma:modelsEquiv}.
\section{Concentration properties}\label{app:decoupling}
The main goal of this section is to prove \eqref{concen} holds.
Recall $\mathcal{E}_c = (\Gamma/L)\sum_{i_c=1}^{N/\Gamma} x_{i_c} (x_{i_c} - s_{i_c})$.

\begin{proposition}[Concentration of ${\cal E}_c$]\label{lemma:concentration}
For any fixed $a >\epsilon > 0$ the following holds:
\begin{align} \label{conclemma}
\int_{\epsilon}^a dh \mathbb{E}\big[\big\langle \big( \mathcal{E}_c - \EE[\langle \mathcal{E}_c \rangle_{t,h}]\big)^2\big\rangle_{t,h} \big] =\mathcal{O}(L^{-1/10}).
\end{align}
\end{proposition}

First let us clarify why this implies \eqref{concen}. The integrand in \eqref{conclemma} is bounded uniformly in $L$ (in our setting this is obvious because the prior has bounded support). Therefore by Lebesgue's dominated convergence theorem $\lim_{L\to \infty} \int_{\epsilon}^a dh \mathbb{E}[\langle ( \mathcal{E}_c - \EE[\langle \mathcal{E}_c \rangle_{t,h}])^2\rangle_{t,h} ]=\int_{\epsilon}^a dh \lim_{L\to \infty}\mathbb{E}[\langle ( \mathcal{E}_c - \EE[\langle \mathcal{E}_c \rangle_{t,h}])^2\rangle_{t,h} ] =0$ which implies that \eqref{concen} holds for a.e. $h\in [\epsilon, a]$. Here $\epsilon>0$ is as small as we wish and $a$ can be arbitrarily large, thus for our purposes we can assert that \eqref{concen} holds for a.e. $h>0$.

In order to make the proof more pedagogic, we will restrict to the case $\Gamma\!=\!1$, that is for model 
\eqref{eq:int_hamiltonian}. All steps straightforwardly generalize to the SC case $\Gamma\!>\!1$, 
model \eqref{eq:int_hamiltonian_SC}, at the expense of more heavy notations.
The proof relies on concentration properties of $\mathcal{L}$ given by \eqref{eq:def_L}.
Once these are established, the concentration of 
\begin{align}
{\cal E} \defeq \frac{1}{L}\sum_{i=1}^{N} \bar x_{i} x_{i}  = \frac{1}{L}\sum_{i=1}^{N} (x_{i} - s_{i}) x_{i}
\label{eq:defEps}
\end{align}
follows as a consequence.
\subsection{Concentration of $\mathcal{L}$ on $\langle \mathcal{L} \rangle_{t,h}$}
We first show the following lemma which expresses concentration of $\mathcal{L}$ around its posterior mean.
\begin{lemma}[Concentration of $\mathcal{L}$] \label{lemma:concentration_1}
For any fixed $a >\epsilon > 0$, the following holds:
\begin{align}
\int_{\epsilon}^{a} dh \mathbb{E} \big[\big\langle \big|{\cal L} - \langle {\cal L} \rangle_{t,h} \big| \big\rangle_{t,h}\big] = \mathcal{O}(L^{-1/2}). \label{eq:concentrationL_0}
\end{align}
\end{lemma}
\begin{IEEEproof}
Let us evaluate the integral
\begin{align}
\int_{\epsilon}^{a} dh \mathbb{E}\big[ \langle {\cal L}^2 \rangle_{t,h} - \langle {\cal L}\rangle_{t,h}^2 \big] 
= - \frac{1}{L} \int_{\epsilon}^{a} dh \frac{d^2i_{t,h}}{dh^2} + {\cal O}(L^{-1}) = \frac{1}{L}\frac{di_{t,h}}{dh}\Big|_{h=\epsilon} - \frac{1}{L}\frac{di_{t,h}}{dh}\Big|_{h=a} + {\cal O}(L^{-1}), \label{eq:concentrationL_1}
\end{align}
where the first equality follows from \eqref{eq:seconde_der_h_fInt}. The ${\cal O}(L^{-1})$ term comes from the second term in \eqref{eq:seconde_der_h_fInt} which is ${\cal O}(1)$ (this is easily shown by integrating by parts the noise, and in addition the term with absolute value gives also a finite contribution). The proof is finished by noticing that  
the first derivatives of the MI are $\mathcal{O}(1)$ uniformly in $L$. This last point follows from general arguments on the $L\to \infty$ limit of concave in $h>0$ functions (recall Lemma~\ref{lemma:f_concav_h}). But an explicit check is also possible:
from \eqref{eq:first_der_h_fInt}, \eqref{eq:def_L} one obtains by integration by part w.r.t the standard Gaussian variable $\widehat z_{i}$
\begin{align}
\frac{di_{t,h}}{dh} &= \frac{1}{2L} \sum_{i=1}^{N}  \Big(\EE[\langle X_{i}^2\rangle_{t,h}] - 2\EE[\langle X_{i}S_{i}\rangle_{t,h}] - \EE[\langle X_{i}^2\rangle_{t,h}] + \EE[\langle X_{i}\rangle_{t,h}^2]\Big) + \frac{v}{2} + \frac{s_{\rm max}B }{\sqrt{2\pi h}} 
\nonumber\\ &
= -\frac{1}{2L} \sum_{i=1}^{N} \EE[\langle X_{i}\rangle_{t,h}^2]+ \frac{v}{2} + \frac{s_{\rm max}B }{\sqrt{2\pi h}} 
= 
\frac{1}{2L} \sum_{i=1}^{N} \EE[(\langle X_{i}\rangle_{t,h} - S_i)^2]+ \frac{s_{\rm max}B }{\sqrt{2\pi h}} 
= \mathcal{O}(1), \label{dfintdh_order1}
\end{align} 
where the second equality is due to the Nishimori identities, and the last uses that the signal is bounded. Finally, we 
obtain \eqref{eq:concentrationL_0} from \eqref{eq:concentrationL_1}, \eqref{dfintdh_order1} and Cauchy-Schwarz.
\end{IEEEproof}
\subsection{Concentration of $\langle \mathcal{L}\rangle_{t,h}$ on $\EE[\langle \mathcal{L}\rangle_{t,h}]$} \label{concetrationLdisorder}
We will use a concentration statement for the free energy at fixed measurement realization. Recall that by definition $f_{t,h}(\mathring{\by}) \defeq - \ln (\mathcal{Z}_{t,h}(\mathring{\by}))/L$ with $\mathring{\by}$ the concatenation of all measurements and measurement matrix.
\begin{proposition}[Concentration of the free energy]\label{cor:fs_minus_meanfs_small}
For any $0<\eta <1/4$ we have 
\begin{align}
\mathbb{E}\big[\big|f_{t,h}(\mathring \by) - 
\mathbb{E}[f_{t,h}(\mathring \bY)] \big| \big] = \mathcal{O}(L^{- \eta})\,.
\end{align}
\end{proposition}
A similar result has already been obtained in \cite{korada2010} for the CDMA problem. Here the proof has to be slightly generalised and uses the Ledoux-Talagrand and McDiarmid concentration theorems in conjunction (see appendix \ref{appendix_concentration}).

Let us use the shorthand notations
$f_h \defeq f_{t,h}(\mathring \by)$ and $\bar f_h\defeq \mathbb{E}[f_{t,h}(\mathring \bY)]$ which emphasize that we will look at small $h$-perturbations.
By \eqref{eq:first_der_h_fInt} 
%
\begin{align}
&\frac{d f_h}{dh}- \frac{d\bar f_h}{dh} = \langle {\cal L} \rangle_{t,h} -\mathbb{E}[ \langle {\cal L} \rangle_{t,h} ] + a_1 + a_2 , \label{eq:q_f_relation}\\
&a_1 = \frac{1}{2L} \sum_{i=1}^L s_i^2 - \frac{v}{2}, \qquad a_2 = \frac{s_{\rm max}}{2\sqrt{h}L} \sum_{i=1}^N |\widehat z_i| -\frac{s_{\rm max}B }{\sqrt{2\pi h}}. \label{as}
\end{align}
The concavity of the free energy in $h$ (Lemma~\ref{lemma:f_concav_h}) allows to write the following inequalities for any $\delta >0$:
\begin{align}
&\frac{d f_{h}}{dh} - \frac{d\bar f_{h}}{dh} \le \frac{f_{h-\delta}-f_{h}}{\delta} - \frac{d\bar f_{h}}{dh} \le \frac{f_{h-\delta} - \bar f_{h-\delta}}{\delta} 
- \frac{f_{h} - \bar f_{h}}{\delta} + \frac{d\bar f_{h-\delta}}{dh} - \frac{d\bar f_{h}}{dh},  \label{eq:firstBound_df}\\
&\frac{d f_{h}}{dh} - \frac{d\bar f_{h}}{dh} \ge \frac{f_{h+\delta} - \bar f_{h+\delta}}{\delta} - \frac{f_{h} - \bar f_{h}}{\delta} 
+ \frac{d\bar f_{h+\delta}}{dh} - \frac{d\bar f_{h}}{dh}. \label{eq:secondBound_df}
\end{align}
Note that the difference between the derivatives appearing here cannot be considered small because at a first order transition point the derivatives have jump discontinuities. We now have all the necessary tools to show the second concentration.
\begin{lemma}[Concentration of $\langle \mathcal{L}\rangle_{t,h}$] \label{lemma:concentration_2}
For any fixed $a >\epsilon > 0$ the following holds for any $0<\eta <1/4$:
\begin{align} \label{eq:concentrationL_2}
\int_{\epsilon}^a dh \mathbb{E}\big[\big| \langle \mathcal{L} \rangle_{t,h} - \EE[\langle \mathcal{L} \rangle_{t,h} ] \big| \big] = {\cal O}(L^{-\eta/2}).
\end{align}
\end{lemma}
\begin{IEEEproof}
Note that due to the concavity Lemma~\ref{lemma:f_concav_h}, we have
\begin{align}
-C_h^-&\defeq \frac{d\bar f_{h+\delta}}{dh} - \frac{d\bar f_{h}}{dh} \le 0, \qquad C_h^+\defeq\frac{d\bar f_{h-\delta}}{dh} - \frac{d\bar f_{h}}{dh} \ge 0. \label{175}
\end{align}
Using \eqref{eq:q_f_relation}, \eqref{eq:firstBound_df}, \eqref{eq:secondBound_df}, \eqref{175} we can write
\begin{align}
&\frac{f_{h+\delta} \!-\! \bar f_{h+\delta}}{\delta}\! -\! \frac{f_{h} \!-\! \bar f_{h}}{\delta} 
\!-\!C_h^- \le \mathbb{E}[\langle {\cal L} \rangle_{t,h}] \!-\! \langle {\cal L} \rangle_{t,h} \!+\! a_1 \!+\! a_2 \le \frac{f_{h-\delta} 
\!-\! \bar f_{h-\delta}}{\delta} \!-\! \frac{f_{h} \!-\! \bar f_{h}}{\delta}\!+\! C_h^+ 
\end{align}
which implies
\begin{align}
&| \langle {\cal L} \rangle_{t,h} - \mathbb{E}[\langle {\cal L} \rangle_{t,h}] | \le  \sum_{u\in\{h+\delta,h,h-\delta\}} \frac{|f_{u} - \bar f_{u}|}{\delta} + C_h^+ + C_h^- +|a_1| + |a_2|. 
\label{eq:absBound1}
\end{align}
We will now average over 
all quenched random variables and use Corollary~\ref{cor:fs_minus_meanfs_small}. We remark that by the central limit theorem combined with Cauchy-Schwarz, 
and as the noise and signal both have i.i.d components with finite first and second 
moments, $\mathbb{E}[|a_{1,2}|]=\mathcal{O}(L^{-1/2})$. Therefore 
\begin{align}
&\mathbb{E}\big[\big|\langle {\cal L} \rangle_{t,h} - \mathbb{E}[\langle {\cal L} \rangle_{t,h}]|\big] \le \delta^{-1}\mathcal{O}(L^{-\eta}) 
+ C_h^+ + C_h^- + O(L^{-1/2}). 
\label{eq:absBound}
\end{align}
Then  integrating \eqref{eq:absBound} and using \eqref{175} we get
\begin{align}
\int_\epsilon^a dh\mathbb{E}\big[\big|\mathbb{E}[\langle {\cal L} \rangle_{t,h}] - 
\langle {\cal L} \rangle_{t,h}\big|\big]\!\le \!(\bar f_{a-\delta} \!-\! \bar f_{a})\! -\! 
(\bar f_{\epsilon-\delta} \!-\! \bar f_{\epsilon}) \!-\! (\bar f_{a+\delta} \!-\! \bar f_{a}) \!+\! (\bar f_{\epsilon+\delta} \!-\! \bar f_{\epsilon}) \!+\! \delta^{-1}\mathcal{O}(L^{-\eta}) \!+\! O(L^{-1/2}). 
\label{eq:integrqtes}
\end{align}
By the mean value theorem $\bar f_{a-\delta} - \bar f_{a} = -\delta \frac{d\bar f_{h}}{dh}|_{\widetilde h}$
for a suitable $\widetilde h\in[a-\delta,a]$. Since the first derivative of the free energy is $\mathcal{O}(1)$ (the argument is the same as for the MI, 
use \eqref{eq:first_der_h_fInt} or \eqref{dfintdh_order1})
we have $\bar f_{a-\delta} - \bar f_{a}= \delta\,\mathcal{O}(1)$.
We proceed similarly for the other average free energy differences.
Now choose $\delta=L^{-\eta/2}$. All this implies with \eqref{eq:integrqtes} that
\begin{align}
&\int_\epsilon^a \!dh\mathbb{E}\big[\big|\mathbb{E}[\langle {\cal L} \rangle_{t,h}] \!-\! \langle {\cal L} \rangle_{t,h}\big|\big] 
\!=\! \mathcal{O}(L^{-\eta/2}) + \mathcal{O}(L^{-1/2}). 
\label{lastBeforeRes}
\end{align}
Since $0<\eta<1/4$ the leading term is $\mathcal{O}(L^{-\eta/2})$ and this gives the result \eqref{eq:concentrationL_2}.
\end{IEEEproof}
\subsection{Concentration of ${\cal E}$ on $\EE[\langle \mathcal{E}\rangle_{t,h}]$:
Proof of Proposition \ref{lemma:concentration}}\label{subsec:concentration_E}
It will be convenient to use the following {\it overlap} $q_{\bx,\bx'} \defeq (1/L)\sum_{i=1}^N x_i x_i'$.
%
%
From \eqref{eq:defEps}, ${\cal E} = q_{\bx,\bx} -q_{\bx,\bs}$. We have that for any function $g$  
such that $\vert g(x)\vert \le 1$,
\begin{align}
\int_{\epsilon}^a dh \big | \mathbb{E}[\langle\mathcal{L} g\rangle_{t,h}]  - \EE[\langle \mathcal{L}\rangle_{t,h}] 
\EE[\langle g \rangle_{t,h} ] \big|
&
=
\int_{\epsilon}^a dh  \big | \mathbb{E}\big[\langle(\mathcal{L} - \EE[\langle \mathcal{L}\rangle_{t,h}]) g \rangle_{t,h}\big]\big |
\nonumber \\ &
\leq
\int_{\epsilon}^a dh \mathbb{E}\big[\big\langle\big| \mathcal{L}  - \EE[\langle \mathcal{L} \rangle_{t,h} ]  \big| \big\rangle_{t,h} \big] ={\cal O}(L^{-\eta/2}),
\label{eq:two_terms}
\end{align}
where the last equality is obtained combining the triangle inequality with Lemma~\ref{lemma:concentration_1} and Lemma~\ref{lemma:concentration_2}. Now consider $g = {\cal E}/{\cal E}_{\rm max}$ where
${\cal E}_{\rm max}= 2Bs_{\rm max}^2$ for a discrete prior. At the end of the section we show that
\begin{align}
\mathbb{E}[\langle\mathcal{L}{\cal E}\rangle_{t,h}] - \mathbb{E}[\langle\mathcal{L}\rangle_{t,h}]\mathbb{E}[\langle\mathcal{E}\rangle_{t,h}] &=\frac{1}{2}\big(\mathbb{E}[\langle{\cal E}^2\rangle_{t,h} ] - \mathbb{E}[\langle{\cal E}\rangle_{t,h} ]^2+ \mathcal{T}_1+\mathcal{T}_2\big), 
\label{eq:LE_minusLE}\\ 
\mathcal{T}_1 &=\mathbb{E}[\langle{\cal E}\rangle_{t,h} ]\mathbb{E}[\langle q_{\bx,\bx}\rangle_{t,h} ] - \mathbb{E}[\langle{\cal E} q_{\bx,\bx}\rangle_{t,h} ], 
 \label{T1}\\
\mathcal{T}_2 &= \mathbb{E}[\langle{\cal E} q_{\bx,\bx'}\rangle_{t,h} ]- \mathbb{E}[\langle{\cal E} q_{\bx,\bs}\rangle_{t,h} ]. 
\label{eq102}
\end{align}

Let us show that $|\mathcal{T}_1|$ is small. To do so we first notice the following property.
\begin{lemma}[Concentration of self-overlap] \label{remark:concentrationSelfOverlap}
The self-overlap concentrates, i.e., $$\EE[\langle (q_{\bx,\bx}-\EE[\langle q_{\bx,\bx}\rangle_{t,h}])^2\rangle_{t,h}] = {\cal O}(L^{-1}).$$
\end{lemma} 
\begin{IEEEproof}
By the Nishimori identity \eqref{eq:NishId_0} we have $\EE[\langle q_{\bx,\bx}^n\rangle_{t,h}]=\EE[q_{\bs,\bs}^n]$. In particular $\EE[\langle (q_{\bx,\bx}-\EE[\langle q_{\bx,\bx}\rangle_{t,h}])^2\rangle_{t,h}]=\EE[\langle q_{\bx,\bx}^2\rangle_{t,h}] - \EE[\langle q_{\bx,\bx}\rangle_{t,h}]^2 = \EE[q_{\bs,\bs}^2] - \EE[q_{\bs,\bs}]^2$. To conclude the proof, we use that the signal components are i.i.d with finite mean and variance, and thus the central limit theorem implies that $q_{\bs,\bs}$ tends to a Gaussian random variable with finite mean and a variance $\EE[q_{\bs,\bs}^2] - \EE[q_{\bs,\bs}]^2={\cal O}(L^{-1})$.
\end{IEEEproof}

Lemma~\ref{remark:concentrationSelfOverlap} allows to prove that $\vert\mathcal{T}_1\vert = \mathcal{O}(L^{-1/2})$. From \eqref{T1}, using Cauchy-Schwarz and as $\EE[\langle{\cal E}^2\rangle_{t,h}]={\cal O}(1)$ (the signal is bounded) we obtain
\begin{align}
|\mathcal{T}_1| &= |\mathbb{E}[\langle{\cal E} (\mathbb{E}[\langle q_{\bx,\bx}\rangle_{t,h} ] \!-\! q_{\bx,\bx}) \rangle_{t,h} ]|
\le \sqrt{\EE[\langle{\cal E}^2\rangle_{t,h}] \mathbb{E}[\langle(q_{\bx,\bx} \!-\!\mathbb{E}[\langle q_{\bx,\bx}\rangle_{t,h} ])^2 \rangle_{t,h} ]} = {\cal O}(L^{-1/2}).
\label{smallnessT1}
\end{align}

Let us now consider $\mathcal{T}_2$ and show it is positive. From \eqref{eq102} 
\begin{align}
\mathcal{T}_2 &= \mathbb{E}[\langle (q_{\bx,\bx} - q_{\bx,\bs}) q_{\bx,\bx'}\rangle_{t,h} ]- \mathbb{E}[\langle(q_{\bx,\bx} - q_{\bx,\bs}) q_{\bx,\bs}\rangle_{t,h} ] = \mathbb{E}[\langle q_{\bx,\bs}^2\rangle_{t,h} ]- \mathbb{E}[\langle q_{\bx,\bs} q_{\bx,\bx'}\rangle_{t,h} ], \label{200}
\end{align}
where we used the Nishimori identity $\mathbb{E}[\langle q_{\bx,\bx}q_{\bx,\bx'}\rangle_{t,h}]=\mathbb{E}[\langle q_{\bx,\bx}q_{\bx,\bs}\rangle_{t,h}]$. Now by convexity of a parabola,
\begin{align}
\mathcal{T}_2 \ge \mathbb{E}[\langle q_{\bx,\bs}\rangle_{t,h}^2 ]- \mathbb{E}[\langle q_{\bx,\bs} q_{\bx,\bx'}\rangle_{t,h} ] &= \frac{1}{L^2}\sum_{i,j=1}^N\Big(\mathbb{E}[\langle X_i\rangle_{t,h}\langle X_j\rangle_{t,h} S_iS_j] - \mathbb{E}[S_i \langle X_j'X_iX_j\rangle_{t,h}]\Big)\nonumber \\
&= \frac{1}{L^2}\sum_{i,j=1}^N\Big(\mathbb{E}[\langle X_i X_j' X_i''X_j''\rangle_{t,h}] - \mathbb{E}[\langle X_i''X_j' X_iX_j\rangle_{t,h}]\Big) = 0, \label{186}
\end{align}
where in order to reach the last line we replaced the signal by an independent replica $\bx''$ thanks to the Nishimori identity and used that all replicas play the same role for the last equality. Using \eqref{eq:LE_minusLE}, \eqref{smallnessT1}, \eqref{186},
\begin{align}
\mathbb{E}[\langle\mathcal{L}\mathcal{E}\rangle_{t,h}] - \mathbb{E}[\langle\mathcal{L}\rangle_{t,h}]\mathbb{E}[\langle \mathcal{E}\rangle_{t,h}] \ge \frac{1}{2}\Big(\mathbb{E}[\langle{\cal E}^2\rangle_{t,h} ] - \mathbb{E}[\langle{\cal E}\rangle_{t,h} ]^2\Big) + \mathcal{O}(L^{-1/2}). \label{eq:LE_minusLE_2}
\end{align}
Finally, combining this last inequality with \eqref{eq:two_terms} implies for fixed $a>\epsilon>0$,
\begin{align}
\int_{\epsilon}^a dh (\mathbb{E}[\langle{\cal E}^2\rangle_{t,h}]  -  \EE[\langle {\cal E} \rangle_{t,h} ]^2 ) = \mathcal{O}(L^{-\eta/2})\,.
\end{align}
Now we choose the (sub-optimal) value $\eta = 1/5$ (recall one must have $0<\eta<1/4$). This ends the proof of Proposition~\ref{lemma:concentration}.

It remains to prove \eqref{eq:LE_minusLE}--\eqref{eq102}. Consider $\mathbb{E}[\langle\mathcal{L}{\cal E}\rangle_{t,h}] - \mathbb{E}[\langle\mathcal{L}\rangle_{t,h}]\mathbb{E}[\langle\mathcal{E}\rangle_{t,h}]$. From \eqref{eq:def_L} using integration by parts w.r.t $\widehat z_i$ one obtains for the first term
\begin{align}
\mathbb{E}[\langle\mathcal{L}{\cal E}\rangle_{t,h}] &= \frac{1}{2}\Big(\mathbb{E}[\langle{\cal E}^2\rangle_{t,h} ] - \mathbb{E}[\langle{\cal E} q_{\bx,\bs} \rangle_{t,h}] - \mathbb{E}\Big[\Big\langle{\cal E} \frac{1}{\sqrt{h}L} \sum_{i=1}^N X_i\widehat Z_i\Big\rangle_{t,h} \Big]\Big) \nonumber \\
&=\frac{1}{2}\big(\mathbb{E}[\langle{\cal E}^2\rangle_{t,h} ] - \mathbb{E}[\langle{\cal E} q_{\bx,\bs} \rangle_{t,h}] - \mathbb{E}[\langle{\cal E} q_{\bx,\bx}\rangle_{t,h} ] + \mathbb{E}[\langle{\cal E} q_{\bx,\bx'}\rangle_{t,h} ]\big). \label{192}
\end{align}
Note here that as ${\cal E}$ is a function of $\bx$, $\bs$ then $\mathbb{E}[\langle{\cal E} q_{\bx,\bs} \rangle_{t,h}]\neq\mathbb{E}[\langle{\cal E} q_{\bx,\bx'} \rangle_{t,h}]$.

Now consider the second term. Using again an integration by part and the Nishimori identity $\mathbb{E}[\langle q_{\bx,\bx'}\rangle_{t,h} ]=\mathbb{E}[\langle q_{\bx,\bs}\rangle_{t,h} ]$, 
\begin{align}
\mathbb{E}[\langle\mathcal{L}\rangle_{t,h}]\mathbb{E}[\langle\mathcal{E}\rangle_{t,h}] & = \frac{1}{2}\Big(\mathbb{E}[\langle{\cal E}\rangle_{t,h} ]^2 - \mathbb{E}[\langle{\cal E}\rangle_{t,h} ] \mathbb{E}[\langle q_{\bx,\bs}\rangle_{t,h}] - \mathbb{E}[\langle{\cal E}\rangle_{t,h} ] \mathbb{E}\Big[\Big\langle \frac{1}{\sqrt{h}L} \sum_{i=1}^N X_i\widehat Z_i\Big\rangle_{t,h} \Big]\Big) \nonumber \\
&=\frac{1}{2}\big(\mathbb{E}[\langle{\cal E}\rangle_{t,h} ]^2 - \mathbb{E}[\langle{\cal E}\rangle_{t,h} ] \mathbb{E}[\langle q_{\bx,\bs}\rangle_{t,h}] - \mathbb{E}[\langle{\cal E}\rangle_{t,h} ]\mathbb{E}[\langle q_{\bx,\bx}\rangle_{t,h} ] + \mathbb{E}[\langle{\cal E}\rangle_{t,h} ]\mathbb{E}[\langle q_{\bx,\bx'}\rangle_{t,h} ]\big) \nonumber \\
&=\frac{1}{2}\big(\mathbb{E}[\langle{\cal E}\rangle_{t,h} ]^2 - \mathbb{E}[\langle{\cal E}\rangle_{t,h} ]\mathbb{E}[\langle q_{\bx,\bx}\rangle_{t,h} ] \big).\label{193}
\end{align}
The difference of \eqref{192} and \eqref{193} yields \eqref{eq:LE_minusLE}, \eqref{T1}, \eqref{eq102}.

\section{MMSE relation for the CS model: Proof of Theorem \ref{thmMMSE}} \label{secMMSErelation}
The goal of this section is to prove relation \eqref{xymmse} in Theorem \ref{thmMMSE}. The general ideas that go in this proof are similar to those of sec.~\ref{app:T}. There, we used a decoupling property  based on the concentration of 
${\cal E}_c$ for a.e. $h$ (see the proof of \eqref{y2}). Here instead we will prove that 
\be \label{Q}
{\cal Q} \defeq \frac{1}{L} \sum_{\mu=1}^M [\bm{\phi} \bar\bx]_\mu[\bm{\phi} \bx]_\mu=\frac{1}{L} \sum_{\mu=1}^M \big([\bm{\phi} \bx]_\mu - [\bm{\phi} \bs]_\mu\big) [\bm{\phi} \bx]_\mu
\ee
concentrates for a.e. $\Delta$. The advantage of doing so is that we do not need to add an $h$-perturbation to the CS model and as a consequence we obtain \eqref{xymmse} 
for a.e. $\Delta$ instead of \eqref{mmse-hh} in Lemma \ref{lemma:MSEequivalence}.
The reader may then wonder why we introduce the detour through the $h$-perturbation instead of directly showing \eqref{xymmse} for a.e. $\Delta$. 
The reason is that in order to prove the invariance of the MI under spatial coupling we need 
the general relation \eqref{eq:ymmserth_Ecth} involving the measurement MMSE associated to a sub-extensive set $\mathcal{S}_r$ and the present proof does not go through (as such) for sub-extensive sets.

\subsection{Concentration of $\mathcal{Q}$ on $\EE[\langle Q\rangle]$}

The concentration proof of ${\cal Q}$ closely follows the one of Proposition~\ref{lemma:concentration}. 
We first need to show the concentration of another intermediate
quantity ${\cal M}$, analogously to the proof of the concentration of ${\cal E}$ which requires first to 
show the concentration of ${\cal L}$ (see sec.~\ref{app:decoupling}). This quantity $\mathcal{M}$ naturally appears in the first and second derivatives
of the MI w.r.t $\Delta$. 
We consider the CS model \eqref{eq:CSmodel} with posterior given by \eqref{eq:posteriorCS}. Posterior averages are as usual denoted by $\langle\! -\!\rangle$. 
A calculation of the first and second derivatives of the MI yields
\begin{align}
\frac{di^{\rm cs}}{d\Delta^{-1}} &=  \EE[ \langle \mathcal{M}\rangle] + \frac{1}{2L} \sum_{\mu=1}^M \EE[[{\bm \Phi} \bS]_\mu^2], \label{d1}\\
\frac{d^2i^{\rm cs}}{d(\Delta^{-1})^2} &= -L\EE\big[ \langle \mathcal{M}^2\rangle - \langle \mathcal{M}\rangle^2\big] + 
\frac{\Delta^{3/2}}{4L} \sum_{\mu=1}^M\EE\big[\langle [\bm{\Phi}\bX]_\mu\rangle Z_\mu\big],
\label{d2_}
\end{align}
where 
\begin{align}
\mathcal{M} &\defeq \frac{1}{L} \sum_{\mu=1}^M \Big(\frac{[{\bm \phi} \bx]_\mu^2}{2} - [{\bm \phi} \bx]_\mu[{\bm \phi} \bs]_\mu - \frac{\sqrt{\Delta}}{2} [{\bm \phi} \bx]_\mu z_\mu\Big)\,.
\label{eq:def_M}
\end{align}

\begin{lemma}[Concentration of ${\cal M}$] \label{concM1}
For any fixed $a>\epsilon >0$ we have 
\begin{align}
\int_{\epsilon}^{a} d\Delta^{-1} \mathbb{E}\big[\big\langle \big({\cal M}- \langle {\cal M} \rangle \big)^2 \big\rangle\big] = \mathcal{O}(L^{-1})\,. \label{eq:concentrationM_0}
\end{align}
\end{lemma}

\begin{IEEEproof}
 The proof is similar to the one of Lemma \ref{lemma:concentration_1} so we only give a sketch. Integrating \eqref{d2_}
 \begin{align}
  L\int_{\epsilon}^{a} d\Delta^{-1} \mathbb{E} \big[\big\langle \big({\cal M}- \langle {\cal M} \rangle \big)^2 \big\rangle\big] 
  = \frac{di^{\rm cs}}{d\Delta^{-1}}\big|_{\Delta=\epsilon} - \frac{di^{\rm cs}}{d\Delta^{-1}}\big|_{\Delta=a}
  +\frac{\Delta^{3/2}}{4L} \sum_{\mu=1}^M\EE\big[\langle [\bm{\Phi}\bX]_\mu\rangle Z_\mu\big]\,.
 \end{align}
The first derivatives  of the MI is $\mathcal{O}(1)$ (uniformly in $L$). This is clear 
from the I-MMSE relation \eqref{y-immse} as ${\rm ymmse}$ must be $\mathcal{O}(1)$. However we can also show it directly from the r.h.s of 
\eqref{d1} by Nishimori's identity and Cauchy-Schwarz. 
Cauchy-Schwarz is used for decoupling $\bs$ and $\bx$ when they appear together in averages, and the  Nishimori identity 
is used for reducing all averages to moments of Gaussian random variables.
Doing so, only terms of the form \eqref{eq78} appear which are all bounded independently of $L$.
To show that the last term is $\mathcal{O}(1)$ we first integrate by parts $Z_\mu$ and then proceed through the Nishimori identity and the Cauchy-Schwarz inequality
as just indicated (note that this term is identical to the last one entering in $\mathbb{E}[\langle {\cal M}\rangle]$).
\end{IEEEproof}

Now we prove the concentration of $\langle\mathcal{M}\rangle$ on $\mathbb{E}[\langle\mathcal{M}\rangle]$, namely
\begin{lemma}[Concentration of $\langle\mathcal{M}\rangle$]\label{concM2}
 For any $a>\epsilon >0$ we have
 \begin{align}
\int_{\epsilon}^a d\Delta^{-1} \mathbb{E}\big[(\mathbb{E}[\langle {\cal M} \rangle] - \langle {\cal M} \rangle)^2\big] = \mathcal{O}(L^{-1/10}). 
\label{concentrationQ}
\end{align}
\end{lemma}

\begin{IEEEproof}
The proof is very similar to the arguments developed in sec.~\ref{concetrationLdisorder}.
Consider the free energy  $f^{\rm cs}(\by,\bm{\phi}) \defeq- \ln(\mathcal{Z}^{\rm cs}(\by,\bm{\phi}))/L$. 
Note that Proposition \ref{cor:fs_minus_meanfs_small} apply since the CS model is a 
special case of the perturbed interpolated model where $h\!=\!0$ and $t\!=\!1$. These results are used below. We focus here
on the rest of the proof which is analogous to that of Lemma \ref{lemma:concentration_2}.
The shorthand notations
$f_{\Delta^{-1}} \defeq f^{\rm cs}(\by,\bm{\phi})$ and $\bar f_{\Delta^{-1}} \defeq \mathbb{E}[f^{\rm cs}(\bY,\bm{\Phi})]$
are convenient and emphasize that we are interested in small perturbations of $\Delta^{-1}$. 
We have
\begin{align}
&\frac{d f_{\Delta^{-1}}}{d\Delta^{-1}}- \frac{d \bar f_{\Delta^{-1}}}{d\Delta^{-1}} 
= \langle {\cal M} \rangle -\mathbb{E}[ \langle {\cal M} \rangle ] + a,\\
&a \defeq \frac{1}{2L}\sum_{\mu=1}^M
([{\bm \phi} \bs]_\mu^2 - \mathbb{E}[[{\bm \Phi} \bS]_\mu^2]). \label{eqa}
\end{align}
%
By concavity of the MI (and thus of the free energy) in $\Delta^{-1}$ (see e.g. \cite{5165186}),
\begin{align}
- C_{\Delta^{-1}}^-&\defeq \frac{d\bar f_{\Delta^{-1}+\delta}}{d\Delta^{-1}} - \frac{d\bar f_{\Delta^{-1}}}{d\Delta^{-1}} \le 0, 
\quad C_{\Delta^{-1}}^+\defeq\frac{d\bar f_{\Delta^{-1}-\delta}}{d\Delta^{-1}} - \frac{d\bar f_{\Delta^{-1}}}{d\Delta^{-1}} \ge 0\, . \label{175_2}
\end{align}
Set $C_{\Delta^{-1}} \defeq C_{\Delta^{-1}}^+ + C_{\Delta^{-1}}^-\ge 0$. 
Using this remark with \eqref{eqa}, by proceeding exactly as in \eqref{175}--\eqref{eq:absBound1} we obtain an inequality of the type \eqref{eq:absBound}. Taking the square and averaging we get
\begin{align}
\mathbb{E}[(\mathbb{E}[\langle {\cal M} \rangle] - \langle {\cal M} \rangle)^2] & \le  \mathbb{E}\Big[\Big(\delta^{-1}\sum_{u\in\mathcal{K}} 
|f_{u} - \bar f_u| + C_{\Delta^{-1}} +|a|\Big)^2\Big] \leq 5 \Big(\delta^{-2}\sum_{u\in\mathcal{K}} \mathbb{E}[(f_{u} - \bar f_u)^2] + C_{\Delta^{-1}}^2 + \mathbb{E}[a^2] \Big),
\label{eq:absBound_3}
\end{align}
where ${\cal K}\defeq\{\Delta^{-1}+\delta,\Delta^{-1},\Delta^{-1}-\delta\}$. We used the convexity of the squarre
$(\sum_{i=1}^p v_i)^2\leq p\sum_{i=1}^p v_i^2$ (here $p=5$) to get the second equality. Now note that 
$\mathbb{E}[a^2] = {\cal O}(L^{-1})$. 
Indeed conditional on $\bs$, $[{\bm \Phi} \bs]_\mu$ and $[{\bm \Phi} \bs]_\nu$ are i.i.d Gaussian random variables with zero mean and finite variance, and by the central limit theorem $a$ tends in distribution to 
a zero mean Gaussian random variable with variance $\mathbb{E}_{ \bY|\bs}[a^2]={\cal O}(L^{-1})$, and thus after averaging over $\bS$, 
we have $\mathbb{E}[a^2]={\cal O}(L^{-1})$ too. Furthermore by the usual arguments
$d \bar f_{\Delta^{-1}}/d\Delta^{-1}$ is $\mathcal{O}(1)$, and thus $C_{\Delta^{-1}}$ as well, uniformly in $L\to \infty$. 
Note also that $C_{\Delta^{-1}} \geq 0$.
These remarks imply that $C_{\Delta^{-1}}^2 = C_{\Delta^{-1}} \mathcal{O}(1)$. So the right hand side of 
\eqref{eq:absBound_3} can be replaced by 
\begin{align}
\mathcal{O}\big(\delta^{-2}\sum_{u\in\mathcal{K}} \mathbb{E}[(f_{u} - \bar f_u)^2]\big) + C_{\Delta^{-1}} \mathcal{O}(1) + \mathcal{O}(L^{-1}).
\label{bigOinterm}
\end{align}
It remains to exploit the concentration of the free energy.  
From 
Proposition~\ref{cor:mcdiarmid}  it easily follows that
$\mathbb{E}\big[\big(f_{\Delta^{-1}} - \bar f_{\Delta^{-1}} \big)^2 \big] = \mathcal{O}(L^{-2\eta})$
with $0<\eta<1/4$.
So taking 
$\delta=L^{-\eta/2}$  we obtain for \eqref{bigOinterm} 
$\mathcal{O}(L^{-\eta}) + C_{\Delta^{-1}} \mathcal{O}(1)+ \mathcal{O}(L^{-1})$.
Integrating \eqref{eq:absBound_3} for finite $a>\epsilon>0$, we get
\begin{align}
&\int_{\epsilon}^a d\Delta^{-1} \mathbb{E}\big[(\mathbb{E}[\langle {\cal M} \rangle] - \langle {\cal M} \rangle)^2\big]
=\mathcal{O}(L^{-\eta}) + \mathcal{O}(1)\int_{\epsilon}^a d\Delta^{-1}C_{\Delta^{-1}}\,.
\label{EQ217}
\end{align}
The integral equals
differences of average free energies which, by the mean value theorem, are all $\mathcal{O}(\delta) = \mathcal{O}(L^{-\eta/2})$. Finally if we take $\eta = 1/5$ this last estimate is ${\cal O}(L^{-1/10})$. 
This ends the proof of the lemma.
\end{IEEEproof}

We are now ready to show the concentration of $\mathcal{Q}$. The proof proceeds similarly as in sec.~\ref{subsec:concentration_E}. 

\begin{lemma}[Concentration of ${\cal Q}$]\label{concenQ}
For any $a>\epsilon>0$ we have
 \begin{align}
\int_{\epsilon}^a d\Delta^{-1} \mathbb{E}\big[\big\langle\big({\cal Q}  -  \EE[\langle {\cal Q} \rangle ]\big)^2 \big\rangle\big] = \mathcal{O}(L^{-1/20}). 
\label{concentrationQ}
\end{align}
\end{lemma}

\begin{IEEEproof}
Let $g$ be a function s.t $\EE[\langle g^2\rangle] = {\cal O}(1)$ that we choose later on. By Cauchy-Schwarz applied to $\int_{\epsilon}^a d\Delta^{-1} \mathbb{E}[\langle\!-\!\rangle]$,
\begin{align}
\int_{\epsilon}^a d\Delta^{-1} \big|\mathbb{E}[\langle \mathcal{M} g \rangle ] - \EE[\langle \mathcal{M} \rangle ]\EE[\langle g \rangle] \big|
\leq 
\sqrt{\int_{\epsilon}^a d\Delta^{-1} \mathbb{E}[\langle g^2 \rangle] \int_{\epsilon}^a d\Delta^{-1} \mathbb{E}\big[\big\langle\big(\mathcal{M}  - \EE[\langle \mathcal{M} \rangle ]  \big)^2 \big\rangle \big]}.
\end{align} 
From Lemma \ref{concM1}, Lemma \ref{concM2} and the triangle inequality (for the $L_2$-norm) this implies
\begin{align}
&\int_{\epsilon}^a d\Delta^{-1} \big|\mathbb{E}[\langle \mathcal{M} g \rangle ] - \EE[\langle \mathcal{M} \rangle ]\EE[\langle g \rangle] \big| = {\cal O}(L^{-1/20}). \label{L18forM}
\end{align} 
Now we apply this satement to $g=\cal Q$. A useful guide for the rest of the proof is to note
the ``symmetry'' between $\cal Q$ in \eqref{Q} and $\cal E$ in \eqref{eq:defEps} and between 
$\cal L$ in  \eqref{eq:def_L} and $\cal M$ in \eqref{eq:def_M} under the changes $x_i \leftrightarrow [{\bm \phi} \bx]_\mu$, $s_i \leftrightarrow [{\bm \phi} \bs]_\mu$, $h \leftrightarrow \Delta^{-1}$, $\widehat z_i \leftrightarrow z_\mu$. First, one checks $\EE[\langle {\cal Q}^2\rangle] = {\cal O}(1)$: This is shown by combining Cauchy-Schwarz with the Nishimori identity which leads to terms of the form \eqref{eq78}, and from the discussion below \eqref{eq78} it follows that they are ${\cal O}(1)$. Second, we manipulate $\mathbb{E}[\langle \mathcal{M} g \rangle ] - \EE[\langle \mathcal{M} \rangle ]\EE[\langle g \rangle]$. Define the overlap $\widetilde q_{\bx,\bx'} \defeq (1/L)\sum_{\mu=1}^M [{\bm \phi}\bx]_\mu [{\bm \phi}\bx']_\mu$. Nishimori identities imply similar equalities as in \eqref{192}, \eqref{193}, and here we get
\begin{align}
\mathbb{E}[\langle\mathcal{M}{\cal Q}\rangle] - \mathbb{E}[\langle\mathcal{M}\rangle]\mathbb{E}[\langle\mathcal{Q}\rangle] &=\frac{1}{2}\Big(\mathbb{E}[\langle{\cal Q}^2\rangle ] - \mathbb{E}[\langle{\cal Q}\rangle]^2+ \mathcal{T}_1'+\mathcal{T}_2'\Big), \label{eq:MQ_minusMQ}\\ 
\mathcal{T}_1' &=\mathbb{E}[\langle{\cal Q}\rangle]\mathbb{E}[\langle \widetilde q_{\bx,\bx}\rangle ] - \mathbb{E}[\langle{\cal Q} \widetilde q_{\bx,\bx}\rangle ],  \\
\mathcal{T}_2' &= \mathbb{E}[\langle{\cal Q} \widetilde q_{\bx,\bx'}\rangle ]- \mathbb{E}[\langle{\cal Q}\widetilde q_{\bx,\bs}\rangle ], \label{fdsvds}
\end{align}
where recall that $\bx'$ is an independent replica. We now claim that $|\mathcal{T}_1'|$ is small. Lemma~\ref{remark:concentrationSelfOverlap} extends to the self-overlap $\widetilde q_{\bx,\bx}$ which thus concentrates. In addition, as already mentionned above $\EE[\langle {\cal Q}^2\rangle] = {\cal O}(1)$. Thus applying \eqref{smallnessT1} with $q_{\bx,\bx}$ replaced by $\widetilde q_{\bx,\bx}$ and ${\cal E}$ replaced by ${\cal Q}$ 
we obtain $|\mathcal{T}_1'|={\cal O}(L^{-1/2})$.
We now consider $\mathcal{T}_2'$. By the same steps used to obtain \eqref{200}, \eqref{186} but applied to new overlaps we reach that $\mathcal{T}_2'\ge 0$. 
From $|\mathcal{T}_1'|={\cal O}(L^{-1/2})$ and $\mathcal{T}_2'\ge 0$ together with 
\eqref{eq:MQ_minusMQ} we have 
\begin{align}
\mathbb{E}[\langle\mathcal{M}{\cal Q}\rangle] - \mathbb{E}[\langle\mathcal{M}\rangle]\mathbb{E}[\langle\mathcal{Q}\rangle] &\ge \frac{1}{2}\Big(\mathbb{E}[\langle{\cal Q}^2\rangle ] - \mathbb{E}[\langle{\cal Q}\rangle]^2\Big) + {\cal O}(L^{-1/2}). 
\label{eq:MQ_minusMQ_end}
\end{align}
The result of the lemma then follows from \eqref{L18forM} with $g={\cal Q}$ and \eqref{eq:MQ_minusMQ_end}.
\end{IEEEproof}

\subsection{Proof of the MMSE relation \eqref{xymmse}}

We now have all the necessary tools for proving \eqref{xymmse}. We can specialise the proof 
in sec.~\ref{app:T} for the non SC case $\Gamma=1$ and with $\mathcal{S}\to \{1,\ldots,M\}$,  and $\gamma \to \Delta^{-1}$. The steps leading to \eqref{eq:allpieces}, \eqref{eq:Y1}, \eqref{eq:Y2beforeDecoup} yield 
\begin{align}
{\rm ymmse} = {\cal Y}_{1} - {\cal Y}_{2}
\label{start}
\end{align}
with
\begin{align}
{\cal Y}_{1}\defeq \EE\Big[M^{-1}\sum_{\mu=1}^M Z_{\mu}^2 \langle {\cal E}\rangle\Big], \qquad
{\cal Y}_{2}\defeq \Delta^{-1/2}\, \EE\Big[M^{-1}\sum_{\mu=1}^M Z_{\mu}\Big\langle [\bm{\Phi} \bar \bX]_{\mu} {\cal E}\Big\rangle\Big], \label{eq:Y2beforeDecoup_2}
\end{align}
where ${\cal E}$ is given by \eqref{eq:defEps}. 

By the law of large numbers and the Nishimori identity to recognize the MSE through $\EE[\langle {\cal E}\rangle] = {\rm mmse}$, we get that
\be
{\cal Y}_{1} = {\rm mmse}+\smallO_L(1). 
\label{eq:Y1_3}
\ee

We now turn to ${\cal Y}_2$.
We will first re-express it in such a way that $\mathcal{Q}$ appears explicitly. An integration by parts w.r.t 
the noise yields (this is analogous to \eqref{eq:ippNoise})
\begin{align}
\alpha B\Delta {\cal Y}_{2} = \frac{1}{L}\sum_{\mu=1}^M \EE\big[\langle{\cal E} [\bm{\Phi} \bar \bX]_{\mu}^2  \rangle - \langle {\cal E}[\bm{\Phi} \bar \bX]_{\mu}  \rangle\langle [\bm{\Phi} \bar \bX]_{\mu}\rangle\big].
\end{align}
Define $\bar \bx' = \bx' - \bs$ where $\bx'$ is an independent replica. Then
\begin{align}
\alpha B \Delta {\cal Y}_{2} &=  \frac{1}{L}\sum_{\mu=1}^M \EE\big[\big\langle {\cal E}[\bm{\Phi} \bar \bX]_{\mu} \big( [\bm{\Phi} \bar \bX]_{\mu} - [\bm{\Phi} \bar \bX']_{\mu}\big)\big\rangle\big] =\frac{1}{L}\sum_{\mu=1}^M \EE\big[\big\langle{\cal E} [\bm{\Phi} \bar \bX]_{\mu}  [\bm{\Phi} \bX]_{\mu}\big\rangle\big] - \frac{1}{L}\sum_{\mu=1}^M \EE\big[\big\langle {\cal E}[\bm{\Phi} \bar \bX]_{\mu} [\bm{\Phi} \bX']_{\mu}\big\rangle\big] 
\nonumber \\ &
= \EE[\langle {\cal E}{\cal Q}\rangle] + {\cal O}(L^{-1/2}), 
\label{233}
\end{align}
where we recognized ${\cal Q}$ given by \eqref{Q} to get the last line and claimed that the second term involving two replicas satisfies 
\begin{align}
\frac{1}{L}\sum_{\mu=1}^M \EE\big[\big\langle {\cal E}[\bm{\Phi} \bar \bX]_{\mu} [\bm{\Phi} \bX']_{\mu}\big\rangle\big] = 
{\cal O}(L^{-1/2})\,.
\label{claimed}
\end{align}
We defer the proof of \eqref{claimed} to the end of this section. Using for now \eqref{233} we have 
\begin{align}
\alpha B \Delta {\cal Y}_{2} & = \EE[\langle {\cal E}\rangle] \EE[\langle{\cal Q}\rangle]  + 
\EE[\langle {\cal E}({\cal Q} - \EE[\langle{\cal Q}\rangle])\rangle]  + {\cal O}(L^{-1/2}),
\end{align}
and by Cauchy-Swcharz, $\EE[\langle {\cal E}^2\rangle]={\cal O}(1)$, and Lemma \ref{concenQ} we get that for a.e $\Delta$
\begin{align}
\EE[\langle {\cal E}({\cal Q} - \EE[\langle{\cal Q}\rangle])\rangle]
\leq 
\sqrt{\EE[\langle ({\cal Q} \!-\! \EE[\langle {\cal Q}\rangle])^2 \rangle]\EE[\langle {\cal E}^2\rangle]} = \smallO_L(1).
\end{align}
Recalling again that $\EE[\langle {\cal E}\rangle] = {\rm mmse}$ we obtain 
\begin{align}
\alpha B \Delta \mathcal{Y}_2 = \EE[\langle{\cal Q}\rangle] \, {\rm mmse} + \smallO_L(1)\,.
\label{Qprev}
\end{align}
In addition expressing $\mathcal{Q}$ in terms of the overlap,
\begin{align}
\EE[\langle {\cal Q}\rangle] & = \EE[\langle\widetilde q_{\bx,\bx} - \widetilde q_{\bx,\bs} \rangle ] 
= \EE[\widetilde q_{\bs,\bs} - \langle \widetilde q_{\bx,\bx'}\rangle ]= \EE[\langle\widetilde q_{\bs,\bs} - 2\widetilde q_{\bx,\bs} + \widetilde q_{\bx,\bx'}\rangle ] 
=
\alpha B \,{\rm ymmse}\,.
\label{Qover}
\end{align}
The second equality uses the Nishimori identity $\EE[\langle \widetilde q_{\bx,\bx}\rangle ] = \EE[ \widetilde q_{\bs,\bs}]$, 
the third one $\EE[\langle \widetilde q_{\bx,\bx'}\rangle ] = \EE[\langle \widetilde q_{\bx,\bs}\rangle ]$, and the last one 
follows from \eqref{defymmse}. From \eqref{Qprev} and \eqref{Qover} we finally get for a.e. $\Delta$
\begin{align}
\mathcal{Y}_2 = \frac{{\rm mmse}}{\Delta}\, {\rm ymmse} + \smallO_L(1) \,.
\label{final}
\end{align}

Now we can combine \eqref{start}, \eqref{eq:Y1_3}, \eqref{final} to obtain for a.e $\Delta$
\begin{align}
{\rm ymmse} = {\rm mmse} - {\rm mmse}\, \frac{\rm ymmse}{\Delta} + \smallO_L(1).
\end{align}
This proves relation \eqref{xymmse}.

It remains to justify the claim \eqref{claimed}. Recalling $q_{\bx,\bx'} \defeq (1/L)\sum_{i=1}^N x_i x_i'$ we have from \eqref{eq:defEps} 
that ${\cal E}=q_{\bx,\bx} - q_{\bx,\bs}$. Using $\widetilde q_{\bx,\bx'} \defeq (1/L)\sum_{\mu=1}^M [{\bm \phi}\bx]_\mu [{\bm \phi}\bx']_\mu$ as well we have
\begin{align}
\frac{1}{L}\sum_{\mu=1}^M \EE\big[\big\langle {\cal E}[\bm{\Phi} \bar \bX]_{\mu} [\bm{\Phi} \bX']_{\mu}\big\rangle\big] = \bigl(\EE[\langle q_{\bx,\bx} \widetilde q_{\bx,\bx'}\rangle] - \EE[\langle q_{\bx,\bx} \widetilde q_{\bs,\bx'}\rangle]\bigr) + \bigl(\EE[\langle q_{\bx,\bs}\widetilde q_{\bs,\bx'} \rangle] - \EE[\langle q_{\bx,\bs}\widetilde q_{\bx,\bx'} \rangle]\bigr). \label{234}
\end{align}
We will show that the first difference on the r.h.s is $\mathcal{O}(L^{-1/2})$ and that the second vanishes.
We start with the second difference. By the Nishimori identity, we can replace $\bs$ by an independent replica $\bx''$ and get 
$
\EE[\langle q_{\bx,\bs}\widetilde q_{\bs,\bx'} \rangle] - \EE[\langle q_{\bx,\bs}\widetilde q_{\bx,\bx'} \rangle] = \EE[\langle q_{\bx,\bx''}\widetilde q_{\bx'',\bx'} \rangle] - \EE[\langle q_{\bx,\bx''}\widetilde q_{\bx,\bx'} \rangle]
$.
But independent replicas are dummy variables in posterior averages $\langle\! -\! \rangle$ and can be interchanged $\bx \leftrightarrow \bx''$, so  
$\EE[\langle q_{\bx,\bx''}\widetilde q_{\bx,\bx'} \rangle] = \EE[\langle q_{\bx'',\bx}\widetilde q_{\bx'',\bx'} \rangle]$.
Also $q_{\bx'',\bx} = q_{\bx,\bx''}$, thus the second difference in \eqref{234} vanishes.
We now turn to the first difference in \eqref{234}. 
By the Nishimori identity $\EE[\langle\widetilde q_{\bx,\bx'}\rangle]= \EE[\langle\widetilde q_{\bs,\bx'}\rangle]$, thus
\begin{align}
\EE[\langle q_{\bx,\bx} \widetilde q_{\bx,\bx'}\rangle] - \EE[\langle q_{\bx,\bx} \widetilde q_{\bs,\bx'}\rangle] =  \EE\big[\big\langle \widetilde q_{\bx,\bx'}\big(q_{\bx,\bx} - \EE[\langle q_{\bx,\bx}\rangle]\big)\big\rangle\big] - \EE\big[\big\langle \widetilde q_{\bs,\bx'}\big(q_{\bx,\bx} - \EE[\langle q_{\bx,\bx}\rangle]\big)\big\rangle\big].
\end{align}
By the triangle inequality and Cauchy-Schwarz
\begin{align}
\big|\EE[\langle q_{\bx,\bx} \widetilde q_{\bx,\bx'}\rangle] \!-\! \EE[\langle q_{\bx,\bx} \widetilde q_{\bs,\bx'}\rangle]\big|
&
 \!\le\! \Big(\sqrt{\EE[\langle \widetilde q_{\bx,\bx'}^2\rangle]} + \sqrt{\EE[\langle \widetilde q_{\bs,\bx'}^2\rangle]}\Big) \sqrt{\EE\big[\big\langle\big(q_{\bx,\bx}\! -\! \EE[\langle q_{\bx,\bx}\rangle]\big)^2\big\rangle\big]}  
 \nonumber \\ &
 = \mathcal{O}(1) \sqrt{\EE\big[\big\langle\big(q_{\bx,\bx}\! -\! \EE[\langle q_{\bx,\bx}\rangle]\big)^2\big\rangle\big]}, 
 \label{ConcToUse}
\end{align}
where in the last step we used the usual arguments combining the Nishimori identity, Cauchy-Schwarz and the discussion below \eqref{eq78} to assert that $\EE[\langle \widetilde q_{\bx,\bx'}^2\rangle]=\EE[\langle \widetilde q_{\bs,\bx'}^2\rangle]={\cal O}(1)$. 
Now we use the concentration of $q_{\bx,\bx}$ in Lemma~\ref{remark:concentrationSelfOverlap} (that applies here as the Nishimori identiy is verified) to conclude from \eqref{ConcToUse}, 
$
|\EE[\langle q_{\bx,\bx} \widetilde q_{\bx,\bx'}\rangle] \!-\! \EE[\langle q_{\bx,\bx} \widetilde q_{\bs,\bx'}\rangle]| = {\cal O}(L^{-1/2})
$.
This ends the proof of claim \eqref{claimed}.

\section{Optimality of AMP}\label{sec:AMP}
At this stage, the main Theorem~\ref{th:replica} has been proven. In this section we show how to combine Corollary~\ref{cor:MMSE} with the state evolution analysis of \cite{bayati2011dynamics} for proving Theorem~\ref{thm:optAMP}. The proof is valid outside of the hard phase for $B=1$ and with a discrete prior s.t the RS potential $i^{\rm RS}(E;\Delta)$ in \eqref{eq:rs_mutual_info} has at most three stationary points. 
In order to prove this, we first need to show that the MSE of AMP verifies an identity of the form \eqref{xymmse}. 

The results we state here are only rigorously valid for $B=1$ as the state evolution analysis of \cite{bayati2011dynamics} is done in this case. Nevetheless, we conjecture that they are true for any finite $B$.
\subsection{MMSE relation for AMP}
Recall the definition of the measurement and usual MSE of AMP given in \eqref{yMSEAMP} and  \eqref{MSEamp}.
Their limits as $t\to \infty$ exist \cite{bayati2011dynamics} and are denoted by $E^{(\infty)}$ and ${\rm ymse}_{\rm AMP}^{(\infty)}$.
\begin{lemma}[MMSE relation for AMP] \label{lemmammserelationAMP}
We have almost surely (a.s)
\begin{align} \label{relationMSEAMP}
{\rm ymse}_{\rm AMP}^{(\infty)} = \frac{E^{(\infty)}}{1+E^{(\infty)}/\Delta}.
\end{align}
\end{lemma}
\begin{IEEEproof}
Set
\be \label{wt}
w^{(t-1)}_L\defeq\frac{1}{N\alpha}\sum_{i=1}^N [\eta^\prime(\bm{\phi}_0 \bz\!\!~^{(t-1)} + \widehat{\bs}\!\!~^{(t-1)};\tau_{t-1}^2)]_i.
\ee
This quantity appears in the Onsager term in \eqref{eq:AMP1}.
A general concentration result based on initial results in \cite{bayati2011dynamics} and needed here 
is Equation (4.11) in \cite{lassoMontanariBayati}. This states that the following limit exists and is equal to 
\be \label{limwl}
\lim_{L\to\infty}w^{(t-1)}_L = w^{(t-1)} = \frac{1}{\alpha B}\sum_{i=1}^B \mathbb{E}\big[[\eta^\prime(\widetilde\bS \!+\! \widetilde\bZ \tau_{t-1};\tau_{t-1}^2)]_i\big]
\ee
a.s, where $\widetilde \bS \sim P_0$, $\widetilde \bZ \sim \mathcal{N}(0,\mathbf{I}_B)$.

We start from \eqref{eq:AMP1} and replace the measurements by \eqref{eq:CSmodel} (note the rescaling factor $\sqrt{\alpha B}$ between the CS model and the definition of AMP).
Then we isolate the asymptotic measurement MSE of AMP and take the limit $t\to \infty$. We get
\begin{align}
\frac{1}{\alpha B}{\rm ymse}_{\rm AMP}^{(\infty)}
& =
\lim_{t\to \infty}\lim_{L\to\infty}\frac{1}{M}\|\bm{\phi}_0(\bs\!\!~^{(t)}-\bs)\|^2 
\nonumber \\ &
= \lim_{t\to \infty}\lim_{L\to\infty}\frac{1}{M}\|\sqrt{\frac{\Delta}{\alpha B}}\bz - \bz\!\!~^{(t)} + \bz\!\!~^{(t-1)} w^{(t)}_L\|^2 \nonumber\\
&=\frac{\Delta}{\alpha B} + (w^{(\infty)} - 1)^2\lim_{t\to \infty}\lim_{L\to\infty}\frac{1}{M}\|\bz^{(t-1)}\|^2 + 2(w^{(\infty)} -1) \lim_{t\to \infty}\lim_{L\to\infty}\frac{1}{M} \langle \bz\!\!~^{(t-1)},\bz \rangle \label{196}
\end{align}
a.s. Here $\langle \bx,\by \rangle = \sum_{i=1}^M x_i y_i$ and we used $\lim_{L\to\infty}\|\bz\|^2/M=1$ a.s by the central limit theorem. 
Recall the definition \eqref{MSEamp}. We use Equation (3.19) in \cite{bayati2011dynamics} which allows us to write that a.s,
\begin{align}
\frac{E^{(\infty)}}{\alpha B}
=\lim_{t\to \infty}\lim_{L\to\infty}\frac{1}{M} \|\sqrt{\frac{\Delta}{\alpha B}}\bz - \bz\!\!~^{(t-1)}\|^2 
\end{align}
which implies 
\begin{align}
2\lim_{t\to \infty}\lim_{L\to\infty}\frac{1}{M}\langle \bz\!\!~^{(t-1)},\bz \rangle 
= \frac{\Delta}{\alpha B} - \frac{E^{(\infty)}}{\alpha B} + \lim_{t\to \infty}\lim_{L\to\infty}\frac{1}{M}\|\bz^{(t-1)}\|^2.
\label{197}
\end{align}
Two other useful facts are Equation (2.1) in \cite{bayati2011dynamics} and Lemma 4.1 of \cite{lassoMontanariBayati} which respectively become in the present case,
\begin{align}
\tau_{t-1}^2 &= \frac{\Delta+ E^{(t-1)}}{\alpha B} = \Sigma(E^{(t-1)};\Delta)^2, 
\label{198_}\\
\tau_{t-1}^2 &=\lim_{L\to\infty}\frac{1}{M}\|\bz^{(t-1)}\|^2,
\end{align}
a.s (we used \eqref{eq:defSigma2} in the first equality). 
Using these two relations together with \eqref{197} allows to re-express \eqref{196} after simple algebra as
\begin{align}
{\rm ymse}_{\rm AMP}^{(\infty)} = (w^{(\infty)} - 1)^2(\Delta + E^{(\infty)})  + 2\Delta w^{(\infty)} -\Delta. \label{201}
\end{align}
We now claim that in the Bayesian optimal setting where the denoiser is defined by \eqref{denoiser}, we can express $w^{(\infty)}$ as a function of the MSE per section of AMP trough
\be\label{winf}
w^{(\infty)} = \frac{E^{(\infty)}}{E^{(\infty)} + \Delta}\,.
\ee
We will prove this formula in the next subsection. Plugging this formula in \eqref{201} directly implies \eqref{relationMSEAMP}.
\end{IEEEproof}
\subsection{Proof of the identity \eqref{winf}}
We compute explicitely \eqref{limwl} using \eqref{denoiser}. By computing the gradient we get
\begin{align}
\EE\big[[\eta^\prime(\widetilde\bS + \widetilde\bZ \tau_{t-1};\tau_{t-1}^2)]_i\big]
= \EE\big[\frac{1}{\tau_{t-1}^2} (\langle x_i^2\rangle_{t-1} - \langle x_i\rangle_{t-1}^2)\big],
\label{intermediate}
\end{align}
where $\mathbb{E}$ is w.r.t $\widetilde\bS\sim P_0$, $\widetilde{\bZ}\sim\mathcal{N}(0, \mathbf{I}_B)$ and where the posterior average $\langle\! -\!\rangle_{\tau_{t-1}}$ is
\begin{align}
\langle A(\bX) \rangle_{\tau_{t-1}} = \frac{\int d\bx\, A(\bx)  P_0(\bx) \exp\Big(- \frac{\Vert\widetilde\bS + \widetilde\bZ \tau_{t-1} - \bx\Vert^2}{2\tau_{t-1}^2}\Big)}{\int d\bx\,  P_0(\bx) \exp\Big(- \frac{\Vert\widetilde\bS + \widetilde\bZ \tau_{t-1} - \bx\Vert^2}{2\tau_{t-1}^2}\Big)}.
\end{align}
For a discrete prior these integrals are in fact finite sums so, because of \eqref{198_}, it is clear that
$\lim_{t\to \infty} \langle A(\bX) \rangle_{\tau_{t-1}} = \langle A(\bX) \rangle_{\Sigma(E^{(\infty)};\Delta)}$. 
With a bounded signal we can then easily apply the dominated convergence theorem to \eqref{intermediate} to conclude
\begin{align}
\lim_{t\to \infty} \EE\big[[\eta^\prime(\widetilde\bS + \widetilde\bZ \tau_{t-1};\tau_{t-1}^2)]_i\big]
= \frac{1}{\Sigma(E^{(\infty)};\Delta)^2}\EE\big[\langle x_i^2\rangle_{\Sigma(E^{(\infty)};\Delta)} - \langle x_i\rangle_{\Sigma(E^{(\infty)};\Delta)}^2\big].
\label{dereta}
\end{align}
The expected variance in this formula is nothing else than the MMSE of the effective AWGN channel 
$\bR=\widetilde\bS + \widetilde\bZ \Sigma(E^{(\infty)};\Delta)$. This can be seen explicitly by an application of the 
Nishimori identity \eqref{NishiVar}. 
%
%
This leads with \eqref{limwl} and \eqref{198_} to
\begin{align}
w^{(\infty)} = \frac{1}{\alpha B}\sum_{i=1}^B \EE\big[[\eta^\prime(\bR;\tau^2)]_i\big] = \frac{1}{\Delta +E^{(\infty)}} \mathbb{E}\big[\|\widetilde\bS\!-\! \mathbb{E}[\bX\vert \widetilde\bS \!+\! \widetilde\bZ\Sigma(E^{(\infty)};\Delta) ]\|^2\big]\,.
\end{align}
From \eqref{defmmsefunction}, \eqref{recursion-uncoupled-SE} we recognize that this MMSE actually corresponds to the fixed point of state evolution and is thus the asymptotic MSE of AMP $E^{(\infty)}$. Thus we find \eqref{winf}.
\subsection{Proof of Theorem \ref{thm:optAMP}: Optimality of AMP}\label{secProofOptAMP}
If $\Delta < \Delta_{\rm AMP}$ or $\Delta > \Delta_{\rm RS}$ we can assert, using the definitions of these thresholds and Remark~\ref{remark:extrema_irs_fpSE}, that the MSE of AMP at its fixed point $t=\infty$ is also the global minimum of the potential $E^{(\infty)} = \widetilde E$ given by \eqref{eq:defeTildeE}. Thus we replace $E^{(\infty)}$ by $\widetilde E$ in Lemma~\ref{lemmammserelationAMP}. This combined with Corollary~\ref{cor:MMSE} ends the proof of \eqref{thmOptAMPy}.

Then combining Theorem~\ref{thmMMSE} with \eqref{yMMMSE_Etilde} allows to identify \eqref{MMMSE_Etilde}, and thus to prove \eqref{thmOptAMP} as $E^{(\infty)} = \widetilde E$.
\appendices
\section{I-MMSE formula} \label{app:immse}
We give for completeness a short calculation to derive the I-MMSE formula \eqref{y-immse} for our (structured) vector setting.
Detailed proofs can be found in \cite{GuoShamaiVerdu_IMMSE} and here we do not go through the technical justifications required to exchange integrals and differentiate under the integral sign. 

Thanks to \eqref{eq:true_mutual_info} the MI is represented as follows
\begin{align}
i^{\rm cs} & = -\frac{\alpha B}{2} - \frac{1}{L} \mathbb{E}_{\bm\Phi}\Big[\int d\by {\cal Z}^{\rm cs}(\by,\bm{\Phi})(2\pi\Delta)^{-M/2}\ln ({\cal Z}^{\rm cs}(\by,\bm{\Phi})) \Big]
\nonumber \\ &
= -\frac{\alpha B}{2} - \frac{1}{L} \mathbb{E}_{\bm\Phi}\Big[ 
\int d\by \int \prod_{l=1}^L d\bs_l P_0(\bs_l)\,  \frac{e^{-\frac{1}{2\Delta}\Vert \bm\Phi\bs - \by\Vert^2}}{(2\pi\Delta)^{M/2}} 
\ln\Big\{ \int \prod_{l=1}^L d\bx_l P_0(\bx_l)\, e^{-\frac{1}{2\Delta}\Vert \bm\Phi\bx - \by\Vert^2}  \Big\} \Big].
\end{align}
We exchange the $\by$ and $\bs$ integrals. Then 
for fixed $\bs$ we perform the change of variables $\by \to \bz\sqrt\Delta + \bm\Phi\bs$. This yields 
\begin{align}
i^{\rm cs} = - \frac{\alpha B}{2} - \frac{1}{L} \mathbb{E}_{\bm\Phi, \bS, \bZ}\Big[ 
\ln\Big\{ \int \prod_{l=1}^L d\bx_l P_0(\bx_l)\, e^{-\frac{1}{2}\Vert \frac{\bm\Phi\bx}{\sqrt\Delta} - \frac{\bm\Phi\bS}{\sqrt{\Delta}} - \bZ\Vert^2}  \Big\} \Big],
\end{align}  
where $\bZ\sim{\cal N}(0,\mathbf{I}_M)$. We now perform the derivative $d/d\Delta^{-1} = (\sqrt{\Delta}/2) d/d \Delta^{-1/2}$.
We use the statistical mechanical notation for the ``posterior average'', namely for any quantity $A(\bx)$,
\begin{align}
\langle A(\bX) \rangle \defeq \frac{\int \prod_{l=1}^Ld\bx_l P_0(\bx_l) A(\bx)
e^{-\frac{1}{2}\Vert \frac{\bm\Phi\bx}{\sqrt\Delta} - \frac{\bm\Phi\bs}{\sqrt{\Delta}} - \bz\Vert^2}}
{\int \prod_{l=1}^Ld\bx_l P_0(\bx_l) 
e^{-\frac{1}{2}\Vert \frac{\bm\Phi\bx}{\sqrt\Delta} - \frac{\bm\Phi\bs}{\sqrt\Delta} - \bz\Vert^2}}.
\end{align}
Note that if $A$ does not depend on $\bx$ then $\langle A \rangle = A$.
We get 
\begin{align}
\frac{d i^{\rm cs}}{d\Delta^{-1}}  & = \frac{\sqrt\Delta}{2L} 
\mathbb{E}_{\bm\Phi, \bS, \bZ}\Big[ \Big\langle \Big(\frac{\bm\Phi\bX}{\sqrt\Delta} - \frac{\bm\Phi\bS}{\sqrt \Delta} - \bZ\Big)\cdot
(\bm\Phi\bX - \bm\Phi\bS)\Big\rangle \Big]
\nonumber \\ &
= \frac{1}{2L} 
\mathbb{E}_{\bm\Phi, \bS, \bZ}[ \langle \Vert \bm\Phi\bX - \bm\Phi\bS\Vert^2 \rangle ]
-
\frac{\sqrt\Delta}{2L} 
\mathbb{E}_{\bm\Phi, \bS, \bZ}[ \bZ \cdot \langle
\bm\Phi\bX - \bm\Phi\bS \rangle ].
\end{align}
Integrating the second term by parts w.r.t the Gaussian noise $\bz$ we get that this term is equal to
\begin{align}
\frac{\sqrt\Delta}{2L} 
\mathbb{E}_{\bm\Phi, \bS, \bZ}[ \bZ\!\cdot\!\langle
\bm\Phi\bX \!-\! \bm\Phi\bS \rangle ]\! &=\! \frac{\sqrt\Delta}{2L}\mathbb{E}_{\bm\Phi, \bS, \bZ}\Big[
\Big\langle (\bm\Phi\bX \!-\! \bm\Phi\bS)\!\cdot\! \Big(\frac{\bm\Phi\bX}{\sqrt\Delta} \!-\! \frac{\bm\Phi\bS}{\sqrt\Delta}\! -\! \bZ\Big)\Big\rangle 
\! -\! 
\big\langle \bm\Phi\bX \!-\! \bm\Phi\bS\big\rangle\! \cdot\! \Big\langle \frac{\bm\Phi\bX}{\sqrt\Delta} \!-\! \frac{\bm\Phi\bS}{\sqrt\Delta} \!-\! \bZ\Big\rangle
\Big] 
\nonumber \\
&=
\frac{1}{2L}\mathbb{E}_{\bm\Phi, \bS, \bZ}[
\langle \Vert \bm\Phi\bX \!-\! \bm\Phi\bS\Vert^2\rangle 
 \!-\! 
\Vert\langle \bm\Phi\bX \!-\! \bm\Phi\bS\rangle\Vert^2] .
\end{align}
We therefore obtain 
\begin{align}
\frac{d i^{\rm cs}}{d\Delta^{-1}} & = \frac{1}{2L} \mathbb{E}_{\bm\Phi, \bS, \bZ}[\Vert \langle \bm\Phi\bX - \bm\Phi\bS\rangle\Vert^2] 
=
\frac{\alpha B}{2} \frac{1}{M} \mathbb{E}_{\bm\Phi, \bS, \bZ}[\Vert \bm\Phi\langle \bX\rangle  - \bm\Phi\bS\Vert^2] = \frac{\alpha B}{2} {\rm ymmse}.
\end{align}
In the last step we recognized that 
$\langle \bX \rangle$ is nothing else than the MMSE estimator $\mathbb{E}[\bX\vert \bm\phi \bs +\bz\sqrt\Delta,\bm{\phi}]$
entering in the definition of the measurement MMSE \eqref{defymmse}.

\section{Nishimori identities} \label{app:Nishimori}
We collect here a certain number of Nishimori identities that are used throughout the paper. The basic identity from which other ones follow
is \eqref{eq:nishCond_0} or equivalently \eqref{eq:nishCond} below. In fact this identity is just an expression 
of Bayes law. 

\subsubsection{\bf Basic identity}
Assume the vector $\bs$ is distributed according to a prior $P_0(\bs)$ and its observation $\by$ is drawn from the conditional distribution $P(\by|\bs)$. 
Furthermore, assume $\bx$ is drawn from the posterior distribution $P(\bx|\by) = P_0(\bx)P(\by|\bx)/P(\by)$. Then for any function $g(\bx, \bs)$, using the Bayes formula,
\begin{align}
\mathbb{E}_{\bS}\mathbb{E}_{\bY|\bS}\mathbb{E}_{\bX|\bY} [g(\bX, \bS)] = \mathbb{E}_{\bY} \mathbb{E}_{\bX'|\bY}\mathbb{E}_{\bX|\bY} [g(\bX, \bX')], 
\label{eq:nishCond_0minus}
\end{align}
which is equivalent to 
\begin{align}
\mathbb{E}_{\bS,\bY}\mathbb{E}_{\bX|\bY} [g(\bX, \bS)] = 
 \mathbb{E}_{\bY} \mathbb{E}_{\bX'|\bY}\mathbb{E}_{\bX|\bY} [g(\bX, \bX')],
\label{eq:nishCond_0}
\end{align}
where $\bX, \bX'$ are independent random vectors distributed according to the posterior distribution: we speak in this case about two  ``replicas''. Recalling that $\langle\! -\! \rangle$ is the posterior expectation $\EE_{\bX\vert\by}$ and $\EE$ is the expectation w.r.t the quenched variables $\bY$, $\bS$, relation \eqref{eq:nishCond_0} becomes
\begin{align}
\EE[\langle g(\bX, \bS) \rangle] = \EE[\langle g(\bX, \bX')\rangle], \label{eq:nishCond}
\end{align}
where on the right hand side $\langle \!-\!\rangle$ stands for the average w.r.t the product distribution
$P(\bx,\bx'|\by)=P(\bx|\by)P(\bx'|\by)$.
We call this identity the \emph{Nishimori identity}. Of course it remains valid for functions depending on more that one replica. 
In particular for $g(\bx,\bs) = u(\bx) v(\bs)$ it implies 
\begin{align}
\EE[\langle u(\bX)\rangle v(\bS)] = \EE[\langle u(\bX) \rangle \langle v(\bX') \rangle].
\end{align}
Taking $u =1$ we have the very useful identity
\begin{align}
\EE[v(\bS)] = \EE[\langle v(\bX) \rangle]. 
\label{eq:NishId_0}
\end{align}
%
%

\subsubsection{\bf Second identity}
We show that the expectation of the variance of the MMSE estimator equals the MMSE itself. Indeed,
\begin{align} \label{NishiVar}
\EE[\|\bS - \langle \bX \rangle\|^2] &= \EE[\|\bS\|^2]  
-2 \EE[\bS \cdot \langle \bX\rangle]
+ \EE[\|\langle \bX \rangle\|^2]  = \EE[\langle \|\bX \|^2\rangle] - \EE[\|\langle \bX \rangle\|^2],
\end{align}
where we used $\EE[\|\bS\|^2]=\EE[\langle \|\bX \|^2\rangle]$ by \eqref{eq:NishId_0} and $\EE[\bS \cdot \langle \bX\rangle]=\EE[\|\langle \bX \rangle\|^2]$ by \eqref{eq:nishCond}.

\subsubsection{\bf Third identity}
We now show
\be
\EE\Big[\Big\langle \Big(\sum_{i=1}^N \Phi_{\mu i} (X_i-S_i)\Big)^2\Big\rangle\Big] = 2 \EE\Big[\Big(\sum_{i=1}^N \Phi_{\mu i} (\langle X_i \rangle - S_i)\Big)^2\Big]. \label{eq:Nish1}
\ee
The more elementary identity
\be
\EE[\langle (X_i-S_i)^2\rangle] = 2 \EE[(\langle X_i\rangle-S_i)^2]
\label{nishisquare}
\ee
is derived similarly. The proof goes as follows. Consider the function 
\begin{align}
g_1(\bx,\bs)\defeq \Big(\sum_{i=1}^N \phi_{\mu i} (x_i-s_i)\Big)^2 = \sum_{i,j=1}^N \phi_{\mu i}  \phi_{\mu j} (x_i-s_i) (x_j-s_j)
\end{align}
and apply \eqref{eq:nishCond}. We have $\EE[\langle g_1(\bX,\bS)\rangle] = \EE[\langle g_1(\bX,\bX')\rangle]$ with
\begin{align}
\EE[\langle g_1(\bX,\bX')\rangle] &=  \EE \Big[\sum_{i,j=1}^N \Phi_{\mu i} \Phi_{\mu j} \langle X_i X_j + X_i'X_j'-X_iX_j'-X_i'X_j\rangle \Big] \nonumber \\
&=  2\EE \Big[\sum_{i,j=1}^N \Phi_{\mu i} \Phi_{\mu j}  (\langle X_i X_j\rangle - \langle X_i \rangle \langle X_j\rangle) \Big].
\label{g1}
\end{align}
Now consider 
\begin{align}
g_2(\bx,\bs)\defeq \Big(\sum_{i=1}^N \phi_{\mu i} (\langle x_i\rangle -s_i)\Big)^2 =   \sum_{i,j=1}^N \phi_{\mu i}  \phi_{\mu j} (\langle x_i\rangle-s_i) (\langle x_j\rangle - s_j). 
\end{align}
Applying \eqref{eq:nishCond} again we find 
\begin{align}
\EE[\langle g_2(\bX,\bS)\rangle] &= \EE\Big[\sum_{i,j=1}^N \Phi_{\mu i}  \Phi_{\mu j} 
(\langle X_i\rangle \langle X_j\rangle -   \langle X_i^\prime\rangle \langle X_j\rangle - \langle X_i\rangle \langle X_j^\prime\rangle + \langle X_i^\prime X_j^\prime\rangle) \Big] \nonumber \\
&= \EE \Big[\sum_{i,j=1}^N \Phi_{\mu i} \Phi_{\mu j}  (\langle X_i X_j\rangle - \langle X_i \rangle \langle X_j\rangle) \Big].
\label{g2}
\end{align}
From \eqref{g1}, \eqref{g2} we have $\EE [\langle g_1(\bX,\bS)\rangle] = 2\EE[\langle g_2(\bX,\bS)\rangle]$ which is 
\eqref{eq:Nish1}.
\subsubsection{\bf Fourth identity}
Consider a CS model with inverse noise variance $\gamma$. Recall $\bar x_i \defeq x_i - s_i$, $\bar x_i' = x_i' - s_i$ and set
\begin{align}
u_\mu \defeq \sqrt{\gamma} [\bm{\phi} \bar \bx]_\mu - z_{\mu} = \sqrt{\gamma}([\bm{\phi} \bx]_\mu-y_\mu).
\end{align}
Note also that for a CS model with inverse noise variance $\gamma$, 
$z_\mu = \sqrt{\gamma}(y_\mu - [\bm{\phi} \bs]_\mu)$. 
Then we have 
\begin{align}
\EE[Z_{\mu}\langle U_\mu \bar X_i \bar X_i'\rangle] = \EE[Z_{\mu}\langle U_\mu \bar X_i\rangle \langle \bar X_i\rangle] = -\EE[ Z_{\mu} S_i \langle U_\mu \bar X_i\rangle]. \label{eq:onetermonly}
\end{align}
Indeed, from these previous relations
\begin{align}
\EE[Z_{\mu} \langle U_\mu \bar X_i X_i'\rangle] &= \gamma\EE[\langle(Y_\mu \!-\! [\bm{\Phi} \bS]_\mu)([\bm{\Phi} \bX]_\mu\!-\!Y_\mu) X_i' (X_i\!-\!S_i)\rangle] \nonumber\\
&= \gamma\EE[\langle X_i \rangle \langle([\bm{\Phi} \bS]_\mu\!-\!Y_\mu)([\bm{\Phi} \bX]_\mu\!-\!Y_\mu) (S_i\!-\!X_i)\rangle]\nonumber \\
&=\gamma\EE[\langle X_i \rangle S_i ([\bm{\Phi} \bS]_\mu\!-\!Y_\mu) \langle[\bm{\Phi} \bX]_\mu\!-\!Y_\mu\rangle] \!-\! \gamma\EE[\langle X_i \rangle \langle X_i ([\bm{\Phi} \bX]_\mu\!-\!Y_\mu) \rangle ([\bm{\Phi} \bS]_\mu\!-\!Y_\mu) ] = 0,
\end{align}
using the Nishimori identity for the last step. From \eqref{eq:onetermonly} one obtains the useful identity
\begin{align}
\EE[Z_{\mu}\langle U_\mu \bar X_i \bar X_i'\rangle] &= - \EE[ Z_{\mu}S_i\langle(\sqrt{\gamma} [\bm{\Phi}\bar \bX]_\mu - Z_{\mu}) \bar X_i\rangle  ] = \EE[ Z_{\mu}^2 S_i\langle \bar X_i\rangle  ] - \sqrt{\gamma}\,\EE[ Z_{\mu}S_i\langle [\bm{\Phi}\bar \bX]_\mu \bar X_i\rangle]. \label{eq:niceId}
\end{align}
\section{Differentiation of $i^{\rm RS}$ with respect to $\Delta^{-1}$} \label{app:Differentiation}
In this section we prove \eqref{eq:dirsddelta}. Recall that $\widetilde{E}(\Delta)$ is defined as the (global) minimiser of 
$i^{\rm RS}(E;\Delta)$ when it is unique. It is possible to show that in the first order phase transition scenario 
$\widetilde{E}(\Delta)$ is analytic except at $\Delta_{\rm RS}$\footnote{
This can be done through a direct application of the real analytic implicit function theorem
to the function
$f(E;\Delta) \defeq \partial i^{\rm RS}/\partial E$.}. Therefore in particular $d\widetilde{E}/d\Delta$ is bounded for $\Delta \neq \Delta_{\rm RS}$. Thus since $\widetilde{E}(\Delta)$ is a solution of
$\partial i^{\rm RS}(E;\Delta)/\partial E=0$, we have for
$\Delta \neq \Delta_{\rm RS}$
\begin{align}
\frac{d i^{\rm RS}(\widetilde{E}(\Delta);\Delta)}{d \Delta^{-1}} = \frac{\partial i^{\rm RS}(\widetilde{E};\Delta)}{\partial \Delta^{-1}} + \frac{\partial i^{\rm RS}( E;\Delta)}{\partial E}\Big\vert_{\widetilde E} \frac{d\widetilde E}{d\Delta}
= \frac{\partial i^{\rm RS}(\widetilde{E};\Delta)}{\partial \Delta^{-1}} .
\end{align}
Now, from \eqref{eq:rs_mutual_info} we have
\begin{align}
\frac{\partial i^{\rm RS}(\widetilde E;\Delta)}{\partial \Delta^{-1}} = \frac{\partial \psi(\widetilde E;\Delta)}{\partial \Delta^{-1}}  + \frac{\partial i(\widetilde\bS;\widetilde\bY)}{\partial \Delta^{-1}}, \label{eq:184}
\end{align}
where $i(\widetilde\bS;\widetilde\bY)$ is given by \eqref{eq:i_denoising}. Using simple differentiation along with the chain rule, one gets  
\begin{align}
\frac{\partial \psi(\widetilde E;\Delta)}{\partial \Delta^{-1}} &= \frac{\alpha B}{2}\Big(\frac{\widetilde E}{1+\widetilde E/\Delta} - \frac{\widetilde E}{(1+\widetilde E/\Delta)^2}\Big) = \frac{\alpha B}{2}\frac{{\widetilde E}^2}{\Delta(1+\widetilde E/\Delta)^2}, \label{eq:185}\\
\frac{\partial i(\widetilde\bS;\widetilde\bY)}{\partial \Delta^{-1}}&=\frac{\alpha B}{(1+\widetilde E/\Delta)^2} \frac{\partial i(\widetilde\bS;\widetilde\bY)}{\partial \Sigma^{-2}}\Big\vert_{\widetilde E} = \frac{\alpha B}{2}\frac{\widetilde E}{(1+\widetilde E/\Delta)^2}. \label{eq:186}
\end{align}
Identity \eqref{eq:dirsddelta} follows directly from \eqref{eq:184}, \eqref{eq:185} and \eqref{eq:186}.

We point out it is tedious to check the last equality in \eqref{eq:186} by a direct computation of $\partial i(\widetilde\bS;\widetilde\bY)/ \partial \Sigma^{-2}$. But one can use the following trick. We know that $\widetilde E$ is a stationary point of $i^{\rm RS}(E;\Delta)$ so from \eqref{eq:rs_mutual_info} and \eqref{eq:defSigma2},
\begin{align}
\frac{\partial i(\widetilde\bS;\widetilde\bY)}{\partial E}\Big|_{\widetilde E} = -\frac{\alpha B}{2} \frac{\widetilde E}{(\Delta + \widetilde E)^2}. \label{eq:lastToUse}
\end{align}
Then, by the chain rule one obtains
\begin{align}
\frac{\partial i(\widetilde\bS;\widetilde\bY)}{\partial \Sigma^{-2}}\Big|_{\widetilde E} = \frac{\partial i(\widetilde\bS;\widetilde\bY)}{\partial E}\Big|_{\widetilde E} \Big(\frac{\partial \Sigma^{-2}}{\partial  E}\Big|_{\widetilde E}\Big)^{-1} = -\frac{\alpha B}{2} \frac{\widetilde E}{(\Delta + \widetilde E)^2} \Big(-\frac{\alpha B}{(\Delta + \widetilde E)^2}\Big)^{-1} =\frac{\widetilde E}{2}.
\end{align}
\section{The zero noise limit of the mutual information}\label{app:zeronoise}

The replica formula for the MI is \eqref{eq:rs_mutual_info}. We want to show that, for discrete bounded signals,
$\lim_{\Delta \to 0} i^{\rm RS}(\widetilde  E;\Delta)= H(\widetilde\bS)$ the Shannon entropy of $\widetilde \bS\sim P_0$, where recall that $\widetilde E$ is given by \eqref{eq:defeTildeE}.

{\it Assume} first that $\widetilde  E(\Delta)\to 0$ when
$\Delta\to0$. Then, according to \eqref{eq:defSigma2}, in this noiseless limit $\Sigma(\widetilde E;\Delta) \to
0$. The zero noise limit of the denoising problem is then easily
obtained. Indeed the explicit expression of \eqref{eq:i_denoising} reads (where all vectors and norms are $B$-dimensional)
\begin{align}
i(\widetilde\bS;\widetilde\bY) & =  -\sum_{k=1}^K p_k\int d\bz 
 \frac{e^{-\frac{\Vert \bz\Vert^2}{2}}}{(2\pi)^{B/2}} \ln\Big(\sum_{\ell=1}^K p_\ell e^{-\frac{\Vert \ba_\ell - \ba_k - \bz \Sigma\Vert^2}{2\Sigma^2}} \Big) -\frac{B}{2}
 \nonumber \\ &
 = 
 -\sum_{k=1}^K p_k\int d\bz 
 \frac{e^{-\frac{\Vert \bz\Vert^2}{2}}}{(2\pi)^{B/2}} \ln\Big(p_k e^{-\frac{\Vert \bz\Vert^2}{2}} 
 + \sum_{\ell=1, \ell\neq k}^K p_\ell e^{-\frac{\Vert \ba_\ell - \ba_k - \bz \Sigma\Vert^2}{2\Sigma^2}} \Big) -\frac{B}{2}.
 \label{lebesgueargument}
\end{align}
The term $\sum_{\ell=1, \ell\neq k}^K$ is an average of terms smaller than $1$. Thus 
$\ln(p_k \exp(-\Vert \bz\Vert^2/2))\leq \ln(p_k \exp(-\Vert \bz\Vert^2/2)\!+\!1)$
and therefore the absolute value of the integrand is bounded above by 
\begin{align}
 \frac{e^{-\frac{\Vert \bz\Vert^2}{2}}}{(2\pi)^{B/2}} \max\Big\{\frac{\Vert \bz\Vert^2}{2} - \ln(p_k)\, ,\, p_k e^{-\frac{\Vert \bz\Vert^2}{2}}\Big\}.
\end{align}
This estimate is uniform in $\Sigma$ and integrable. Thus by Lebesgue's dominated convergence theorem we can compute the limit of \eqref{lebesgueargument} by exchanging the limit
and the integral. Obviously the $\Sigma \to 0$ limit of the term inside the logarithm is $p_k \exp(-\Vert \bz\Vert^2/2)$ and computing the resulting integral yields 
\begin{align}
 \lim_{\Sigma\to 0} i(\widetilde\bS;\widetilde\bY) = -\sum_{k=1}^K p_k \ln (p_k) = H(\widetilde \bS).
\end{align}

{\it Assume} further that $\widetilde  E(\Delta)\to 0$ fast enough s.t $\lim_{\Delta\to 0}\widetilde E(\Delta)/\Delta = 0$.
Then from \eqref{eq:psi} we also have that  
$\lim_{\Delta\to 0}\psi(\widetilde E;\Delta) = 0$, and the desired result follows, namely that $\lim_{\Delta \to 0} i^{\rm RS}(\widetilde E;\Delta)= H(\widetilde\bS)$. 

We thus only have to verify that $\lim_{\Delta\to 0}\widetilde E(\Delta)/\Delta = 0$.
Recall from Remark \ref{remark:extrema_irs_fpSE} that $\widetilde E(\Delta)$ is a solution of
\begin{align}\label{fpappendix}
E = {\rm mmse}(\Sigma(E; \Delta)^{-2}),\quad
\Sigma(E;\Delta)^{-2} = \frac{\alpha B}{\Delta+ E}.
\end{align}
%
We want to look at the behaviour of this fixed point equation when {\it both} $E\to 0$ and $\Delta\to 0$. When this is the case $\Sigma\to 0$. For
a {\it discrete} prior ${\rm mmse}(\Sigma^{-2}) = {\cal O}(\exp(-c/\Sigma^2))$ where the constant $c>0$ is related to the minimum distance between 
alphabet elements (see next paragraph). Therefore for $E\to 0$, $\Delta\to 0$ and thus $\Sigma\to 0$, the solutions of \eqref{fpappendix}
must satisfy $E/\Sigma^2 \to 0$, in other words $E/(\Delta + E)\to 0$ or $(1+\Delta/E)^{-1} \to 0$. This can only 
happen if $E/\Delta\to 0$. Since for $\Delta < \Delta_{\rm AMP}$ there is a unique fixed point solution we deduce 
that  necessarily $\lim_{\Delta\to 0}\widetilde E(\Delta)/\Delta = 0$.

Note that for the above argument to hold 
${\rm mmse}(\Sigma^{-2}) = \smallO_L(\Sigma^2)$ is enough, and that this holds for discrete priors which have information dimension equal to zero
\cite{wu2010renyi}. 
Nevertheless we sketch here for completeness the proof that ${\rm mmse}(\Sigma^{-2}) = {\cal O}(\exp(-c/\Sigma^2))$ for a discrete prior. First we write down explicitly 
the ${\rm mmse}$:
\begin{align}
 {\rm mmse}(\Sigma^{-2}) & = \sum_{k=1}^K p_k \int d\bz 
\frac{e^{-\frac{\Vert \bz\Vert^2}{2}}}{(2\pi)^{B/2}}
\bigg\Vert \frac{\sum_{\ell=1}^K p_\ell \ba_\ell e^{- \frac{\Vert \ba_\ell -\ba_k -\bz \Sigma\Vert^2}{2\Sigma^2}}}
{\sum_{\ell=1}^K p_\ell e^{- \frac{\Vert \ba_\ell -\ba_k -\bz \Sigma\Vert^2}{2\Sigma^2}}} - \ba_k \bigg\Vert^2
\nonumber \\ &
=
\sum_{k=1}^K p_k \int d\bz
\frac{e^{-\frac{\Vert \bz\Vert^2}{2}}}{(2\pi)^{B/2}}
\bigg\Vert \frac{\sum_{\ell=1, \ell\neq k}^K p_\ell (\ba_\ell - \ba_k) e^{- (\Vert\ba_\ell -\ba_k\Vert^2 - 2\Sigma\bz\cdot (\ba_\ell -\ba_k))/2\Sigma^2}}
{ p_k +    \sum_{\ell=1, \ell\neq k}^K p_\ell e^{- (\Vert\ba_\ell -\ba_k\Vert^2 - 2\Sigma\bz\cdot (\ba_\ell -\ba_k))/2\Sigma^2}}\bigg\Vert^2.
\end{align}
Let $\delta = \min_{j \neq  j^\prime}\Vert \ba_j - \ba_{j^\prime}\Vert$ the minimum distance between any two distinct elements of the signal alphabet. 
We separate the integral in two sets: $\Vert \bz\Vert \leq \delta/(4\Sigma)$ and $\Vert \bz\Vert \geq \delta/(4\Sigma)$. To estimate the contribution of the second set we 
simply note that by the triangle inequality $\Vert \cdots\Vert^2 \leq 4B s_{\max}^2$ so the whole contribution is smaller than 
$4Bs_{\max}^2 (2\pi)^{-B/2}\int_{\Vert \bz\Vert \geq \delta/(4\Sigma)} d\bz
\exp(-\Vert \bz\Vert^2/2)= {\cal O}(\exp(-\delta^2/(64\Sigma^2))$. For the first contribution we note that $\Vert \bz\Vert \leq \delta/(4\Sigma)$ implies
$\Vert\ba_\ell -\ba_k\Vert^2 - 2\Sigma\bz\cdot (\ba_\ell -\ba_k)\geq \delta^2/2$ so that it is upper bounded by 
\begin{align}
 & \sum_{k=1}^K p_k^{\!-\!1} \int_{\Vert \bz\Vert \leq \delta/(4\Sigma)} d\bz
\frac{e^{\!-\!\frac{\Vert \bz\Vert^2}{2}}}{(2\pi)^{B/2}}
\bigg\Vert \frac{\sum_{\ell=1, \ell\neq k}^K p_\ell (\ba_\ell \!-\! \ba_k) e^{\!-\! (\Vert\ba_\ell \!-\!\ba_k\Vert^2 \!-\! 2\Sigma\bz\cdot (\ba_\ell \!-\!\ba_k))/2\Sigma^2}}
{ 1 \!+\!    \sum_{\ell=1, \ell\neq k}^K \frac{p_\ell}{p_k} e^{\!-\! (\Vert\ba_\ell \!-\!\ba_k\Vert^2 \!-\! 2\Sigma\bz\cdot (\ba_\ell \!-\!\ba_k))/2\Sigma^2}}\bigg\Vert^2
\leq 
4 Bs_{\max}^2  e^{\!-\!\delta^2/2\Sigma^2} \sum_{k=1}^K p_k^{\!-\!1}.
\end{align}
This concludes the argument.

In sec.~\ref{sec:integrationArgument} we also use that $\lim_{\Delta\to 0} i^{\rm cs} = H(\widetilde \bS)$. This follows from the fact that 
for discrete priors even a single noiseless measurement
 allows for a perfect reconstruction with high probability
\cite{wu2010renyi,donoho2013information}.

\section{Concentration of the free energy: Proof of Proposition \ref{cor:fs_minus_meanfs_small}}\label{appendix_concentration}
\subsection{Probabilistic tools}
We have to prove concentration w.r.t the various types of quenched variables. We use two probabilistic tools in conjunction, namely the concentration 
inequality of Ledoux and Talagrand \cite{Talagrand1996} for Gaussian random variables $\bm{\Phi}, \bZ, \widetilde{\bZ}, \widehat{\bZ}$ and 
the McDiarmid inequality \cite{McDiarmid,boucheron2004concentration} for the bounded random signal $\bS$. These inequalities are stated here for the convenience of the reader. 
\begin{proposition}[Ledoux-Talagrand inequality]\label{Talagrand1996}
Let $f(U_1, \dots, U_P)$ a function of $P$ independent standardized Gaussian random variables which is Lipshitz w.r.t the Euclidean norm
on the whole of $\mathbb{R}^P$, that is $\vert f(u_1, \dots, u_P)  -  f(u_1^\prime, \dots, u_P^\prime)\vert \leq K_P \|\textbf{u} - \textbf{u}'\|$. Then 
\begin{align}
\mathbb{P}\big(\big|f(u_1, \ldots, u_P) - \mathbb{E}[f(U_1, \dots, U_P)] \big| \geq r\big) \leq e^{-\frac{r^2}{2K_P^2}}\,.
\end{align}
\end{proposition}
\begin{proposition}[McDiarmid inequality] \label{McD}
 Let $f(U_1, \cdots, U_P)$ a function of $P$ i.i.d random variables that satisfies the following bounded difference property: For all $i=1,\ldots, P$, $\vert f(u_1, \dots, u_i, \ldots, u_P) - f(u_1, \dots, u_i^\prime, \ldots, u_P)\vert \leq c_i$
where $c_i>0$ is independent of 
 $u_1,\dots, u_i, u_i^\prime, \dots, u_P$. Then for any $r>0$,
 \begin{align}
 \mathbb{P}\big(\big| f(u_1, \dots, u_P) - \mathbb{E}[f(U_1, \dots, U_P)]\big| \geq r\big) \leq e^{-\frac{2r^2}{\sum_{i=1}^P{c_i^2}}}\,.
 \end{align}
\end{proposition}

In our application the function (the free energy) is not Lipshitz over the whole of $\mathbb{R}^P$. To circumvent this technical problem  we will use a result obtained
in \cite{korada2010} (appendix I.A Theorem 9). 
\begin{proposition}[Lipshitz extension]
Let $f$ be a Lipshitz function over $G\subset \mathbb{R}^P$ with Lipshitz constant $K_P$. By the McShane and Whitney extension theorem \cite{Heinonen2005}
there exists an extension $g$ defined on the whole of $\mathbb{R}^P$ (so $g\vert_{G} = f$) which is Lipshitz with the same constant $K_P$ on the whole
of $\mathbb{R}^P$. 
\end{proposition}

Applying Proposition \ref{Talagrand1996} to $g$ yields the following (see \cite{korada2010} appendix I.A Lemma 7 for a detailed proof).

\begin{proposition}[Concentration of almost Lipshitz functions]\label{concalmostLip}
Let $f$ be a Lipshitz function over $G\subset \mathbb{R}^P$ with Lipshitz constant $K_P$. Assume $0\in G$, $f(0)^2 \leq C^2$, $\mathbb{E}[f^2] \leq C^2$ for some $C>0$. 
Then for $r\geq 6(C + \sqrt{P K_P}) \sqrt{\mathbb{P}(G^c)}$ we have 
\begin{align}
 \mathbb{P}\big(\big| f(u_1,\dots, u_P) - \mathbb{E}[f(U_1,\dots,U_P)]\big| \geq r\big) \leq 2 e^{- \frac{r^2}{16 K_P^2}} + \mathbb{P}(G^c)\,.
\end{align}
\end{proposition}
\subsection{Concentration of the free energy with respect to the Gaussian quenched variables}
We will apply these tools to show concentration properties of the free energy
$f_{t,h}(\mathring{\by}) \defeq -\ln(\mathcal{Z}_{t,h}(\mathring{\by}))/L$.
The proof is decomposed in two parts. First we show thanks to the Ledoux-Talagrand inequality that for $\bs$ fixed $f_{t,h}(\mathring{\by})$ concentrates on $\mathbb{E}[f_{t,h}(\mathring{\bY}) | \bs]$ (so the expectation is over all Gaussian quenched variables $\bm{\Phi}$, $\bZ$, $\widetilde\bZ$, $\widehat\bZ$).
Second, we show thanks to McDiarmid's inequality that $\mathbb{E}[f_{t,h}(\mathring{\bY}) | \bs]$ concentrates on 
$\mathbb{E}[f_{t,h}(\mathring{\bY})]$ (so the last expectation also includes the average over the signal distribution). Let us first fix $\bs$ for all this sub-section.

\begin{proposition}[Concentration of the free energy w.r.t the Gaussian quenched variables]\label{thingprop1}
One can find two positive constants $c_1$ and $c_2$ (depending only on $s_{\rm max}$, $K$, $\Delta$, $\alpha$) s.t for any $r={\Omega}(e^{-c_2 L^{1/2}})$, we have for any 
fixed $\bs$
\begin{align}
\mathbb{P}\big(| f_{t,h}(\mathring \by) - \mathbb{E}[f_{t,h}(\mathring \bY)\mid \bs] | \ge r \,\big|\, \bs\big)\le e^{-c_1 r^2 L^{1/2}}.
\label{tala}
\end{align}
\end{proposition}
\begin{IEEEproof}
The proof is the same as in \cite{korada2010} so we give the main steps only. 
Let 
\begin{align}
{\cal G}\defeq \big\{\bm{\phi}, \bz, \widehat\bz, \widetilde\bz \,|\, \max_\mu\vert z_\mu\vert \leq \sqrt{D_1} \cap \max_i\vert \widehat z_i\vert \leq \sqrt{D_1} \cap  
 \max_i\vert \widetilde z_i\vert \leq \sqrt{D_1} \cap  \text{for all~} \bx,\bs: \Vert \bm{\phi}(\bx - \bs)\Vert^2\leq D_2 L\big\} \, ,
\end{align}
where $D_1$ and $D_2$ will be chosen later as suitable powers of $N$ ($\bx$, $\bs$ are always in the the discrete alphabet). We will see that the free energy is Lipshitz on ${\cal G}$, which will allow to use Proposition~\ref{concalmostLip}. Let us first estimate $\mathbb{P}({\cal G}^c)$, required in this Proposition. If $U$ is a zero mean Gaussian variable then $\mathbb{P}(|U|\geq \sqrt{A}) \leq 4 e^{-A/4}$. Therefore from the union bound 
\begin{align}
 \mathbb{P}\big(\max_\mu\vert z_\mu\vert \geq \sqrt{D_1} \cup \max_i\vert \widehat z_i\vert \geq \sqrt{D_1} \cup 
 \max_i\vert \widetilde z_i\vert \geq \sqrt{D_1}\big) \leq4 N(2 + \alpha) e^{-D_1/4}.
 \label{one}
\end{align}
Now conditional on $\bx$ and $\bs$, $\sum_{i=1}^N \phi_{\mu i} ( x_i - s_i)$, $\mu=1,\dots, M$, are independent Gaussian random variables with zero mean and variance $a^2\!\le\!4B s_{\rm max}^2$. Let $X\!\sim\!{\cal N}(0,a^2)$. The identity $\EE[\exp(X^2\!/\!(16B s_{\rm max}^2)] \!=\! \sqrt{1\!+\!a^2\!/\!(8 B s_{\rm max}^2\!-\!a^2)}$ $\le\sqrt{2}$ thus implies 
\begin{align}
 \mathbb{E}\Big[e^{\frac{\Vert \bm{\Phi}(\bx - \bs)\Vert^2}{16Bs_{\rm max}^2}}\Big] \leq  2^{M/2}\,.
\end{align}
Thus from Markov's inequality for any given $\bx$, $\bs$ in the discrete alphabet
\begin{align}
 \mathbb{P} \big(\Vert \bm{\Phi}(\bx - \bs)\Vert^2\geq D_2N\big) \leq 2^{\alpha L B/2} e^{-\frac{D_2L}{16Bs_{\rm max}^2}}\,.
\end{align}
From the union bound we obtain (recall that here $K$ is the size of the discrete signal alphabet)
\begin{align}
 \mathbb{P}\big(\text{there exist~} \bx,\bs \text{~in the discrete alphabet}: \Vert \bm{\Phi}(\bx - \bs)\Vert^2\geq D_2L\big) \leq K^{2BL} 2^{\alpha L B/2} e^{-\frac{D_2L}{16Bs_{\rm max}^2}}\,.
 \label{two}
\end{align}
Therefore from \eqref{one}, \eqref{two} and the union bound we obtain 
\begin{align} \label{255}
 \mathbb{P}({\cal G}^c) \leq 4 N(2 + \alpha) e^{-D_1/4} + K^{2BL} 2^{\alpha L B/2} e^{-\frac{D_2L}{16Bs_{\rm max}^2}}\,.
\end{align}
This probability will be made small by a suitable choice of $D_1$ and $D_2$.

Now we must show that $f_{t,h}(\mathring \by)$ is Lipshitz on ${\cal G}$. To this end we set $\bm{\phi}^0 = L^{1/2} \bm{\phi}$ in order to work only with 
standardized Gaussian random variables, and consider two sets of quenched variables $\bm{\phi}^{0}$, $\bz$, $\widehat\bz$, $\widetilde\bz$ 
and $\bm{\phi}^{{0}\prime}$, $\bz'$, $\widehat\bz'$, $\widetilde\bz^\prime$ belonging to the set ${\cal G}$. Proceeding exactly as in appendix I.E of \cite{korada2010},
a slightly painful calculation leads to (recall the expression of the Hamiltonian \eqref{eq:int_hamiltonian})
\begin{align}
 \vert {\cal H}_{t,h}(\bx | \mathring{\by}) - {\cal H}_{t,h}(\bx | \mathring{\by}^\prime)\vert \leq ({\cal O}(\sqrt{LD_1}) + {\cal O}(\sqrt{D_2})) (\Vert \bm{\phi}^{0} -\bm{\phi}^{{0}\prime}\Vert_{F} + \Vert \bz-\bz^\prime\Vert + \Vert \widehat\bz-\widehat\bz^\prime\Vert 
 + \Vert \widetilde \bz-\widetilde\bz^\prime\Vert),
\end{align}
where $\Vert - \Vert_F$ is the Frobenius norm of the matrix and $\Vert - \Vert$ the Euclidean norm of the vectors. For the free energy difference we proceed as follows 
\begin{align}
 f_{t,h}(\mathring \by) - f_{t,h}(\mathring \by^\prime) &
 = \frac{1}{L}\ln\Big(\frac{\int d\bx P_0(\bx) e^{-\mathcal{H}_{t,h}(\bx|\mathring{\by}^\prime)}}{\int d\bx P_0(\bx) e^{-\mathcal{H}_{t,h}(\bx|\mathring{\by})}}\Big)
= \frac{1}{L}\ln\Big(\frac{\int d\bx P_0(\bx) e^{-\mathcal{H}_{t,h}(\bx|\mathring{\by}) + (\mathcal{H}_{t,h}(\bx|\mathring{\by})- 
 \mathcal{H}_{t,h}(\bx|\mathring{\by}^\prime)) }}{\int d\bx P_0(\bx) e^{-\mathcal{H}_{t,h}(\bx|\mathring{\by})}}\Big)
 \nonumber \\ &
 \leq 
 \frac{1}{L}\ln\Big(\frac{\int d\bx P_0(\bx) e^{-\mathcal{H}_{t,h}(\bx|\mathring{\by}) 
 + \vert \mathcal{H}_{t,h}(\bx|\mathring{\by})- \mathcal{H}_{t,h}(\bx|\mathring{\by}^\prime)\vert }}{\int d\bx P_0(\bx) e^{-\mathcal{H}_{t,h}(\bx|\mathring{\by})}}\Big)
 \nonumber \\ &
 \leq 
 \frac{1}{L} ({\cal O}(\sqrt{LD_1}) + {\cal O}(\sqrt{D_2})) (\Vert \bm{\phi}^{0} -\bm{\phi}^{{0}\prime}\Vert_{F} + \Vert \bz-\bz^\prime\Vert + \Vert \widehat\bz-\widehat\bz^\prime\Vert 
 + \Vert \widetilde \bz-\widetilde\bz^\prime\Vert). \label{257}
\end{align}
A similar argument yields a corresponding lower bound so that we get
\begin{align}
 \vert f_{t,h}(\mathring \by) - f_{t,h}(\mathring \by^\prime) \vert 
 &\leq 
 \frac{1}{L} ({\cal O}(\sqrt{LD_1}) + {\cal O}(\sqrt{D_2})) (\Vert \bm{\phi}^{0} -\bm{\phi}^{{0}\prime}\Vert_{F} + \Vert \bz-\bz^\prime\Vert + \Vert \widehat\bz-\widehat\bz^\prime\Vert 
 + \Vert \widetilde \bz-\widetilde\bz^\prime\Vert) \nonumber \\
 &\leq 
 \frac{1}{L} ({\cal O}(\sqrt{LD_1}) + {\cal O}(\sqrt{D_2})) \Vert (\bm{\phi}^{0},\bz,\widehat\bz,\widetilde \bz) - (\bm{\phi}^{{0}\prime},\bz^\prime,\widehat\bz^\prime,\widetilde\bz^\prime)\Vert,
 \label{lastlast}
\end{align}
where for the last inequality we used $\sum_{i=1}^n\|\bx_i\| \le \sqrt{n}\|(\bx_1,\ldots,\bx_n)\|$ which follows from the convexity of the parabola ($(\bx_1,\ldots,\bx_n)$ is the concatenation of $\{\bx_n\}$). Now we set $D_1= L^{\gamma}$ and $D_2$ a large enough constant. This gives a Lipshitz constant of order ${\cal O}(L^{-\frac{1-\gamma}{2}})$ for the Gaussian quenched variables in ${\cal G}$. 
It is perhaps worth to stress that the Lipshitz constant in the Ledoux-Talagrand inequality (and thus in Proposition \ref{concalmostLip} as well) must be computed w.r.t the 
Euclidean norm and that the r.h.s of \eqref{lastlast} is nothing else than the Euclidean norm of the $2N + M + NM= BL(2+\alpha+\alpha B L)={\cal O}(L^2)$ components vector formed by all Gaussian quenched
random variables.

Applying Proposition \ref{concalmostLip} using \eqref{255} and $K_P={\cal O}(L^{-\frac{1-\gamma}{2}})$ we get for any $r={\Omega}(L^{(5+\gamma)/4} e^{-L^\gamma /8})={\Omega}(e^{-c_2L^\gamma})$ (for some small enough $c_2$) and fixed $\bs$,
\begin{align}
 \mathbb{P}\big(\vert f_{t,h}(\mathring \by) - \mathbb{E}[f_{t,h}(\mathring \bY)|\bs] \vert \ge r \,\big|\, \bs\big) = {\cal O}(e^{- c r^2 L^{1-\gamma}})+ {\cal O}(L e^{-L^\gamma /4})
\end{align}
for some $c>0$. The choice $\gamma = 1/2$ optimizes this estimate and yields \eqref{tala} for a proper $c_1$. 

We must finally check that $C$ in Proposition \ref{concalmostLip} is ${\cal O}(1)$.
Note that one can prove explicitly that
$f_{t,h}(\mathring{\by})|_{\bm\Phi=\bz=\widetilde \bz=\widehat \bz=\boldsymbol{0}}^2$ and $\mathbb{E}[f_{t,h}(\mathring{\by})^2\vert \bs]$ are finite. 
We sketch the argument which is essentially already found in \cite{korada2010}. For $\bm\Phi=\bz=\widetilde \bz=\widehat \bz=\boldsymbol{0}$ we get a simple free energy on a discrete alphabet with no quenched variables. This is easily shown to be bounded. For the second quantity we proceed as follows. Since the Hamiltonian is positive 
we get the lower bound $f_{t, h}(\mathring{\by}) \geq - B\ln K$ (remember the free energy is defined with a minus in front of the logarithm and $K$ is the cardinality of the alphabet). On the other hand retaining only the term $\bx =\bs$ in the statistical sum
we obtain the upper bound $f_{t, h}(\mathring{\by}) \leq (\sum_\mu z_\mu^2 +\sum_i \widetilde z_i^2 + \widehat z_i^2)/(2L) + \sqrt{h} s_{\rm max}(\sum_i\vert \widehat z_i\vert)/L$. Averaging over all Gaussian variables shows that $\mathbb{E}[f_{t,h}(\mathring{\by})^2\vert \bs]$ must indeed be finite.
\end{IEEEproof}
\subsection{Concentration of the free energy with respect to the signal}
We define the set of signal realizations
%
\begin{equation}\label{Salpah}
{\cal S}_\alpha\defeq\Big\{\bs\, \Big| \, \mathbb{E}\Big[\sum_{\mu=1}^M\Phi_{\mu i}\langle [\bm{\Phi}\bX]_\mu\rangle_{t,h} \, \Big| \,  \bs\Big]^2 < N^{2\alpha}\Big\} \,,
\end{equation}
where $0<\alpha< 1$ is to be fixed later.
\begin{proposition}[Concentration of the free energy w.r.t the signal]\label{thingprop2}
One can find a constant $c_3>0$ (depending only on $s_{\rm max}$, $\alpha$ and $\Delta$) and 
$0<\alpha <1/2$ s.t for any $\bs \in {\cal S}_\alpha$, 
\begin{align}
\mathbb{P}\big(\vert \mathbb{E}[f_{t,h}(\mathring \bY) \mid \bs] - \EE_{\bS\in {\cal S}_\alpha}\mathbb{E}[f_{t,h}(\mathring \bY) \mid \bS]| \ge r\, \big|\,\bs\in  {\cal S}_\alpha\big) \le e^{-c_3 r^2 L^{1-2\alpha}}\,.
\label{thing}
\end{align}
\end{proposition}

\begin{IEEEproof} 
Let $\bs \in {\cal S}_\alpha$. We show a bounded difference property for $\mathbb{E}[f_{t,h}(\mathring{\bY}) \mid \bs]/L$. Consider $s_1,\dots,s_i,s_i^\prime,\dots, s_N$ in ${\cal S}_\alpha$ and estimate the corresponding Hamiltonian variation $\delta\mathcal{H} \!\defeq\! \mathcal{H}_{t,h}(\bx|\mathring{\by}) \!-\! \mathcal{H}_{t,h}(\bx|\mathring{\by}^\prime)$. From \eqref{eq:int_hamiltonian}, $\delta\mathcal{H} \! =\!\delta\mathcal{H}_\gamma \!+\!\delta\mathcal{H}_\lambda \!+\!\delta\mathcal{H}_h$ where
\begin{align}
 \delta\mathcal{H}_\gamma &= 
 \sum_{\mu=1}^M \Big\{\phi_{\mu i} (s_i - s_i^\prime)\Big( \gamma(t)\phi_{\mu i} (s_i+s_i^\prime) + 
 \frac{\gamma(t)}{2}\sum_{j\neq i}\phi_{\mu j}(s_j + s_j^\prime) - \gamma(t)[\bm\phi\bx]_\mu \label{thething}
 + 2\frac{z_\mu}{\sqrt{\gamma(t)}}\Big)\Big\}, \\
 \delta\mathcal{H}_\lambda &= \frac{\lambda(t)}{2}(s_i - s_i^\prime)\Big( (s_i+s_i^\prime) -2x_i + 2\frac{\widetilde z_i}{\sqrt{\lambda(t)}}\Big),
\end{align}
and $\delta\mathcal{H}_h$ is similar to $\delta\mathcal{H}_\lambda$ but with $h$ replacing $\lambda(t)$. These will be used a bit later, but first we need the following remark.
Let $\mathcal{H}=\mathcal{H}_{t,h}(\bx|\mathring{\by})$, $\mathcal{H}^\prime=\mathcal{H}_{t,h}(\bx|\mathring{\by}^\prime)$
and 
\begin{align}
\langle A(\bX) \rangle_{\mathcal{H}} = \frac{1}{\mathcal{Z}}\int d\bx P_0(\bx) A(\bx)e^{-\mathcal{H}},
\quad  \mathcal{Z} = \int d\bx P_0(\bx) e^{-\mathcal{H}}, 
\label{eq:meanA}
\end{align}
and similarly for $\langle A \rangle_{\mathcal{H}'}$, $\mathcal{Z}'$. We note
\begin{align}
f_{t,h}(\mathring \by)&= -\frac{1}{L} \ln(\mathcal{Z}) =  -\frac{1}{L} \ln(\mathcal{Z}' \langle e^{-\delta \mathcal{H}}\rangle_{\mathcal{H}'}) = f_{t,h}(\mathring\by') -\frac{1}{L} \ln( \langle e^{-\delta \mathcal{H}}\rangle_{\mathcal{H}'} ), \\ 
f_{t,h}(\mathring\by') &= -\frac{1}{L} \ln(\mathcal{Z}') = -\frac{1}{L} \ln(\mathcal{Z} \langle e^{\delta\mathcal{H}}\rangle_{\mathcal{H}} ) = f_{t,h}(\mathring\by) -\frac{1}{L} \ln( \langle e^{\delta \mathcal{H}}\rangle_{\mathcal{H}} ),
\end{align}
so using the convexity of the exponential, 
\begin{align} \label{eq:sandwich_tilde_concentration1}
f_{t,h}(\mathring\by') + \frac{\langle \delta \mathcal{H} \rangle_{\mathcal{H}}}{L} \le f_{t,h}(\mathring\by) \le f_{t,h}(\mathring\by') 
+ \frac{\langle \delta \mathcal{H}\rangle_{\mathcal{H}'}}{L}.
\end{align}
Averaging over $\bm\Phi$, $\bZ$, $\widehat\bZ$, $\widetilde \bZ$ we obtain for fixed $\bs$ and $\bs^\prime$ in $\cal{S}_\alpha$
\begin{align} \label{eq:sandwich_tilde_concentration2}
\mathbb{E}[f_{t,h}(\mathring\bY') \mid \bs^\prime] + \frac{1}{L} \mathbb{E}[\langle \delta \mathcal{H} \rangle_{\mathcal{H}} \mid  \bs, \bs^\prime]
\le \mathbb{E}[f_{t,h}(\mathring\bY) \mid \bs] \le \mathbb{E}[f_{t,h}(\mathring\bY') \mid \bs']
+ \frac{1}{L}\mathbb{E}[\langle \delta \mathcal{H}\rangle_{\mathcal{H}'} \mid \bs, \bs^\prime].
\end{align}
Recall $\delta\mathcal{H}  =\delta\mathcal{H}_\gamma +\delta\mathcal{H}_\lambda +\delta\mathcal{H}_h$. Let us estimate $\mathbb{E}[\langle \delta\mathcal{H}_\gamma\rangle_{\mathcal{H}, \mathcal{H}^\prime} \mid \bs, \bs^\prime]$. From \eqref{thething} we deduce
\begin{align}
\big|\mathbb{E}[\langle \delta\mathcal{H}_\gamma\rangle_{\mathcal{H}, \mathcal{H}^\prime} \mid \bs, \bs^\prime]\big| & = \Big|\frac{M}{L}\gamma(t)(s_i^2 - s_i^{\prime 2}) - \gamma(t) \mathbb{E}\Big[\sum_{\mu=1}^M \Phi_{\mu i}\langle [\bm{\Phi} \bX]_\mu \rangle_{\mathcal{H}, \mathcal{H}^\prime}\,\Big|\, \bs, \bs'\Big](s_i - s_i^\prime)\Big|
\nonumber \\ &
\leq 
\frac{M}{L}\gamma(t)(s_i^2 - s_i^{\prime 2}) +
\gamma(t) N^\alpha |s_i - s_i^{\prime}|  = {\cal O}(L^\alpha).
\label{last_}
\end{align}
In the last line we used boundedness of the signal. Similar equations lead to $|\mathbb{E}[\langle \delta\mathcal{H}_\lambda\rangle_{\mathcal{H}, \mathcal{H}^\prime} \mid \bs, \bs^\prime]|\!=\!|\mathbb{E}[\langle \delta\mathcal{H}_h\rangle_{\mathcal{H}, \mathcal{H}^\prime} \mid \bs, \bs^\prime]|\!=\!{\cal O}(1)$. Combining this with \eqref{eq:sandwich_tilde_concentration2} and \eqref{last_} we obtain the bounded difference 
property
\begin{align}
 \vert \mathbb{E}[f_{t,h}(\mathring\bY) \mid \bs]  -  \mathbb{E}[f_{t,h}(\mathring\bY') \mid \bs^\prime]\vert = {\cal O}(L^{\alpha -1})
\end{align}
where we recall that $\bs, \bs'\in {\cal S}_\alpha$. The last step is a direct application of MacDiarmid's inequality (Proposition \ref{McD}).
\end{IEEEproof}
\subsection{Proof of Proposition \ref{cor:fs_minus_meanfs_small}}
Proposition \ref{cor:fs_minus_meanfs_small} is a corollary of the following proposition.
\begin{proposition}[Concentration of the free energy] \label{cor:mcdiarmid}
One can find a constant $c>0$ depending only on the parameters $s_{\rm max}$, $K$, $B$ and $\alpha$ s.t for any $\bs\in{\cal S}_\alpha$ and $0<\eta <1/4$,
\begin{align}
\mathbb{P}\big(|f_{t,h}(\mathring \by) - \EE_{\bS\in {\cal S}_\alpha}\mathbb{E}[f_{t,h}(\mathring \bY) \mid \bS]| \ge L^{-\eta}\, \big|\,\bs\in  {\cal S}_\alpha\big) 
\le e^{-c L^{1/2 - 2\eta}}.
\end{align}
\end{proposition} 
\begin{IEEEproof}
Let $\bs\in{\cal S}_\alpha$. We have that the event $\{\vert f_{t,h}(\mathring{\by}) - \EE_{\bS\in {\cal S}_\alpha}\mathbb{E}[f_{t,h}(\mathring \bY) \mid \bS]\vert \geq r\}$ implies by the triangle inequality the event $\{\vert f_{t,h}(\mathring{\by}) - \mathbb{E}[f_{t,h}(\mathring{\bY}) \mid \bs]\vert \geq r/2\} \cup \{\vert \mathbb{E}[f_{t,h}(\mathring{\bY})\mid \bs] - \EE_{\bS\in {\cal S}_\alpha}\mathbb{E}[f_{t,h}(\mathring{\bY})|\bS]\vert \geq r/2\}$. Thus from the union bound and 
Propositions \ref{thingprop1} and \ref{thingprop2} we obtain 
\begin{align}
&\mathbb{P}\big(|f_{t,h}(\mathring \by) - \EE_{\bS\in {\cal S}_\alpha}\mathbb{E}[f_{t,h}(\mathring \bY) \mid \bS]| \ge r\, \big|\,\bs\in  {\cal S}_\alpha\big) \nonumber \\
\le \,&\mathbb{P}\big(|f_{t,h}(\mathring \by) - \mathbb{E}[f_{t,h}(\mathring \bY) \mid \bs]| \ge r/2\, \big|\,\bs\in  {\cal S}_\alpha\big) + \mathbb{P}\big(|\mathbb{E}[f_{t,h}(\mathring \bY) \mid \bs] - \EE_{\bS\in {\cal S}_\alpha}\mathbb{E}[f_{t,h}(\mathring \bY) \mid \bS]| \ge r/2\, \big|\,\bs\in  {\cal S}_\alpha\big) \nonumber\\
\le \,&e^{-(c_1/4) r^2 L^{1/2}} + e^{- (c_3/4) r^2 L^{1-2\alpha}}.
\end{align}
We finally choose $r= L^{-\eta}$ and $\alpha =\eta$. Then the upper bound is $\mathcal{O}(e^{-c L^{1/2 - 2\eta}})$ for some  $c>0$. This bound goes to $0$ as $L$ for any $0< \eta <1/4$, which ends the proof.
\end{IEEEproof}

Now we can prove Proposition \ref{cor:fs_minus_meanfs_small}. Let $b \defeq |f_{t,h}(\mathring \by) - \EE_{\bS\in {\cal S}_\alpha}\mathbb{E}[f_{t,h}(\mathring \bY) \mid \bS]|$. 
Then using Proposition \ref{cor:mcdiarmid} with $0<\eta <1/4$, 
\begin{align}
\EE_{\bS\in {\cal S}_\alpha}\mathbb{E}[b] &=
\EE_{\bS\in {\cal S}_\alpha}\mathbb{E}[b \mathds{1}(b< L^{-\eta})] + \EE_{\bS\in {\cal S}_\alpha}\mathbb{E}[b\mathds{1}(b\ge L^{-\eta})]
\nonumber\\&
\le L^{-\eta} + c^\prime e^{-\frac{c}{2} L^{1/2 - 2\eta}} = {\cal O}(L^{-\eta}).
 \label{meanb}
\end{align}
where the last term follows by applying Cauchy-Schwarz and bounding $\mathbb{E}[b^2]$ by remarking that $b^2$ can be estimated by a 
polynomial-like function of Gaussian variables $\bz, \widetilde\bz, \widehat\bz$ (see the remarks at the end of the proof of Proposition \ref{thingprop1}).
Let us now estimate $\mathbb{P}({\cal S}_\alpha^c)$ defined by \eqref{Salpah}. Applying successively Markov's inequality, a convexity inequality and the Nishimori identity one obtains
\begin{align}
\mathbb{P}({\cal S}_\alpha^c) &= \mathbb{P}\Big(\mathbb{E}\Big[\Big\langle \sum_{\mu=1}^M\Phi_{\mu i}[\bm{\Phi}\bX]_\mu\Big\rangle_{t,h} \, \Big| \,  \bs\Big]^2 \ge N^{2\alpha}\Big) 
\leq N^{-2\alpha}\mathbb{E}_\bS\Big[\mathbb{E}\Big[\Big\langle\sum_{\mu=1}^M\Phi_{\mu i}[\bm{\Phi}\bX]_\mu\Big\rangle_{t,h} \, \Big| \,  \bS\Big]^2\Big] \nonumber\\
&\leq N^{-2\alpha}\mathbb{E}\Big[\Big\langle\Big(\sum_{\mu=1}^M\Phi_{\mu i}[\bm{\Phi}\bX]_\mu\Big)^2\Big\rangle_{t,h} \Big]
= N^{-2\alpha}\mathbb{E}\Big[\Big(\sum_{\mu=1}^M\Phi_{\mu i}[\bm{\Phi}\bS]_\mu\Big)^2 \Big] \nonumber \\
&= N^{-2\alpha}\mathbb{E}\Big[\sum_{\mu,\nu=1}^M\sum_{j,k=1}^N\Phi_{\mu i}\Phi_{\nu i}\Phi_{\mu j}\Phi_{\nu k} S_j S_k \Big] 
= {\cal O}(N^{-2\alpha}), \label{last}
\end{align}
where the last equality is easy to check carefully using independence of random variables. Finally using the same bounds as above on $f_{t,h}(\mathring \by)$
we deduce that 
$\EE_{\bS\in {\cal S}_\alpha}\mathbb{E}[b] = \mathbb{E}[|f_{t,h}(\mathring \by) - \mathbb{E}[f_{t,h}(\mathring \bY)]|] + {\cal O}(\mathbb{P}({\cal S}_\alpha^c))$
where the last average $\mathbb{E}$ includes all quenched variables, with $\bS$ not anymore restricted to ${\cal S}_\alpha$. 
Then combining this with \eqref{meanb}, \eqref{last} and recalling that $\alpha=\eta$ one 
obtains $\mathbb{E}[|f_{t,h}(\mathring \by) - \mathbb{E}[f_{t,h}(\mathring \bY)]|]= {\cal O}(L^{-\eta})$ which is Proposition \ref{cor:fs_minus_meanfs_small}.

\section{Super-additivity of the mutual information}\label{appendix_thermolimit}

As explained in the main text (Sec. \ref{sec:mutualInfoSCmodel_1}, paragraph \ref{thermo}) the same proof leading to 
\eqref{eq121} also yields the inequality $i^{\rm cs}_L \geq \frac{L_1}{L} i^{\rm cs}_{L_1} + \frac{L_2}{L} i^{\rm cs}_{L_2} + \smallO_L(1)$. However to apply Fekete's lemma (and its stronger versions) we need to improve the error term
from $\smallO_L(1)$ to $\mathcal{O}(L^{-\eta})$. In order to do so we note that an inspection of the arguments leading to \eqref{eq121} shows that they ultimately rest on proposition \ref{lemma:concentration}. As a consequence it is easily seen 
that we can replace $\smallO_L(1)$ by $\mathcal{O}(L^{-\eta})$ for some 
small $\eta >0$, if we look at the {\it average} of the mutual information over a small $h$-interval, $h\in [h_0 - a_L, h_0+ a_L]$ where $h_0\in ]0, 1]$ and $a_L = \mathcal{O}(L^{-\eta/2})$ (in proposition \ref{lemma:concentration} the average is taken on a fixed interval but one can take an $L$ dependent interval at the expense of a worse $\eta > 0$). 

We obtain that for  any
$L_1$, $L_2$, $L$ such that $L= L_1 + L_2$,
\begin{align}\label{supintegrer}
 \frac{1}{2a_L} \int_{h_0- a_L}^{h_0+a_L} dh i^{\rm cs}_L(h) \geq \frac{1}{2a_L} \int_{h_0- a_L}^{h_0+a_L}  dh\frac{L_1}{L} i^{\rm cs}_{L_1}(h) 
 + \frac{1}{2a_L} \int_{h_0- a_L}^{h_0+a_L}  dh\frac{L_2}{L} i^{\rm cs}_{L_2}(h) + \mathcal{O}(L^{-\eta}).
\end{align}
By an explicit calculation of the $h$-derivative in equation \eqref{dfintdh_order1} we know that $i_L^{\rm cs}(h)$ is $h$-Lipshitz with constant $\kappa$ independent of $L$. Therefore we have for the term on the l.h.s
\begin{align}
 \bigg\vert \frac{1}{2a_L} \int_{h_0- a_L}^{h_0+a_L}  dh(i^{\rm cs}_L(h) - i^{\rm cs}_L(h_0))\bigg\vert 
 & \leq \frac{1}{2a_L} \int_{h_0- a_L}^{h_0+a_L}  dh\vert i^{\rm cs}_L(h) - i^{\rm cs}_L(h_0)\vert 
 \nonumber \\ &
 \leq 
 \frac{1}{2a_L} \int_{h_0- a_L}^{h_0+a_L}  dh\kappa \vert h- h_0\vert
 \nonumber \\ &
 \leq \frac{\kappa\, a_L}{2}\, .
\end{align}
For the two terms on the r.h.s we have a similar statement. Combining with \eqref{supintegrer} we find 
\begin{equation}
i^{\rm cs}_L(h_0) \geq \frac{L_1}{L} i^{\rm cs}_{L_1}(h_0) + \frac{L_2}{L} i^{\rm cs}_{L_2}(h_0) + \smallO_L(L^{-\eta})\,.
\end{equation}
which is the desired super-additivity. 

\section{Unicity of $\Delta_{\rm RS}$ under the assumptions on $i_{\rm RS}(E; \Delta)$}\label{appendix_AnalyticityAndDeltaRS}

First we argue that $\widetilde{E}(\Delta)$, and by way of consequence $i^{\rm RS}(\widetilde E(\Delta);\Delta)$, are analytic in $\Delta$ except at $\Delta_{\rm RS}$ where the
${\rm argmin}$ in \eqref{eq:defeTildeE} is not unique.
For models with a discrete prior, \eqref{eq:rs_mutual_info} can be shown to be a {\it real analytic function} in each argument $E > 0$, $\Delta > 0$. Set $f(E, \Delta) = \partial_E i_{\rm RS}(E; \Delta)$ for the partial derivative w.r.t $E$. Consider a stationary point, i.e., a solution $E_*(\Delta)$ of the equation $f(E, \Delta) = 0$. As long as $\partial_E f(E, \Delta)\vert_{E_*(\Delta)} \neq 0$ for $\Delta \in U\subset \mathbb{R}_+$ an open set, i.e., as long as this solution is not a horizontal inflexion point,  the {\it implicit function theorem for real analytic functions} implies that $E_*(\Delta)$ is analytic w.r.t $\Delta\in U$. One concludes that the solutions of the stationary point equation are analytic branches over their domain of definition as long as they do not merge (because if a minimum and a maximum merge a horizontal inflexion point appears). This allows to conclude that $\widetilde E(\Delta)$ must be analytic in $\Delta$ except at $\Delta_{\rm RS}$ where the ${\rm argmin}\, i^{\rm RS}(E;\Delta)$ is not unique. 

When there are at most three stationary points for~\eqref{eq:rs_mutual_info}~the non-unicity of the 
${\rm argmin}$ occurs in two possible ways: (i) it continuously bifurcates into a maximum and minimum pair, or; (ii) it jumps from one analytic branch to the other. Possibility (i) corresponds to a continuous phase transition in the MMSE, and possibility (ii) corresponds to a first order phase transition. Let us now argue that $\Delta_{\rm RS}$ is unique.

We define the {\it gap} as the quantity: 
$+\infty$ when the ${\rm argmin}\, i^{\rm RS}(E;\Delta)$ is unique and there is no other local minimum; as 
$i^{\rm RS}(E_{\rm loc}(\Delta); \Delta) - i^{\rm RS}(\widetilde{E}(\Delta); \Delta)$ when the ${\rm argmin}\, i^{\rm RS}(E;\Delta)$ is unique and there is a local minimum $E_{\rm loc}(\Delta) \neq \widetilde{E}(\Delta)$; as $0$ when the ${\rm argmin}\, i^{\rm RS}(E;\Delta)$ takes two values. 
Now, by explicit calculation  we have for any stationary point $E_*(\Delta)$ an I-MMSE type of relation,
\begin{align}
\frac{d}{d\Delta^{-1}} i^{\rm RS}(E_*(\Delta); \Delta) = \frac{\alpha B}{2} \frac{E_*(\Delta)}{1+ E_*(\Delta)/\Delta}\, .
\end{align}
This relation is derived by explicit calculation in Appendix \ref{app:Differentiation} for the global minimum $\widetilde{E}(\Delta)$, but the calculation is exactly the same for all stationary points. Therefore the variation of the gap with $\Delta$ (when it is neither zero or infinite) is given by
\begin{align}\label{derivativeofgap}
\frac{d}{d\Delta}\biggl(i^{\rm RS}(E_{\rm loc}(\Delta); \Delta) - i^{\rm RS}(\widetilde{E}(\Delta); \Delta)\biggr) & = 
-\frac{\alpha B}{2\Delta^{-2}}\frac{E_{\rm loc}(\Delta) - \widetilde{E}(\Delta)}{(1+E_{\rm loc}(\Delta)/\Delta)(1+ \widetilde{E}(\Delta)/\Delta)}\,.
\end{align}
For small enough $\Delta$ one can argue that the gap is infinite. Suppose that a local minimum appears as we increase $\Delta$. First assume that 
$E_{\rm loc}(\Delta) \geq \widetilde{E}(\Delta)$. Then the last formula shows that the gap must diminish. Suppose it diminishes until a value of $\Delta_{\rm RS}$ where it vanishes. At this point the local and global minima switch roles and for slightly higher $\Delta$ we necessarily have $\widetilde{E}(\Delta)  \geq E_{\rm loc}(\Delta)$ which implies that the gap must increase.
A bit of thought shows that the gap necessarily will increase to infinity (either the local minimum continues to exist and by analyticity the gap cannot stay constant, or the local minimum merges with the local maximum and the gap becomes infinite by definition). Finally consider the situation where as $\Delta >0$ increases the local minimum appears s.t. $E_{\rm loc}(\Delta) \geq \widetilde{E}(\Delta)$. Then the gap must increase and arguing as before we can see that it will increase to infinity.

\section*{Acknowledgments}
We thank Marc Lelarge and Andrea Montanari for clarifications on state evolution for the vectorial case $B \geq 2$ and  Andrea Montanari 
for helpful discussions related to Section \ref{sec:AMP}. We also thank L\'eo Miolane for indicating how to extend our results
to a wider class of prior distributions (see remark \ref{remark-comparing-with-GH}). 
This work was done while Jean Barbier and Mohamad Dia were affiliated with EPFL. Jean Barbier and Mohamad Dia acknowledge funding from the Swiss National Science Foundation (grant
200021-156672). Florent Krzakala thanks the Simons Institute in Berkeley for its hospitality and acknowledges funding from the European Union (FP/2007-2013/ERC grant agreement 307087-SPARCS).
\ifCLASSOPTIONcaptionsoff
\fi
\bibliographystyle{IEEEtran}
\bibliography{refs}

\begin{thebibliography}{10}
\providecommand{\url}[1]{#1}
\csname url@samestyle\endcsname
\providecommand{\newblock}{\relax}
\providecommand{\bibinfo}[2]{#2}
\providecommand{\BIBentrySTDinterwordspacing}{\spaceskip=0pt\relax}
\providecommand{\BIBentryALTinterwordstretchfactor}{4}
\providecommand{\BIBentryALTinterwordspacing}{\spaceskip=\fontdimen2\font plus
\BIBentryALTinterwordstretchfactor\fontdimen3\font minus
  \fontdimen4\font\relax}
\providecommand{\BIBforeignlanguage}[2]{{%
\expandafter\ifx\csname l@#1\endcsname\relax
\typeout{** WARNING: IEEEtran.bst: No hyphenation pattern has been}%
\typeout{** loaded for the language `#1'. Using the pattern for}%
\typeout{** the default language instead.}%
\else
\language=\csname l@#1\endcsname
\fi
#2}}
\providecommand{\BIBdecl}{\relax}
\BIBdecl

\bibitem{candes2006near}
E.~J. Candes and T.~Tao, ``Near-optimal signal recovery from random
  projections: Universal encoding strategies?'' \emph{IEEE Transactions on
  Information Theory}, vol.~52, no.~12, pp. 5406--5425, Dec 2006.

\bibitem{verdu1999spectral}
S.~Verd{\'u} and S.~Shamai, ``Spectral efficiency of cdma with random
  spreading,'' \emph{IEEE Transactions on Information Theory}, vol.~45, no.~2,
  pp. 622--640, Mar 1999.

\bibitem{barron2010sparse}
A.~R. Barron and A.~Joseph, ``Toward fast reliable communication at rates near
  capacity with gaussian noise,'' in \emph{2010 IEEE International Symposium on
  Information Theory}, June 2010, pp. 315--319.

\bibitem{atia2012boolean}
G.~K. Atia and V.~Saligrama, ``Boolean compressed sensing and noisy group
  testing,'' \emph{IEEE Transactions on Information Theory}, vol.~58, no.~3,
  pp. 1880--1901, March 2012.

\bibitem{mezard1990spin}
M.~M{\'e}zard, G.~Parisi, and M.-A. Virasoro, ``Spin glass theory and beyond.''
  1987.

\bibitem{donoho2009message}
D.~L. Donoho, A.~Maleki, and A.~Montanari, ``Message-passing algorithms for
  compressed sensing,'' \emph{Proceedings of the National Academy of Sciences},
  vol. 106, no.~45, pp. 18\,914--18\,919, Nov 2009.

\bibitem{krzakala2012statistical}
F.~Krzakala, M.~M{\'e}zard, F.~Sausset, Y.~Sun, and L.~Zdeborov{\'a},
  ``Statistical-physics-based reconstruction in compressed sensing,''
  \emph{Phys. Rev. X}, vol.~2, p. 021005(18), May 2012.

\bibitem{krzakala2012probabilistic}
------, ``Probabilistic reconstruction in compressed sensing: algorithms, phase
  diagrams, and threshold achieving matrices,'' \emph{Journal of Statistical
  Mechanics: Theory and Experiment}, vol. 2012, no.~08, p. P08009(57), 2012.

\bibitem{tanaka2002statistical}
T.~Tanaka, ``A statistical-mechanics approach to large-system analysis of cdma
  multiuser detectors,'' \emph{IEEE Transactions on Information Theory},
  vol.~48, no.~11, pp. 2888--2910, Nov 2002.

\bibitem{gardner1988space}
E.~Gardner, ``The space of interactions in neural network models,''
  \emph{Journal of physics A: Mathematical and general}, vol.~21, no.~1, pp.
  257--270, 1988.

\bibitem{gardner1988optimal}
E.~Gardner and B.~Derrida, ``Optimal storage properties of neural network
  models,'' \emph{Journal of Physics A: Mathematical and general}, vol.~21,
  no.~1, pp. 271--284, 1988.

\bibitem{mezard1989space}
M.~M{\'e}zard, ``The space of interactions in neural networks: Gardner's
  computation with the cavity method,'' \emph{Journal of Physics A:
  Mathematical and General}, vol.~22, no.~12, pp. 2181--2190, 1989.

\bibitem{guo2003multiuser}
D.~Guo and S.~Verd{\'u}, \emph{Multiuser detection and statistical
  mechanics}.\hskip 1em plus 0.5em minus 0.4em\relax Boston, MA: Springer US,
  2003, p. 229–277.

\bibitem{guo2005randomly}
------, ``Randomly spread cdma: Asymptotics via statistical physics,''
  \emph{IEEE Transactions on Information Theory}, vol.~51, no.~6, pp.
  1983--2010, June 2005.

\bibitem{guo2009single}
D.~Guo, D.~Baron, and S.~Shamai, ``A single-letter characterization of optimal
  noisy compressed sensing,'' in \emph{2009 47th Annual Allerton Conference on
  Communication, Control, and Computing (Allerton)}, Sept 2009, pp. 52--59.

\bibitem{rangan2009asymptotic}
S.~Rangan, V.~Goyal, and A.~K. Fletcher, ``Asymptotic analysis of map
  estimation via the replica method and compressed sensing,'' in \emph{Advances
  in Neural Information Processing Systems 22}, 2009, p. 1545–1553.

\bibitem{kabashima2009typical}
Y.~Kabashima, T.~Wadayama, and T.~Tanaka, ``A typical reconstruction limit for
  compressed sensing based on lp-norm minimization,'' \emph{Journal of
  Statistical Mechanics: Theory and Experiment}, vol. 2009, no.~09, p.
  L09003(12), 2009.

\bibitem{ganguli2010statistical}
S.~Ganguli and H.~Sompolinsky, ``Statistical mechanics of compressed sensing,''
  \emph{Physical review letters}, vol. 104, p. 188701(4), May 2010.

\bibitem{wu2012optimal}
Y.~Wu and S.~Verd{\'u}, ``Optimal phase transitions in compressed sensing,''
  \emph{IEEE Transactions on Information Theory}, vol.~58, no.~10, pp.
  6241--6263, Oct 2012.

\bibitem{tulino2013support}
A.~M. Tulino, G.~Caire, S.~Verd{\'u}, and S.~Shamai, ``Support recovery with
  sparsely sampled free random matrices,'' \emph{IEEE Transactions on
  Information Theory}, vol.~59, no.~7, pp. 4243--4271, July 2013.

\bibitem{thouless1977solution}
D.~J. Thouless, P.~W. Anderson, and R.~G. Palmer, ``Solution of`solvable model
  of a spin glass','' \emph{Philosophical Magazine}, vol.~35, no.~3, p.
  593–601, 1977.

\bibitem{kabashima2003cdma}
Y.~Kabashima, ``A cdma multiuser detection algorithm on the basis of belief
  propagation,'' \emph{Journal of Physics A: Mathematical and General},
  vol.~36, no.~43, pp. 11\,111--11\,121, 2003.

\bibitem{bayati2011dynamics}
M.~Bayati and A.~Montanari, ``The dynamics of message passing on dense graphs,
  with applications to compressed sensing,'' \emph{IEEE Transactions on
  Information Theory}, vol.~57, no.~2, pp. 764--785, Feb 2011.

\bibitem{bayati2015universality}
M.~Bayati, M.~Lelarge, and A.~Montanari, ``Universality in polytope phase
  transitions and message passing algorithms,'' \emph{The Annals of Applied
  Probability}, vol.~25, no.~2, pp. 753--822, 2015.

\bibitem{Bolthausen2014}
E.~Bolthausen, ``An iterative construction of solutions of the tap equations
  for the sherrington--kirkpatrick model,'' \emph{Communications in
  Mathematical Physics}, vol. 325, no.~1, p. 333–366, 2014.

\bibitem{guerra2005introduction}
F.~Guerra, ``An introduction to mean field spin glass theory: methods and
  results,'' \emph{Mathematical Statistical Physics: Lecture Notes of the Les
  Houches Summer School 2005}, pp. 243--271, 2005.

\bibitem{korada2010}
S.~B. Korada and N.~Macris, ``Tight bounds on the capacity of binary input
  random cdma systems,'' \emph{IEEE Transactions on Information Theory},
  vol.~56, no.~11, pp. 5590--5613, Nov 2010.

\bibitem{kudekar2010effect}
S.~Kudekar and H.~D. Pfister, ``The effect of spatial coupling on compressive
  sensing,'' in \emph{2010 48th Annual Allerton Conference on Communication,
  Control, and Computing (Allerton)}, Sept 2010, pp. 347--353.

\bibitem{donoho2013information}
D.~L. Donoho, A.~Javanmard, and A.~Montanari, ``Information-theoretically
  optimal compressed sensing via spatial coupling and approximate message
  passing,'' \emph{IEEE Transactions on Information Theory}, vol.~59, no.~11,
  pp. 7434--7464, Nov 2013.

\bibitem{kudekar2011threshold}
S.~Kudekar, T.~J. Richardson, and R.~L. Urbanke, ``Threshold saturation via
  spatial coupling: Why convolutional ldpc ensembles perform so well over the
  bec,'' \emph{IEEE Transactions on Information Theory}, vol.~57, no.~2, pp.
  803--834, Feb 2011.

\bibitem{hassani2010coupled}
S.~H. Hassani, N.~Macris, and R.~Urbanke, ``Coupled graphical models and their
  thresholds,'' in \emph{2010 IEEE Information Theory Workshop}, Aug 2010, pp.
  1--5.

\bibitem{pfister2012}
A.~Yedla, Y.~Y. Jian, P.~S. Nguyen, and H.~D. Pfister, ``A simple proof of
  threshold saturation for coupled scalar recursions,'' in \emph{2012 7th
  International Symposium on Turbo Codes and Iterative Information Processing
  (ISTC)}, Aug 2012, pp. 51--55.

\bibitem{kumar2014}
S.~Kumar, A.~J. Young, N.~Macris, and H.~D. Pfister, ``Threshold saturation for
  spatially coupled ldpc and ldgm codes on bms channels,'' \emph{IEEE
  Transactions on Information Theory}, vol.~60, no.~12, pp. 7389--7415, Dec
  2014.

\bibitem{barbier2014replica}
J.~Barbier and F.~Krzakala, ``Replica analysis and approximate message passing
  decoder for superposition codes,'' in \emph{2014 IEEE International Symposium
  on Information Theory}, June 2014, pp. 1494--1498.

\bibitem{barbier2015approximate}
------, ``Approximate message-passing decoder and capacity-achieving sparse
  superposition codes,'' \emph{IEEE Transactions on Information Theory}, 2017.

\bibitem{barbier2016proof}
J.~Barbier, M.~Dia, and N.~Macris, ``Proof of threshold saturation for
  spatially coupled sparse superposition codes,'' in \emph{2016 IEEE
  International Symposium on Information Theory (ISIT)}, July 2016, pp.
  1173--1177.

\bibitem{krzakala2016mutual}
F.~Krzakala, J.~Xu, and L.~Zdeborov{\'a}, ``Mutual information in rank-one
  matrix estimation,'' in \emph{2016 IEEE Information Theory Workshop (ITW)},
  Sept 2016, pp. 71--75.

\bibitem{XXT}
J.~Barbier, M.~Dia, N.~Macris, F.~Krzakala, T.~Lesieur, and L.~Zdeborová,
  ``Mutual information for symmetric rank-one matrix estimation: A proof of the
  replica formula,'' in \emph{Advances in Neural Information Processing Systems
  29}, 2016, p. 424–432.

\bibitem{HassaniMacris2013}
H.~Hassani, N.~Macris, and R.~Urbanke, ``Threshold saturation in spatially
  coupled constraint satisfaction problems,'' \emph{Journal of Statistical
  Physics}, vol. 150, no.~5, pp. 807--850, March 2013.

\bibitem{BarbierDMK16}
J.~Barbier, M.~Dia, N.~Macris, and F.~Krzakala, ``The mutual information in
  random linear estimation,'' in \emph{2016 54th Annual Allerton Conference on
  Communication, Control, and Computing (Allerton)}, 2016.

\bibitem{private}
G.~Reeves and H.~D. Pfister, ``The replica-symmetric prediction for compressed
  sensing with gaussian matrices is exact,'' in \emph{2016 IEEE International
  Symposium on Information Theory (ISIT)}, July 2016, pp. 665--669.

\bibitem{DBLP:journals/corr/ReevesP16}
\BIBentryALTinterwordspacing
------, ``The replica-symmetric prediction for compressed sensing with gaussian
  matrices is exact,'' \emph{CoRR}, vol. abs/1607.02524, 2016. [Online].
  Available: \url{http://arxiv.org/abs/1607.02524}
\BIBentrySTDinterwordspacing

\bibitem{GuoShamaiVerdu_IMMSE}
D.~Guo, S.~Shamai, and S.~Verd{\'u}, ``Mutual information and minimum
  mean-square error in gaussian channels,'' \emph{IEEE Transactions on
  Information Theory}, vol.~51, no.~4, pp. 1261--1282, April 2005.

\bibitem{phdBarbier}
\BIBentryALTinterwordspacing
J.~Barbier, ``Statistical physics and approximate message-passing algorithms
  for sparse linear estimation problems in signal processing and coding
  theory,'' Ph.D. dissertation, Université Paris Diderot, 2015. [Online].
  Available: \url{http://arxiv.org/abs/1511.01650}
\BIBentrySTDinterwordspacing

\bibitem{reeves2012sampling}
G.~Reeves and M.~Gastpar, ``The sampling rate-distortion tradeoff for sparsity
  pattern recovery in compressed sensing,'' \emph{IEEE Transactions on
  Information Theory}, vol.~58, no.~5, pp. 3065--3092, 2012.

\bibitem{reeves2012compressed}
------, ``Compressed sensing phase transitions: Rigorous bounds versus replica
  predictions,'' in \emph{2012 46th Annual Conference on Information Sciences
  and Systems (CISS)}.\hskip 1em plus 0.5em minus 0.4em\relax IEEE, 2012, pp.
  1--6.

\bibitem{barbierSchulkeKrzakala}
J.~Barbier, C.~Schülke, and F.~Krzakala, ``Approximate message-passing with
  spatially coupled structured operators, with applications to compressed
  sensing and sparse superposition codes,'' \emph{Journal of Statistical
  Mechanics: Theory and Experiment}, vol. 2015, no.~5, 2015.

\bibitem{rush2015capacity}
C.~Rush, A.~Greig, and R.~Venkataramanan, ``Capacity-achieving sparse
  superposition codes via approximate message passing decoding,'' \emph{arXiv
  preprint arXiv:1501.05892}, 2015.

\bibitem{5165186}
M.~Payaro and D.~P. Palomar, ``Hessian and concavity of mutual information,
  differential entropy, and entropy power in linear vector gaussian channels,''
  \emph{IEEE Transactions on Information Theory}, vol.~55, no.~8, pp.
  3613--3628, Aug 2009.

\bibitem{DBLP:journals/corr/MaRB17}
\BIBentryALTinterwordspacing
Y.~Ma, C.~Rush, and D.~Baron, ``Analysis of approximate message passing with a
  class of non-separable denoisers,'' \emph{CoRR}, vol. abs/1705.03126, 2017.
  [Online]. Available: \url{http://arxiv.org/abs/1705.03126}
\BIBentrySTDinterwordspacing

\bibitem{DBLP:journals/corr/abs-1708-03950}
\BIBentryALTinterwordspacing
R.~Berthier, A.~Montanari, and P.~Nguyen, ``State evolution for approximate
  message passing with non-separable functions,'' \emph{CoRR}, vol.
  abs/1708.03950, 2017. [Online]. Available:
  \url{http://arxiv.org/abs/1708.03950}
\BIBentrySTDinterwordspacing

\bibitem{Barbier_IMMSE_subExt}
J.~{Barbier} and N.~{Macris}, ``{I-MMSE relations in random linear estimation
  and a sub-extensive interpolation method},'' \emph{arXiv:1704.04158}, Apr.
  2017.

\bibitem{barbier_stoInt}
\BIBentryALTinterwordspacing
J.~Barbier and N.~Macris, ``The adaptive interpolation method: a simple scheme
  to prove replica formulas in bayesian inference,'' \emph{Probability Theory
  and Related Fields}, Oct 2018. [Online]. Available:
  \url{https://doi.org/10.1007/s00440-018-0879-0}
\BIBentrySTDinterwordspacing

\bibitem{2019arXiv190106516B}
J.~{Barbier} and N.~{Macris}, ``{The adaptive interpolation method for proving
  replica formulas. Applications to the Curie-Weiss and Wigner spike models},''
  \emph{arXiv e-prints}, p. arXiv:1901.06516, Jan 2019.

\bibitem{barbier2017phase}
J.~Barbier, F.~Krzakala, N.~Macris, L.~Miolane, and L.~Zdeborov{\'a}, ``Optimal
  errors and phase transitions in high-dimensional generalized linear models,''
  \emph{Proceedings of the National Academy of Sciences}, vol. 116, no.~12, pp.
  5451--5460, 2019.

\bibitem{guerra2002thermodynamic}
F.~Guerra and F.~L. Toninelli, ``The thermodynamic limit in mean field spin
  glass models,'' \emph{Communications in Mathematical Physics}, vol. 230,
  no.~1, pp. 71--79, 2002.

\bibitem{MontanariTse06}
A.~Montanari and D.~Tse, ``Analysis of belief propagation for non-linear
  problems: The example of cdma (or: How to prove tanaka's formula),'' in
  \emph{2006 IEEE Information Theory Workshop - ITW '06 Punta del Este}, March
  2006, pp. 160--164.

\bibitem{barbier2016threshold}
J.~Barbier, M.~Dia, and N.~Macris, ``Threshold saturation of spatially coupled
  sparse superposition codes for all memoryless channels,'' in \emph{2016 IEEE
  Information Theory Workshop (ITW)}, Sept 2016, pp. 76--80.

\bibitem{FranzLeone}
S.~Franz and M.~Leone, ``Replica bounds for optimization problems and diluted
  spin systems,'' \emph{Journal of Statistical Physics}, vol. 111, no.~3, pp.
  535--564, 2003.

\bibitem{korada2009exact}
S.~B. Korada and N.~Macris, ``Exact solution of the gauge symmetric p-spin
  glass model on a complete graph,'' \emph{Journal of Statistical Physics},
  vol. 136, no.~2, pp. 205--230, 2009.

\bibitem{PanchenkoTalagrand2004}
D.~Panchenko and M.~Talagrand, ``Bounds for diluted mean-fields spin glass
  models,'' \emph{Probability Theory and Related Fields}, vol. 130, no.~3, p.
  319–336, 2004.

\bibitem{Montanari2005}
A.~Montanari, ``Tight bounds for ldpc and ldgm codes under map decoding,''
  \emph{IEEE Transactions on Information Theory}, vol.~51, no.~9, pp.
  3221--3246, Sept 2005.

\bibitem{Macris2007}
N.~Macris, ``Griffith-kelly-sherman correlation inequalities: A useful tool in
  the theory of error correcting codes,'' \emph{IEEE Transactions on
  Information Theory}, vol.~53, no.~2, pp. 664--683, Feb 2007.

\bibitem{MacrisKudekar2009}
S.~Kudekar and N.~Macris, ``Sharp bounds for optimal decoding of low-density
  parity-check codes,'' \emph{IEEE Transactions on Information Theory},
  vol.~55, no.~10, pp. 4635--4650, Oct 2009.

\bibitem{bayati2013}
M.~Bayati, D.~Gamarnik, and P.~Tetali, ``Combinatorial approach to the
  interpolation method and scaling limits in sparse random graphs,'' \emph{The
  Annals of Probability}, vol.~41, no.~6, pp. 4080--4115, Nov 2013.

\bibitem{6284230}
A.~Giurgiu, N.~Macris, and R.~Urbanke, ``How to prove the maxwell conjecture
  via spatial coupling: a proof of concept,'' in \emph{Information Theory
  Proceedings (ISIT), 2012 IEEE International Symposium on}, July 2012, pp.
  458--462.

\bibitem{MacrisGiurgiuUrbanke2016}
------, ``Spatial coupling as a proof technique and three applications,''
  \emph{IEEE Transactions on Information Theory}, vol.~62, no.~10, pp.
  5281--5295, Oct 2016.

\bibitem{Ruelle}
D.~Ruelle, \emph{Statistical Mechanics}.\hskip 1em plus 0.5em minus 0.4em\relax
  W.A. Benjamin Inc, 1969.

\bibitem{lassoMontanariBayati}
M.~Bayati and A.~Montanari, ``The lasso risk for gaussian matrices,''
  \emph{IEEE Transactions on Information Theory}, vol.~58, no.~4, pp.
  1997--2017, April 2012.

\bibitem{wu2010renyi}
Y.~Wu and S.~Verd{\'u}, ``R{\'e}nyi information dimension: Fundamental limits
  of almost lossless analog compression,'' \emph{IEEE Transactions on
  Information Theory}, vol.~56, no.~8, pp. 3721--3748, Aug 2010.

\bibitem{Talagrand1996}
M.~Talagrand, ``A new look at independence,'' \emph{The Annals of Probability},
  vol.~24, pp. 1--34, 1996.

\bibitem{McDiarmid}
C.~McDiarmid, ``On the method of bounded differences,'' in \emph{Surveys in
  Combinatorics}, ser. London Mathematical Society Lecture Note Series, no.
  141.\hskip 1em plus 0.5em minus 0.4em\relax Cambridge University Press, Aug.
  1989, pp. 148--188.

\bibitem{boucheron2004concentration}
S.~Boucheron, G.~Lugosi, and O.~Bousquet, ``Concentration inequalities,'' in
  \emph{Advanced Lectures on Machine Learning}.\hskip 1em plus 0.5em minus
  0.4em\relax Springer, 2004, pp. 208--240.

\bibitem{Heinonen2005}
J.~Heinonen, \emph{Lectures on lipschitz analysis}.\hskip 1em plus 0.5em minus
  0.4em\relax Dept Math and Stat, University of Jyvaskyla, 2005, vol. 100.

\end{thebibliography}
\end{document}